\documentclass[pdftex,twocolumn,epjc3]{svjour3}

\RequirePackage{graphicx}   
\RequirePackage{latexsym}   
\RequirePackage{amsmath,amssymb,mathrsfs}

\RequirePackage{mathptmx}      
\RequirePackage[numbers,sort&compress]{natbib}
\RequirePackage[colorlinks,citecolor=blue,urlcolor=blue,linkcolor=blue]{hyperref}

\journalname{Eur. Phys. J. C}

\usepackage[english]{babel}
\usepackage{appendix}
\usepackage[utf8]{inputenc}
\usepackage{grffile} 
\usepackage{tabularx}
\usepackage{flushend}
\usepackage{xcolor}
\usepackage{subcaption}
\usepackage{float}

\def\snn{${\sqrt s_{\rm NN}}$}
\def\ela{E$_{\rm lab}$}
\def\mev{$\mathrm{MeV/fm^3}$}
\def\gev{$\mathrm{GeV/fm^3}$}
\def\ws{https://theory.gsi.de/~smash/hc}

\newcommand{\be}{\begin{equation}}
    \newcommand{\ee}{\end{equation}}
\newcommand{\ba}{\begin{eqnarray}}
    \newcommand{\ea}{\end{eqnarray}}

\begin{document}
\title{The applicability of hydrodynamics in heavy ion collisions\\at $\sqrt{s_{\rm NN}}$= 2.4-7.7 GeV} 
\author{Gabriele Inghirami\thanksref{emg,gsi}
    \and
Hannah Elfner\thanksref{emh,gsi,itp,fias,hfhf}
}

\thankstext{emg}{e-mail: g.inghirami@gsi.de}
\thankstext{emh}{e-mail: h.elfner@gsi.de}

\institute{GSI Helmholtzzentrum f\"ur Schwerionenforschung GmbH, Planckstr. 1, 64291 Darmstadt , Germany \label{gsi}
    \and
           Institut f\"ur Theoretische Physik, Goethe Universit\"at Frankfurt, Max-von-Laue-Straße 1, 60438 Frankfurt am Main, Germany \label{itp}
    \and
           Frankfurt Institute for Advanced Studies, Ruth-Moufang-Straße 1, 60438 Frankfurt am Main, Germany \label{fias}
    \and
           Helmholtz Research Academy Hesse for FAIR (HFHF), GSI Helmholtz Center, Campus Frankfurt, Max-von-Laue-Straße 12, 60438 Frankfurt am Main, Germany \label{hfhf}
        }

\date{\today}

\maketitle

\begin{abstract}
  To assess the degree of equilibration of the matter created in heavy-ion reactions at low to intermediate beam energies, a hadronic transport approach (SMASH) is employed. By using a coarse-graining  method, we compute the energy momentum tensor of the system at fixed time steps and evaluate the degree of isotropy of the diagonal terms and the relative magnitude of the off-diagonal terms. This study focuses mostly on Au+Au collisions in the energy range \snn = 2.4-7.7 GeV, but central collisions of lighter ions like C+C, Ar+KCl and Ag+Ag are considered as well. We find that the conditions concerning local equilibration for a hydrodynamic description are reasonably satisfied in a large portion of the system for a significant amount of time (several fm/c) when considering the average evolution of many events, yet they are rarely fulfilled on an event by event basis. This is relevant for the application of hybrid approaches at low beam energies as they are or will be reached by the HADES experiment at GSI, the future CBM experiment at FAIR as well as the beam energy scan program at RHIC. 
\end{abstract}

\section{Introduction}
Heavy ion collisions are a mesmerizing tool to deepen our understanding of how matter interacts, clumps and collectively behaves at nuclear and hadronic scales. Experimental programs, like HADES at GSI\cite{HADES:2009aat}, BES at RHIC\cite{Luo:2015doi} and the upcoming NICA at JINR\cite{Kostromin:2021rlk} and CBM at FAIR\cite{Senger:2020pzs}, aim at exploring the QCD phase diagram\cite{Guenther:2020jwe,Rischke:2003mt} in the high net baryon density region, around the critical point. With the aid of transport approaches like UrQMD\cite{Bass:1998ca,Bleicher:1999xi}, PHSD\cite{Cassing:2008sv,Cassing:2009vt} or SMASH\cite{Weil:2016zrk}, it is possible to derive phenomenological predictions from the theory that can be quantitatively tested in experiments. Hydrodynamic modeling\cite{Kolb:2000sd} has been also introduced in heavy ion collisions and, since the emerging of robust evidences\cite{BRAHMS:2004adc,PHENIX:2004vcz,PHOBOS:2004zne,STAR:2005gfr} that a medium was formed at RHIC, the quark-gluon plasma (QGP), with liquid-like properties and very low viscosity over entropy density ratio\cite{Shen:2015msa}, it has become an established framework to study the properties of the QGP\cite{Holopainen:2010gz,Pang:2016vdc,Becattini:2015ska,Karpenko:2016jyx}, constantly improved with many refinements over the years\cite{Romatschke:2009im}. Hybrid approaches combining the advantages of transport and hydrodynamic approaches have been developed, as well (see \cite{Petersen:2014yqa} for a review and references therein). After the determination of the initial state with various approaches\cite{Schenke:2012wb,Broniowski:2007nz,Moreland:2014oya}, the evolution of the fireball created in heavy ion collision is described by hydrodynamics, followed by a hadronic transport approach \cite{Schafer:2021rfk} to perform hadronic rescattering and particle decays. Recently hydrodynamics has been used also to interpret collective behavior observed at LHC in small systems like p-Pb or even p-p\cite{Song:2017wtw,Nagle:2018nvi}. In the low collision energy regime (\snn $\approx$ 2-10 \gev) 3 fluid hydrodynamics\cite{Ivanov:2005yw} has been implemented and deployed with results achieving a reasonable description of experimental data \cite{Ivanov:2020udj}. Indeed, as of today hydrodynamic modeling plays a crucial role in relativistic heavy ion research. On the theory side, strong efforts have been put in establishing the microscopic foundations of relativistic hydrodynamics\cite{Harutyunyan:2021rmb}, in particular from kinetic theory\cite{Betz:2008me,Denicol:2012cn, Molnar:2013lta}, consolidated, over the years, by comparisons with parton cascade codes~\cite{Bouras:2009nn} like BAMPS~\cite{Xu:2004mz} and investigations about the degree of consistency of the hydrodynamic approach~\cite{Bouras:2010hm} in the context of heavy ion collisions~\cite{Molnar:2004yh,Molnar:2009pq,Niemi:2014wta,Chattopadhyay:2019jqj}. 

However, most of these studies focused on high collision energies and in this article we want to partially fill this gap by further elaborating, in a broader and systematic way, the investigations presented in Ref.\cite{Oliinychenko:2015lva} in the low energy range. It is possible to properly validate a macroscopic model like hydrodynamics only if the underlying microscopic model provides a sufficiently reliable description of the natural processes on a smaller scale. Differently to the realm of heavy ion collisions at high energies, like those achieved at RHIC~\cite{Nouicer:2015jrf} or LHC~\cite{Niida:2021wut}, in which QGP medium formation is expected, in the range of energies that we consider most of the time the basic degrees of freedoms are hadrons, therefore a hadronic transport approach provides a suitable benchmark for hydrodynamic models. The aim is to quantitatively investigate how much of the system created in heavy-ion collisions is close to thermal local equilibrium. This is a prerequisite to apply hydrodynamics or hybrid approaches in the very low beam energy domain. For this purpose the SMASH (Simulating Many Accelerated Strongly-interacting Hadrons) approach is employed to assess the properties of the local energy momentum tensor. In contrast to former works examining the degree of equilibration in a rather large central cell lengths 4-5 fm ~\cite{Bravina:1998pi,Bravina:2018vgc,Bravina:2020zua}, we deal with the whole volume of the system at the nodes of a three-dimensional lattice with resolution 1 fm by using a coarse graining method based on smearing kernels. We focus mostly on Au+Au collisions at $E_{lab}$ = 1.23 AGeV in the 0-5\% centrality class, but we also explore the energy dependence up to \snn = 7.7 GeV, we look at variations as a function of centrality and we analyze light and medium ion collisions like C+C, Ar+KCl and Ag+Ag, again in the 0-5\% centrality class and with beam energy between 1.58 and 2 AGeV. The choice of the systems and energies matches the corresponding experiments, mainly by the HADES collaboration.\\
The structure of this article is the following: after an introduction to the SMASH transport approach, we illustrate the criterion that will be used to assess the degree of thermalization of the system, namely the pressure anisotropy and the off-diagonality of the energy momentum tensor in the Landau frame\cite{LandauFluidMechanics2013}. Then, we present and discuss the results. In most cases, we evaluate the quantities by using the average energy momentum tensor of all events together, but we also consider, in a few cases, the final average of the outcomes obtained by evaluating the energy momentum tensor of the system on an event by event basis. To get a simple overall idea of the situation, first we focus on the time evolution of the energy density, the pressure anisotropy and the off-diagonality in the central point of the system, then along the planes parallel to the axes of the frame of reference passing through the center. Afterwards, we display a few two-dimensional histograms of the distributions of the same quantities at a fixed time to provide an overview of their typical values. We continue by showing the time evolution of the volume of the system satisfying a set of constraints based on the aforementioned quantities and its ratio with respect to the total system volume. In the next step, we display some computations of time integrals of the volumes satisfying a few sets of constraints, more extensively summarized in the appendix. We also present some basic topological characteristics of these volumes for a specific case. Next, we summarize our findings in the conclusions. In the appendix, in addition to providing a broader overview of the results organized in several tables, we make basic assessments about the impact of spectator hadrons, the use of test particles and the differences and similarities to former UrQMD results.\\
For obvious reasons of brevity, this article displays only a small subset of the results obtained in the study, with thousands of plots, which are currently available in full on the website \url{\ws}\footnote{As time passes, the web address might change or the data not be available anymore.}, together with a description of the workflow to reproduce them, also included in the ancillary files of the arXiv pre-print.

\section{Methods}
\subsection{SMASH}
SMASH\cite{Weil:2016zrk} is a transport approach that simulates the time evolution of a system of strongly interacting hadrons by using a semi-classical approach, i.e. the hadrons are treated as point-like particles following deterministic trajectories, but they scatter according to cross sections obtained from experimental data or calculated from effective field theories. However, even if a complete implementation of the spin degree of freedom is not yet available, quantum features like nuclear Fermi motion or Pauli blocking are included in the model. SMASH includes almost all known hadrons and resonances up to a rest mass of 2.4 GeV\cite{Zyla:2020zbs}, it takes into account resonance excitations and decays and it handles, by exploiting Pythia\cite{Sjostrand:2014zea,Sjostrand:2006za}, color string formation and fragmentation. The reliability of SMASH is supported by semi-analytical tests\cite{Schafer:2019edr,Hammelmann:2018ath}, by comparison with other codes\cite{Colonna:2021xuh,Zhang:2017esm} and with experimental data\cite{Staudenmaier:2017vtq}.\\
For this work we use SMASH version 2.0.2\cite{dmytro_oliinychenko_2020_4336358} in the cascade mode and we compute the energy momentum tensor in the Landau frame\cite{LandauFluidMechanics2013} at the nodes of a regularly spaced cubic lattice aligned with the beam axis and the reaction plane, with a side length of 75 fm, center coincident with that of the colliding system, space resolution of 1 fm and time intervals of 0.5 fm. The energy momentum tensor $T^{\mu\nu}$ and the four currents $j^{\mu}$ at the position $\mathbf{r}$ are evaluated as:
\begin{equation}
    T^{\mu\nu}=\sum_i \dfrac{p_i^{\mu}p_i^{\nu}}{p_i^0}K(\mathbf{r}-\mathbf{r}_i,\mathbf{p_i})
    \label{eq:Tmunu}
\end{equation}
and 
\begin{equation}
    j^{\mu}=\sum_i \dfrac{p_i^{\mu}}{p_i^0}K(\mathbf{r}-\mathbf{r}_i,\mathbf{p_i}),
    \label{eq:jmu}
\end{equation}
where K is a smearing kernel given by the following expression\cite{Oliinychenko:2015lva}:
\begin{equation}
    K(\mathbf{r}-\mathbf{r}_i,\mathbf{p_i})=\dfrac{\gamma_i}{(2\pi\sigma^2)^{3/2}}\exp\left(-\dfrac{(\mathbf{r}-\mathbf{r}_i)^2+((\mathbf{r}-\mathbf{r}_i)\cdot \mathbf{u})^2}{2\sigma^2}\right)
    \label{eq:K}
\end{equation}
and $p_i^\mu$, $\gamma_i$, $\mathbf{u}_i$, $\mathbf{p_i}$ and $\mathbf{r_i}$ are respectively the four momentum components, the Lorentz factor, the three velocity, the momentum and the spatial position of the $i^{th}$ hadron in the computational frame, while $\sigma$ is a parameter that tunes the smearing of the kernel. Indeed, different values of $\sigma$ can have a significant impact on the results\cite{Oliinychenko:2015lva}, nevertheless we believe that SMASH default $\sigma$ = 1 fm for all particles represents a reasonable choice, given the typical hadron finite size and our lattice resolution. To reduce the computational time, SMASH introduces a cut-off distance and the smearing kernels are set to 0 if $\mathbf{r}-\mathbf{r}_i>4\sigma$.
In this work we neglect the effects of nuclear potentials, therefore the particles contribute to the dynamical evolution of the macroscopic properties of the systems only through elastic and inelastic collisions. Since hydrodynamics only concerns the hot and dense collision region, it makes sense to exclude from the summations in Eqs.~(\ref{eq:Tmunu}) and (\ref{eq:jmu}) the hadrons that do not influence this evolution, i.e. the particles that never collided with other particles, the so called ``spectators'', and include only ``participants'', identified as those particles who had at least one collision with another particle from the beginning of the simulation until the moment of the $T^{\mu\nu}$ evaluation. In \ref{sec:spectators} we provide a limited assessment of the impact of the spectators on the energy momentum tensor components.

\subsection{Conditions for hydrodynamics}
As we already mentioned in the introduction, it is beyond the scope of this article to investigate and elucidate the theoretical foundations of hydrodynamics. As in Ref.~\cite{Oliinychenko:2015lva}, we adopt a practical approach and we evaluate the magnitude of the quantities 
\begin{equation} \label{eq:press_aniso}
    X \equiv \frac{|\langle T^{11}_L \rangle - \langle T^{22}_L\rangle| + |\langle T^{22}_L\rangle - \langle T^{33}_L\rangle| + |\langle T^{33}_L\rangle - \langle T^{11}_L\rangle|}{\langle T^{11}_L\rangle + \langle T^{22}_L\rangle + \langle T^{33}_L\rangle}, 
\end{equation}
\begin{equation} \label{eq:off_dia}
    Y \equiv \frac{3(|\langle T^{12}_L\rangle| + |\langle T^{23}_L\rangle| + |\langle T^{13}_L\rangle|)}{\langle T^{11}_L\rangle + \langle T^{22}_L\rangle + \langle T^{33}_L\rangle},
\end{equation}

\begin{equation} \label{eq:press_aniso_ebe}
    X_{ebe} \equiv \Bigg \langle \frac{|T^{11}_L - T^{22}_L| + |T^{22}_L - T^{33}_L| + |T^{33}_L - T^{11}_L|}{T^{11}_L + T^{22}_L + T^{33}_L} \Bigg \rangle, 
\end{equation}
\begin{equation} \label{eq:off_dia_ebe}
    Y_{ebe} \equiv \Bigg \langle \frac{3(|T^{12}_L| + |T^{23}_L| + |T^{13}_L|)}{T^{11}_L + T^{22}_L + T^{33}_L} \Bigg \rangle,
\end{equation}
during the time evolution of the system, up to t = 40 fm, in which the averages, denoted by the angle brackets $\langle\ldots\rangle$, are taken over the events. The ``L'' subscripts in Eqs.\ref{eq:press_aniso}-\ref{eq:off_dia_ebe} serve as a reminder that the energy momentum tensor $T^{\mu\nu}_L$ is calculated in the Landau frame~\cite{LandauFluidMechanics2013}, but for the sake of brevity we will not use them further. To avoid issues with low energy density regions, we consider only the lattice points in which the denominators of these quantities are larger than 0.1 \mev, a condition approximately corresponding to requiring a positive pressure.\\
We clarify that when we write $X$ (Eq.~\ref{eq:press_aniso}) and $Y$ (Eq.~\ref{eq:off_dia}), we refer to the evaluation of these quantities from the average energy momentum tensor of a series of events, while, when we write $X_{ebe}$ (Eq.~\ref{eq:press_aniso_ebe}) and $Y_{ebe}$ (Eq.~\ref{eq:off_dia_ebe}), we refer to the average of these quantities computed on an event by event basis. $X$ and $X_{ebe}$, called \emph{pressure anisotropy}, measure the degree of pressure isotropisation of the system, while $Y$ and $Y_{ebe}$, called \emph{off-diagonality}, measure the relative magnitude between the spatial off-diagonal and the diagonal components of the energy momentum tensor.\\
If the system is in local thermal equilibrium then $X, Y \simeq 0$ and we can decompose the energy momentum tensor into:
\begin{equation}
    T^{\mu\nu}=\varepsilon u^\mu u^\nu -p(g^{\mu\nu}-u^\mu u^\nu),
    \label{eq:lte}
\end{equation}
where $\varepsilon$ and $p$ are the energy density and the pressure in the local rest frame, respectively, $g^{\mu\nu}$ the metric tensor (with [+,-,-,-] signature) and $u^{\mu}$ the fluid four-velocity components. By using an Equation of State to relate $p$, $\varepsilon$ and possibly the density $\rho$ of a conserved current $j^\mu$ and imposing conservation laws $\partial_\mu T^{\mu\nu}=0$ and $\partial_\mu j^\mu=0$ one gets a closed system of differential equations that constitutes the basis of ideal hydrodynamics\cite{Romatschke:2009im}. When the condition of local thermal equilibrium is relaxed, i.e. $X$ and $Y$ are allowed to be small, but non negligible, the decomposition in Eq.(\ref{eq:lte}) must include additional terms like the shear-stress tensor $\pi^{\mu\nu}$ or, especially in the early phase of high energy ion collisions, distinguish between longitudinal and transverse pressure\cite{Florkowski:2010cf,Alqahtani:2017mhy}. This problem has been already addressed by developing causally consistent and stable approaches\cite{Israel:1976tn,Israel:1979wp}, successfully implemented in various viscous hydrodynamics codes\cite{Song:2007ux,Pang:2014ipa,Schenke:2010rr,Karpenko:2013wva,DelZanna:2013eua}. Nevertheless, the conceptual foundation of the derivation of viscous hydrodynamics from the microscopic kinetic theory relies on series expansions\cite{Betz:2008me,Denicol:2012cn, Molnar:2013lta} that assume moderate deviations from the condition of local thermal equilibrium. 
Adding more terms in the expansion series can improve the accuracy of the model, but it also significantly increases the complexity of numerical codes and their execution time, it requires the knowledge of more transport coefficients and, anyway, it has intrinsic limitations\cite{ Heller:2013fn}. It is very hard to define a precise limit on the magnitude of $X$ ($X_{ebe}$) and $Y$ ($Y_{ebe}$) to decide whether hydrodynamics can be safely applied\cite{Romatschke:2016hle}, as it strongly depends on the details of the hydrodynamic approach and its implementation, so in this study we evaluate how large the volume of the system under a certain threshold is. More precisely, we explore $X,Y< 0.1, 0.3$ and $0.5$ \footnote{The results on \url{\ws} include also 0.2.}, focusing mostly on the intermediate case 0.3, but without endorsing this choice as a ``safe'' or ``limiting'' one. As a general statement, the smaller the threshold the closer to equilibrium is the system, while for larger deviations a significant difference between a hydrodynamic evolution and the SMASH calculation that is valid off equilibrium is expected. Given the importance of the energy density, from now on called $\varepsilon$, for the equation of state~\cite{Borsanyi:2010cj,Bazavov:2009zn,Huovinen:2009yb,Vovchenko:2020lju} and the particlization process~\cite{Vovchenko:2020kwg,McNelis:2019auj,Schwarz:2017bdg}, most of the time we introduce also three energy density thresholds and we restrict our analysis to the regions for which $\varepsilon\ge 1, 100, 500$ \mev.

\section{Results}
Before entering the discussion of the results, let us start with some technical remarks to explain in detail our setup for the calculation. 
For the results that we are going to present we use the center of momentum frame of the colliding ions, with orthogonal Cartesian coordinates x, y and z. The axes x and z span the reaction plane\cite{Voloshin:2008dg}, with z parallel to the beam direction. We use natural units, distance and time are expressed in fm, while energy density and pressure are expressed in \gev or \mev. SMASH sets time t=0 fm at the moment in which two Lorentz contracted spherical nuclei would touch in a perfectly central collision\cite{Weil:2016zrk}. The various quantities are evaluated at the nodes of a lattice with 1 fm uniform spacing along all three spatial directions. When dealing with volumes, we assume that the values of the quantities remain constant within the cubic cells of size 1 $\mathrm{fm^3}$ centered around the lattice points. When integrating in space, we sum up the values at the centers of the cells ($dx=dy=dz=1$ fm), when integrating in time we sum up and multiply by the time interval ($dt=0.5$ fm).\\
For each type of configuration, we perform 1080 events\footnote{This number, 1080, instead of 1000 ($10^3$) or 1024 ($2^{10}$), is due to technical reasons related to available RAM and CPUs.}, a sort of trade-off between stability of the results\cite{Oliinychenko:2015lva} and time needed to run the simulations. In most cases we focus on the centrality class 0-5\%, which is what we mean with ``central'' collisions.

\subsection{Conditions in the center of the system}
\label{sec:conditions_at_center}
Before analyzing the properties of the whole system, it is useful to look at the time evolution in its central region, at coordinates x=y=z=0. The idea is to validate our methodology by looking at rather well known features like the time evolution of the energy density in the center of the fireball. Also, the middle of the system contains the highest density of particles and therefore more homogeneous conditions. 

\subsubsection{Au+Au vs collision energy}
\begin{figure}[ht!]
    \includegraphics[width=\columnwidth]{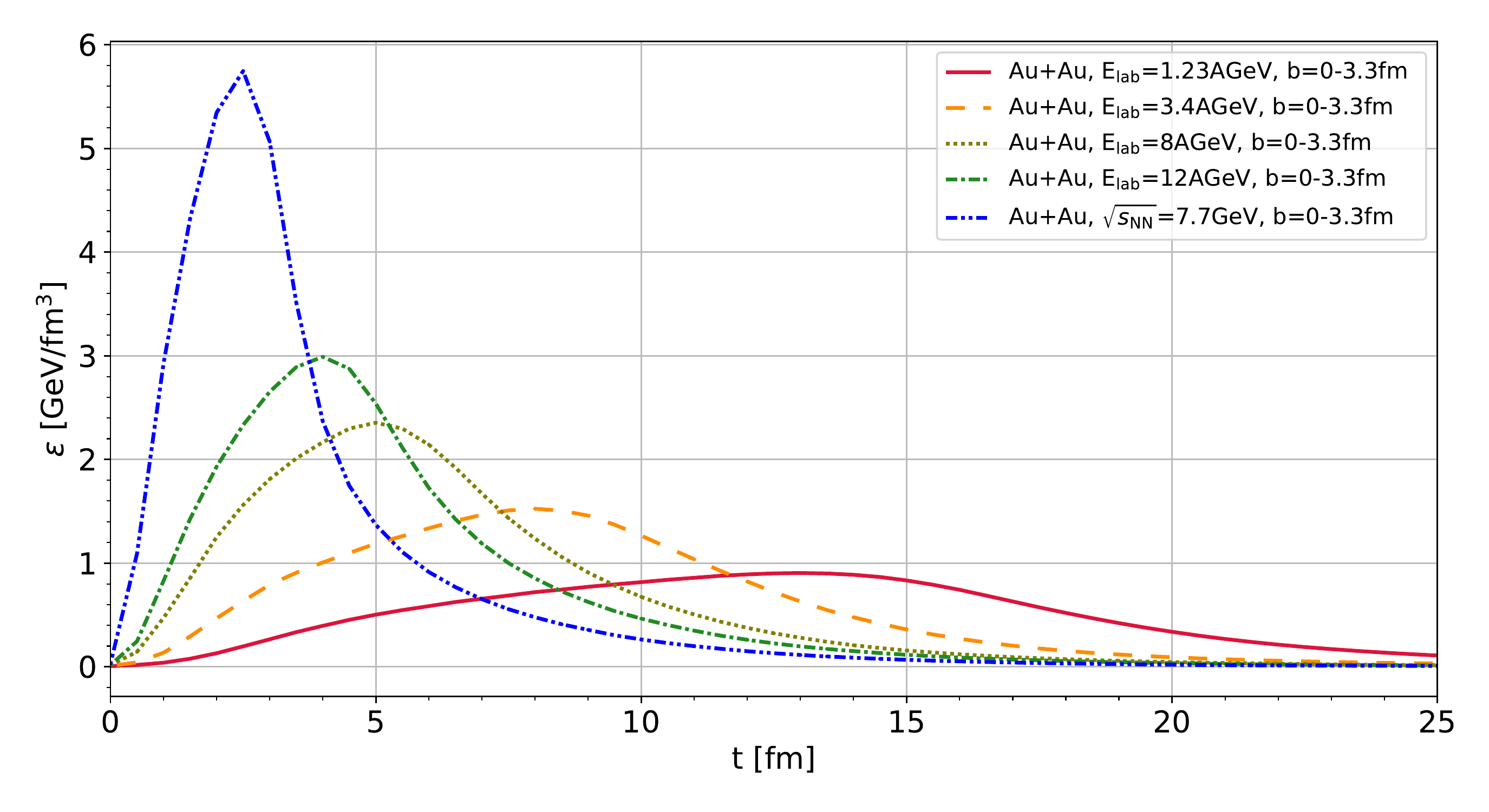}
    \caption{Time evolution of the energy density in the center of the system for Au+Au collisions, 0-5\% centrality class, at \ela = 1.23, 3.4, 8, 12 AGeV and \snn = 7.7 GeV.}
    \label{fig:edens_evo_center_Au_snn_dep}
\end{figure}

\begin{figure}[ht!]
    \includegraphics[width=\columnwidth]{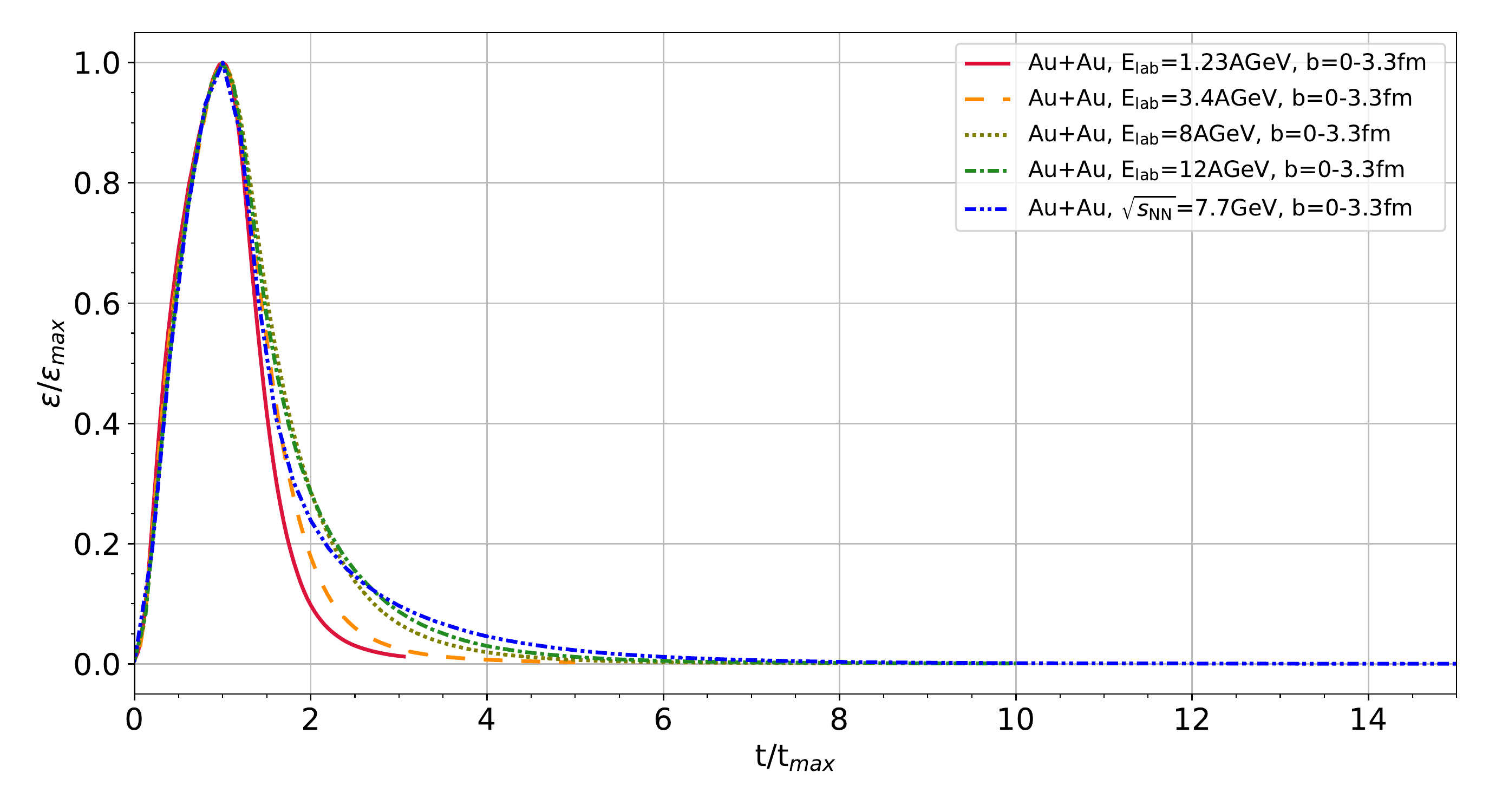}
    \caption{Time evolution of the energy density in the center of the system for Au+Au collisions, 0-5\% centrality class, at \ela = 1.23, 3.4, 8, 12 AGeV and \snn = 7.7 GeV, normalized to the value at its maximum and rescaled to the corresponding time.}
    \label{fig:edens_evo_center_Au_snn_dep_scaled}
\end{figure}

\begin{figure}[ht!]
    \includegraphics[width=\columnwidth]{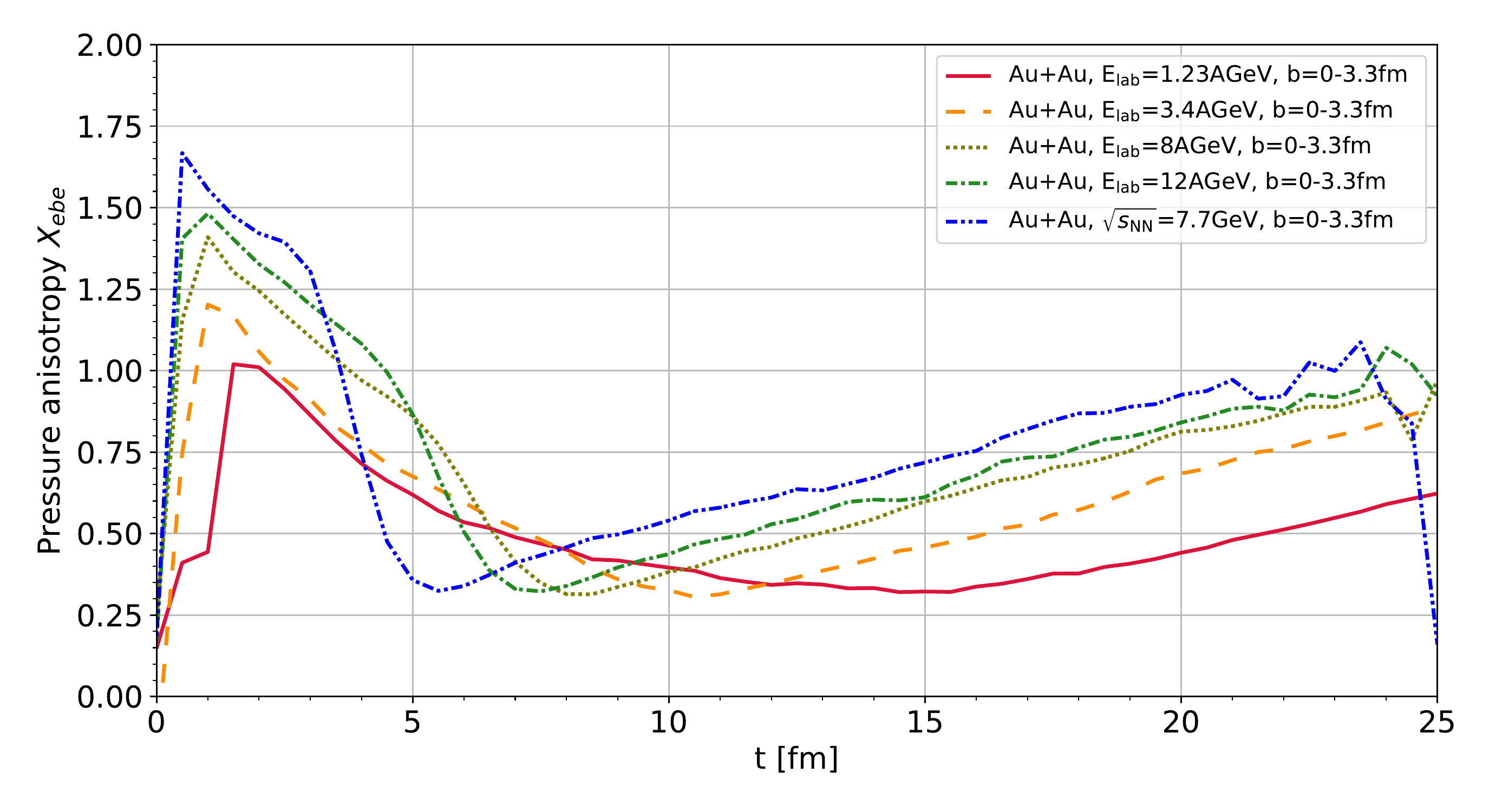}\\
    \includegraphics[width=\columnwidth]{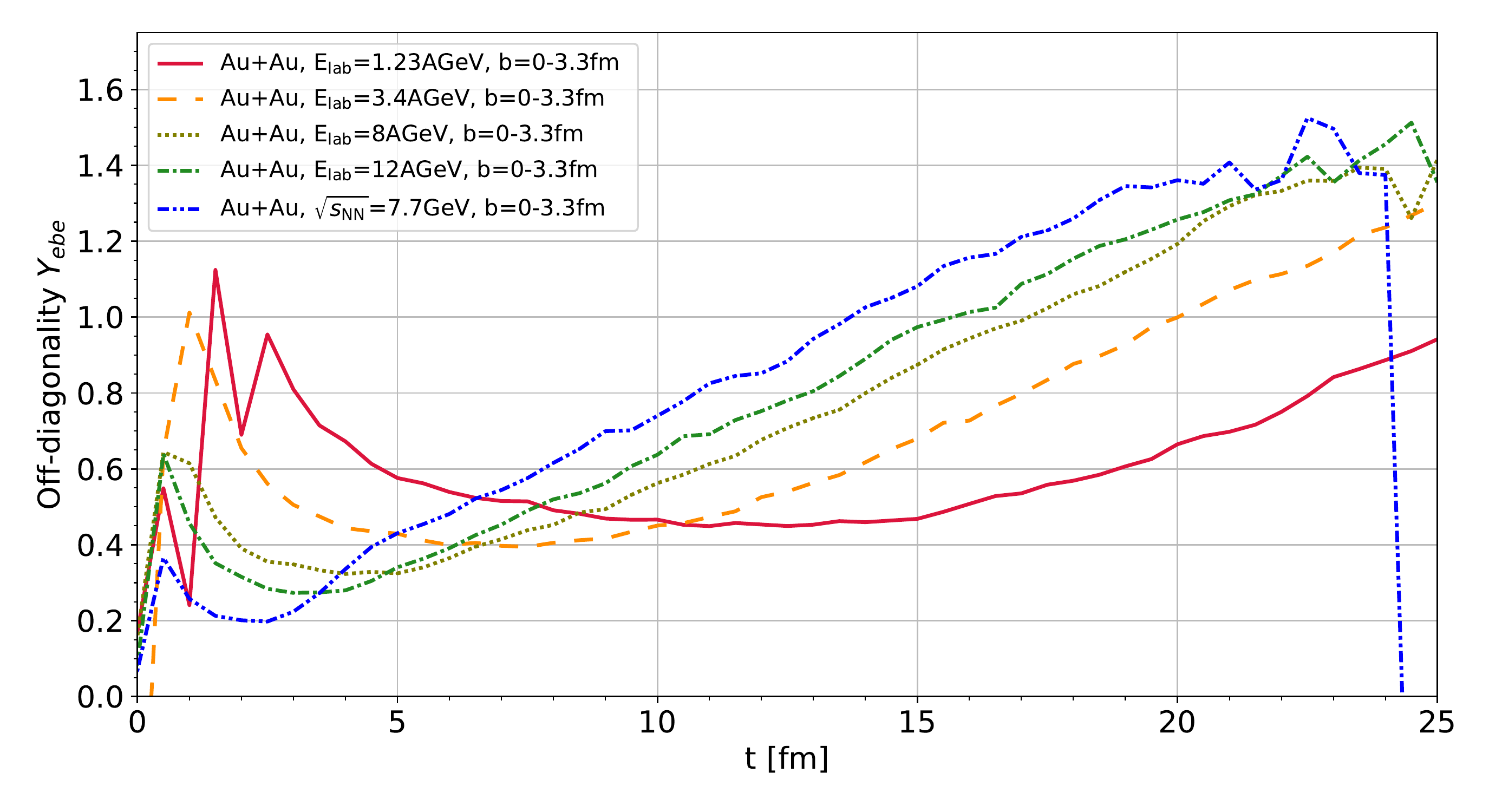}
    \caption{Time evolution of the pressure anisotropy $X_{ebe}$ (upper plot) and off-diagonality $Y_{ebe}$ (lower plot) in the center of the system. The plots refer to Au+Au collisions, 0-5\% centrality class, at \ela = 1.23, 3.4, 8, 12 AGeV and \snn = 7.7 GeV. }
    \label{fig:XY_evo_ebe_Au_Au}
\end{figure}

\begin{figure}[ht!]
    \includegraphics[width=\columnwidth]{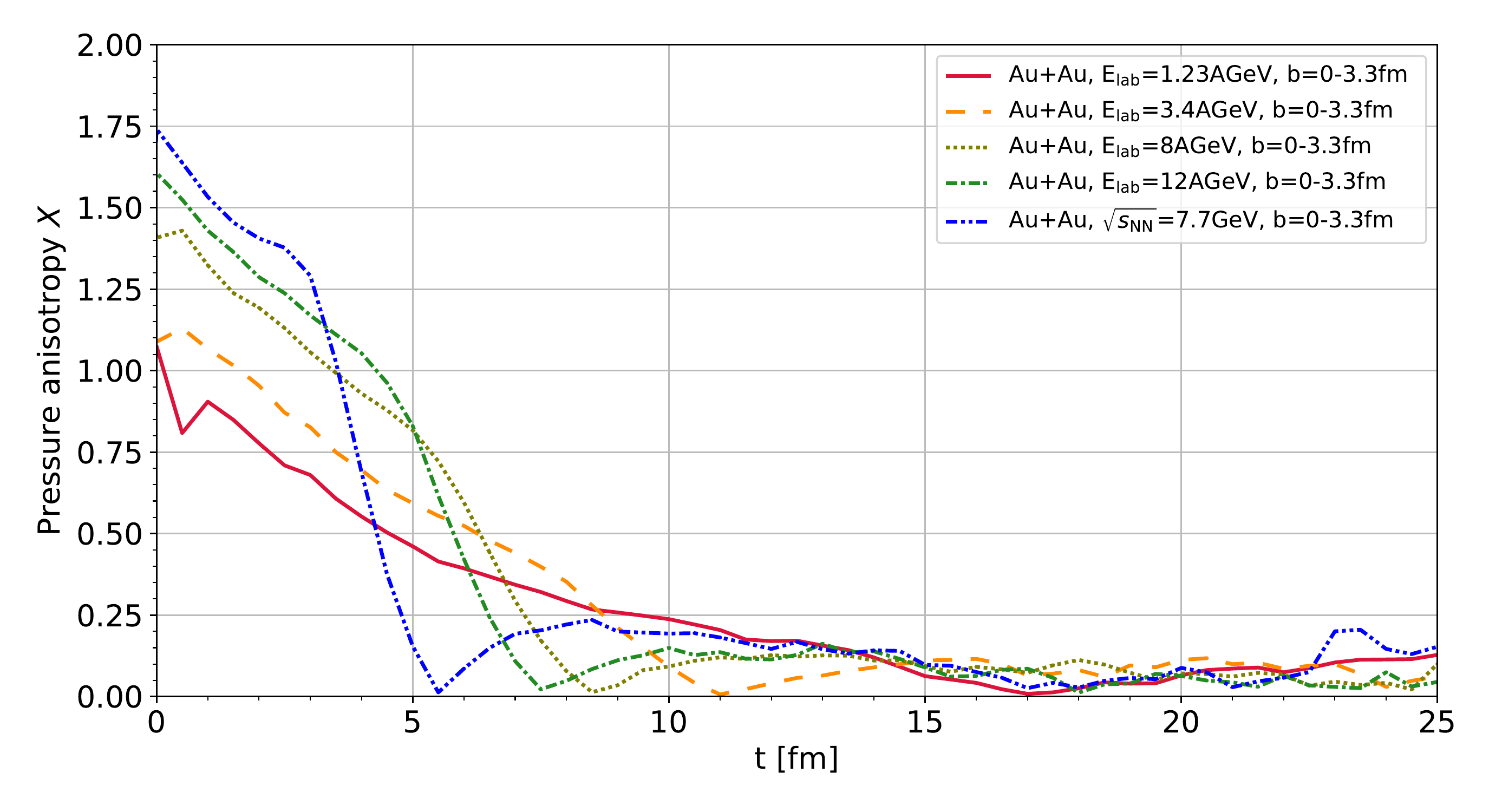}\\
    \includegraphics[width=\columnwidth]{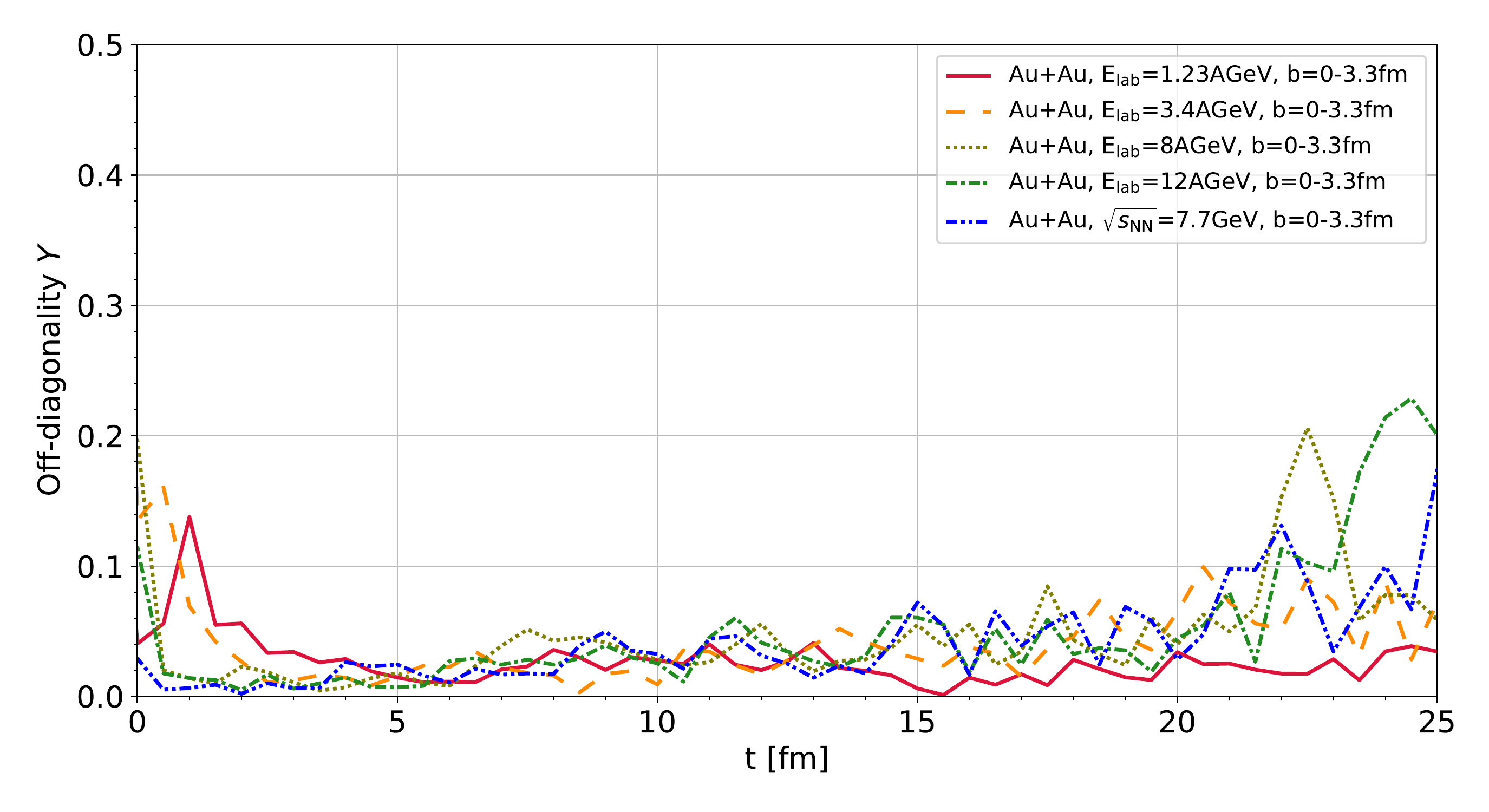}
    \caption{Time evolution of the pressure anisotropy $X$ (upper plot) and off-diagonality $Y$ (lower plot) in the center of the system. The plots refer to Au+Au collisions, 0-5\% centrality class, at \ela = 1.23, 3.4, 8, 12 AGeV and \snn = 7.7 GeV. }
    \label{fig:XY_evo_avgT_Au_Au}
\end{figure}

 As a first step, we evaluate the time evolution of the energy density in the center of the system for Au+Au ions at different collision energies in the 0-5\% centrality class (impact parameter $b$ within the range 0-3.3 fm\cite{HADES:2017def}), displayed in Fig.~\ref{fig:edens_evo_center_Au_snn_dep}. As expected, not only the maximum of the energy density gets higher with increasing collision energy, but the dynamics of the system gets faster, with maxima at earlier times followed by a steeper decrease.
 However, as shown in Fig.~\ref{fig:edens_evo_center_Au_snn_dep_scaled}, if we consider the maximum value of the energy density $\varepsilon_{max}$ and the corresponding time $t_{max}$ and we rescale the plot by using these quantities, we observe that the scaled dynamics of the various systems is approximately the same until it reaches the peak, while at later times it is more differentiated. This behavior suggests that the time evolution of the energy density of the system at its center during the compression phase is driven by the energy deposit of the participant nucleons, which on average are approximately the same in all cases, while during the subsequent expansion phase is determined by the chemical composition of the system, the multiplicity of the particles and their cross sections, which are non linearly dependent on the collision energy and significantly change across the five cases under examination.
 The average pressure anisotropy $X_{ebe}$, shown in the upper half of Fig.~\ref{fig:XY_evo_ebe_Au_Au}, is rather large in first fm/c, it approaches the maximum limit of 2 for very high collision energies (due to the fact that initially the longitudinal pressure dominates over the transverse pressure), then it decreases to a local minimum. The off-diagonality $Y_{ebe}$, shown in the lower half of  Fig.~\ref{fig:XY_evo_ebe_Au_Au}, follows a similar pattern, but with a smaller initial peak and a more rapid growth at later times, when the system becomes very diluted. We restrict the displayed time evolution to t = 25 fm, since afterwards the system is too dilute to obtain meaningful results. In both cases for $X_{ebe}$ and $Y_{ebe}$ we observe that higher collision energies have a larger initial peak and smaller values at late evolution times than lower energies. We remind that we accept as part of the system only the points in which $\sum_{i=1..3} T^{ii}>0.1$ \mev, therefore the denominator in $X_{ebe}$ and $Y_{ebe}$ is always positive. The absolute values at the numerators of these quantities ensure that $X_{ebe}$ and $Y_{ebe}$ are non negative and they also explain some abrupt changes of directions in the plotted curves when approaching the x axis. The same applies for $X$ and $Y$ (Fig.~\ref{fig:XY_evo_avgT_Au_Au}), whose time evolution exhibits a behavior similar to Fig.~\ref{fig:XY_evo_ebe_Au_Au} in its first phase, but strikingly different after the system is somehow on average equilibrated, with values that always remain relatively small for the whole time interval under examination. However, as the collision energy grows, the lower half of Fig.~\ref{fig:XY_evo_ebe_Au_Au} reveals a tendency to leave the equilibrium phase  earlier, which is signaled by increasing values of $Y$. The faster dynamics of the system leads to a faster dilution.\\
 In the rest of the paper we will focus mostly on $X$ and $Y$ because they provide more stable results and better conditions for the applicability of hydrodynamics. Since hydrodynamics is a theory of macroscopic averages, it is consistent to also look at the transport dynamics as a coarse-grained average.

\subsubsection{Au+Au vs centrality}

\begin{figure}[ht!]
    \includegraphics[width=\columnwidth]{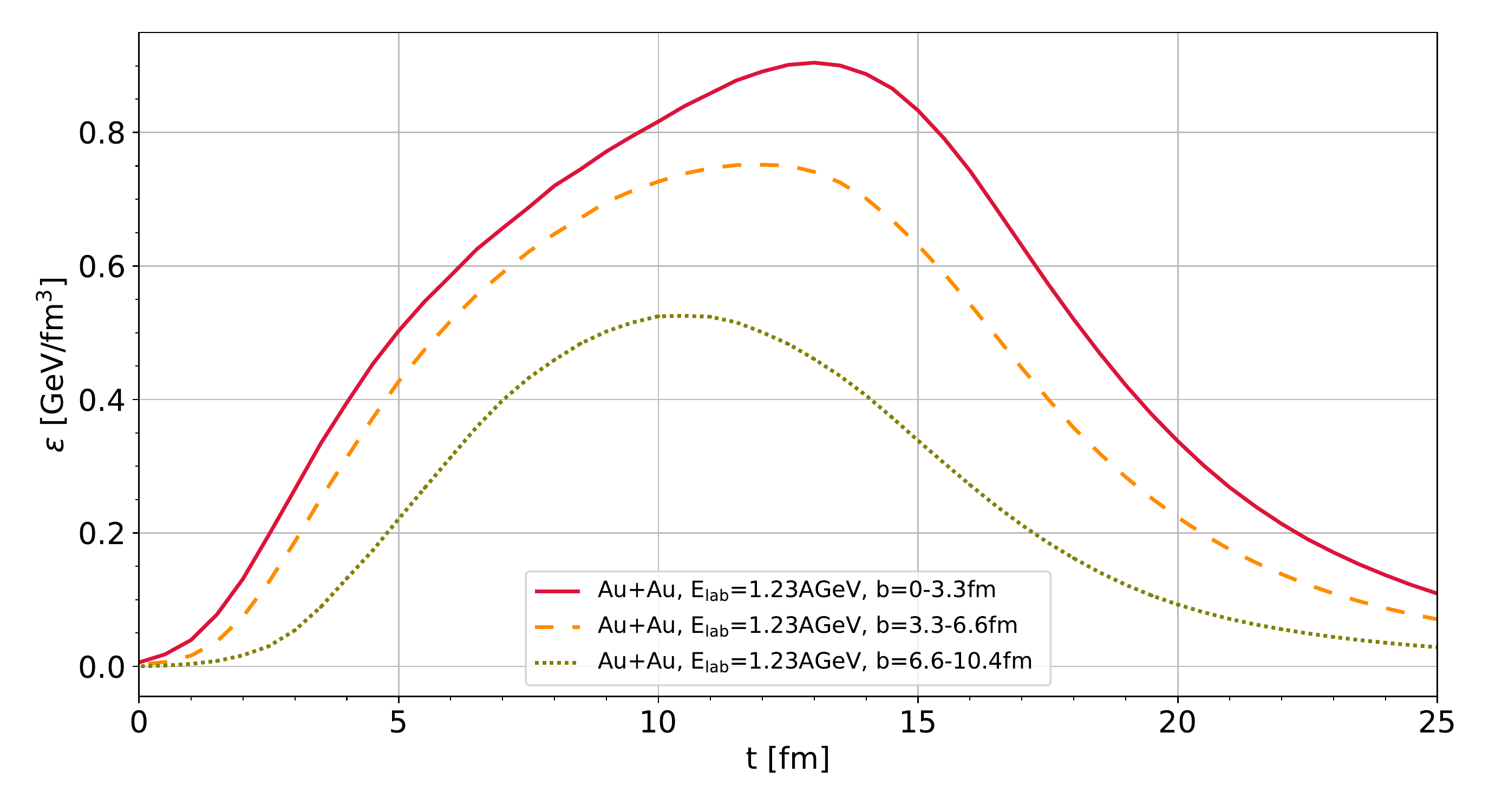}
    \caption{Time evolution of the energy density in the center of the system for Au+Au collisions at \ela = 1.23 AGeV, 0-5\%, 5-20\% and 20-50\% centrality classes.}
    \label{fig:edens_evo_Au_dep}
\end{figure}

\begin{figure}[ht!]
    \includegraphics[width=\columnwidth]{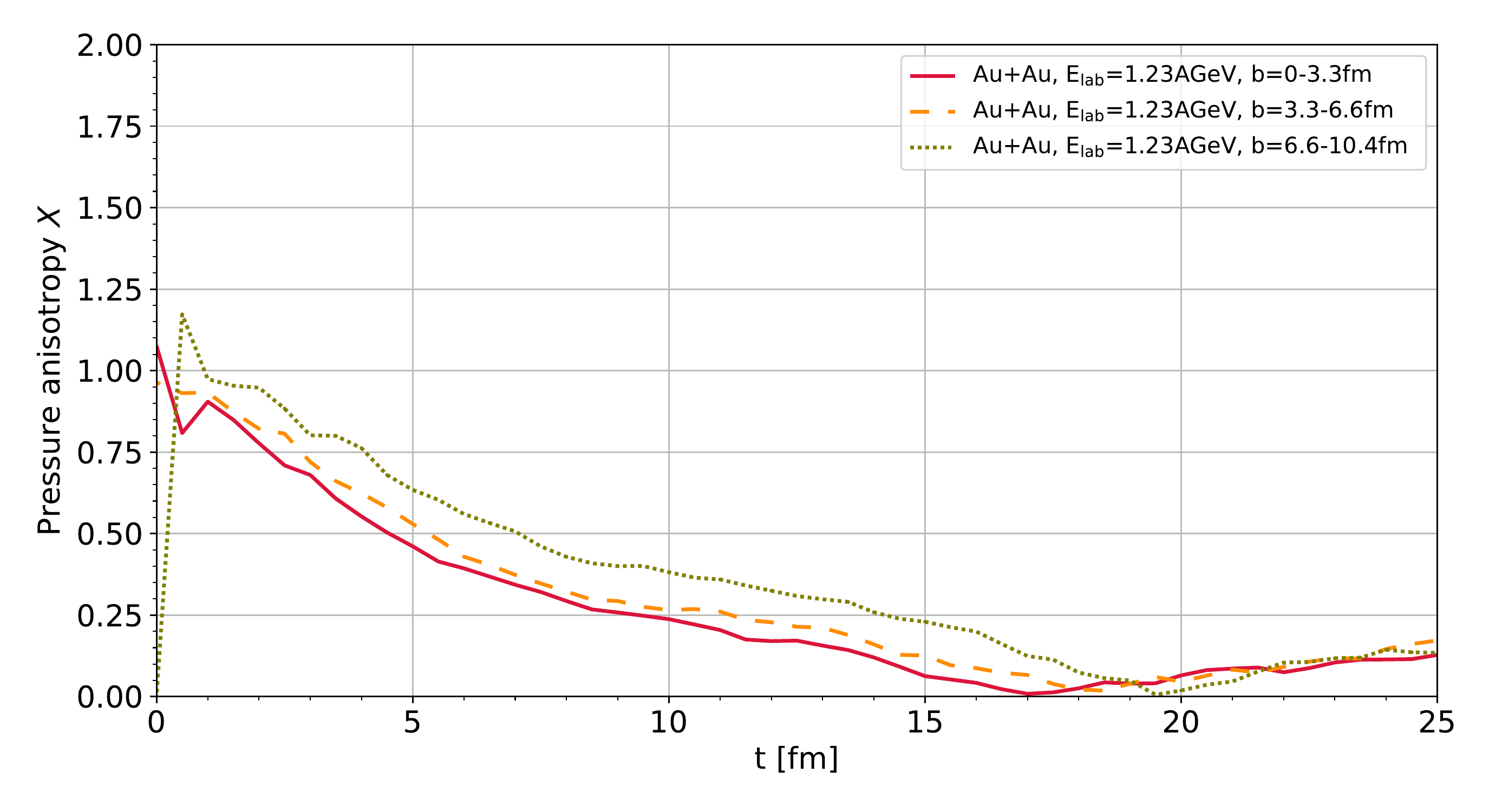}\\
    \includegraphics[width=\columnwidth]{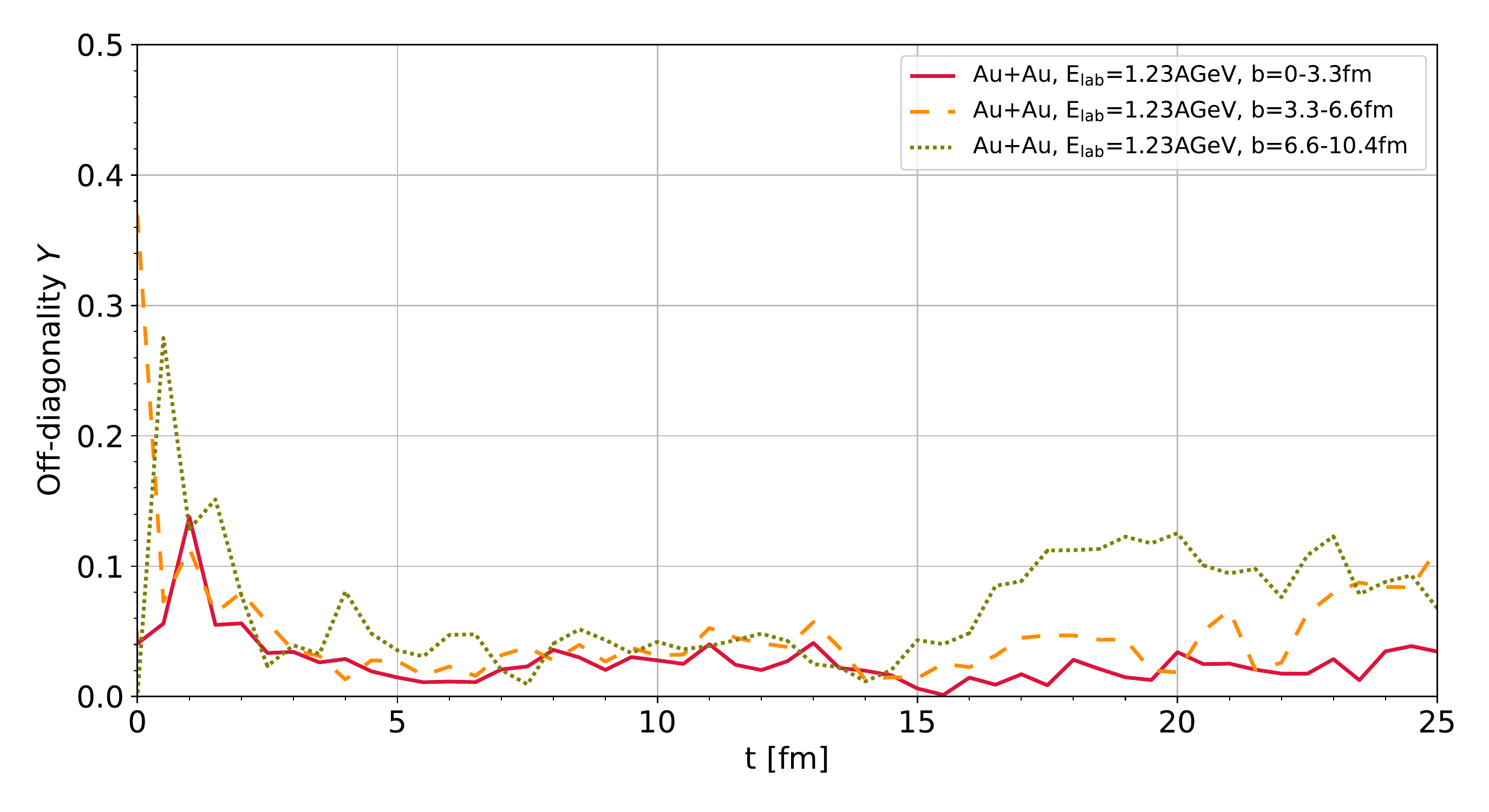}
    \caption{Time evolution of the pressure anisotropy $X$ and  of the off-diagonality $Y$ in the center of the system for Au+Au collisions at \ela = 1.23 AGeV, 0-5\%, 5-20\% and 20-50\% centrality classes.}
    \label{fig:XY_evo_avgT_Au_dep}
\end{figure}

The present study focuses on central collisions, nevertheless it is useful to include a basic evaluation of the role of centrality. We limit our assessment to Au+Au collisions at \ela = 1.23 AGeV and, in  addition to the 0-5\% centrality class, we consider the 5-20\%  and 20-50\% centrality classes (impact parameter $b$ within the range 3.3-6.6 fm and 6.6-10.4 fm, respectively\cite{HADES:2017def}). In Fig.~\ref{fig:edens_evo_Au_dep} we show the time evolution of the energy density in the center of the system for the three centrality classes under examination. We notice that the magnitude of the energy density decreases with increasing impact parameter, as it is naturally expected given the diminishing number of participant nucleons in the overlapping region. Fig.~\ref{fig:XY_evo_avgT_Au_dep} shows the time evolution of $X$ (upper half) and $Y$ (lower half), which do not reveal significant differences in their magnitude. In the case of peripheral collisions, we observe a slightly delayed evolution of the system in the pressure anisotropy due to geometric reasons and a tendency to depart earlier from equilibrium in the off-diagonality due to the formation of a smaller system.

\subsubsection{Au+Au vs C+C, Ar+KCl, Ag+Ag}
\begin{figure}[ht!]
    \includegraphics[width=\columnwidth]{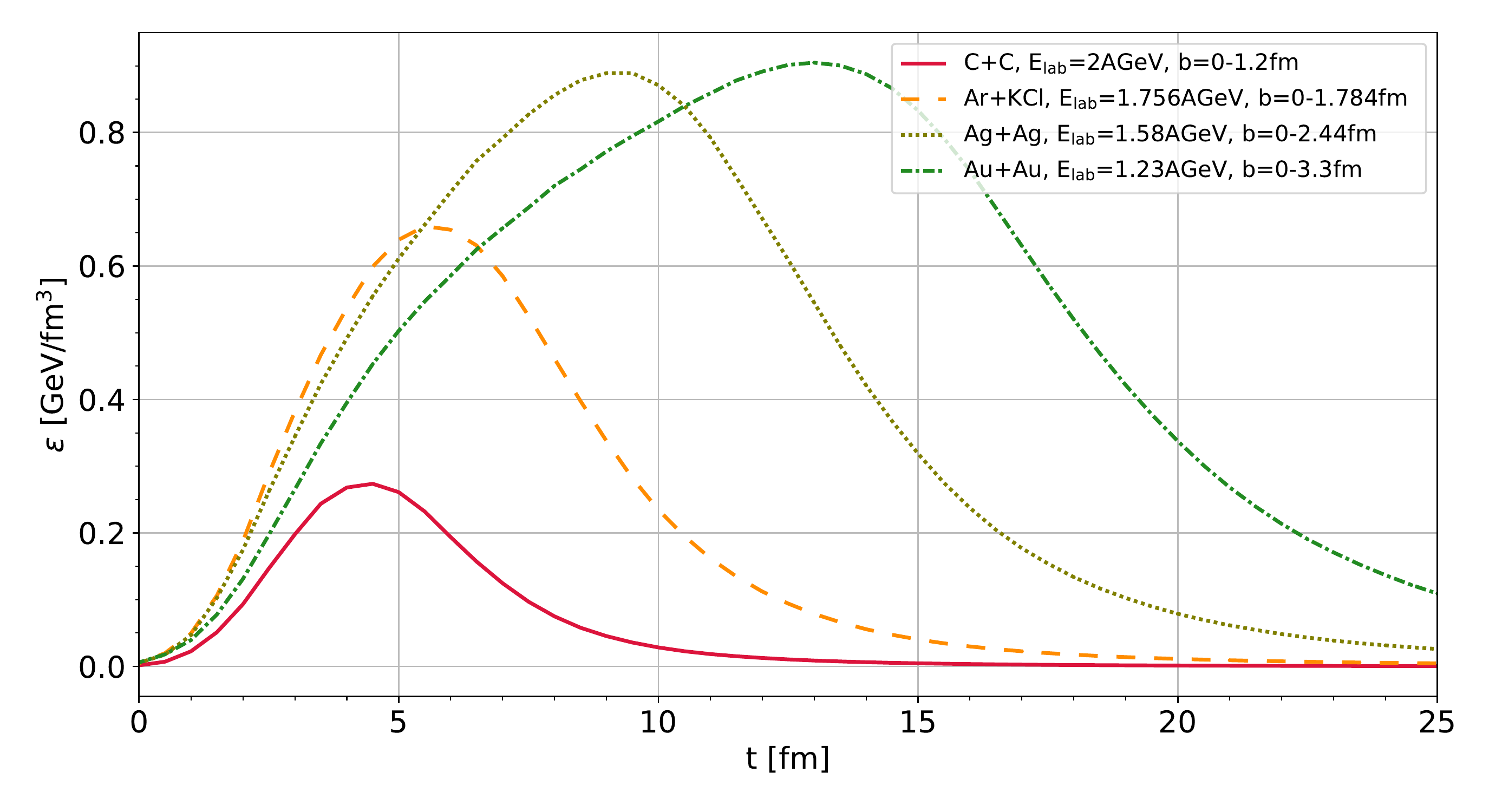}
    \caption{Time evolution of the energy density in the center of the system for the reactions listed in Table ~\ref{table:bE}.}
    \label{fig:edens_evo_center_lh}
\end{figure}

We now compare Au+Au collisions with other reactions studied in experiments by the HADES collaboration: C+C~\cite{HADES:2009jwc}, Ar+KCl~\cite{Staudenmaier:2020xqr} and Ag+Ag~\cite{HADES:2009mtt}. It is expected that smaller systems will be further away from thermal equilibrium, because of the reduced number of binary collisions, that constitute the mechanism to redistribute energy and momentum. We perform this comparison among different systems for the 0-5\% centrality class, determining the corresponding impact parameters for the light ions by using an optical Glauber model calculation, more specifically by exploiting a Mathematica worksheet written by Klaus Reygers. The parameters for these series of simulations are listed in Table~\ref{table:bE}.\\
\begin{table}[h!]
    \centering
    \begin{tabular}{|c|c|c|} 
        \hline
        Ions & b (fm) & \ela\\ 
        \hline\hline
        C+C & 0-1.2 & 2 AGeV\\ 
        Ar+KCl & 0-1.784 & 1.58 AGeV\\
        Ag+Ag & 0-2.44 & 1.756 AGeV\\
        Au+Au & 0-3.3 & 1.23 AGeV\\
        \hline
    \end{tabular}
    \caption{Impact parameters and beam energies adopted in the comparisons between Au+Au and light ion reactions.}
    \label{table:bE}
\end{table}
As in Ref.~\cite{Staudenmaier:2020xqr}, we model the collision between a nucleus of Ar and a molecule of KCl as a collision between two nuclei of Ar$^{37}$.
Fig.~\ref{fig:edens_evo_center_lh} shows that the energy density in the center of the system during its time evolution reaches lower magnitudes in smaller systems like Ar+KCl and especially C+C, in which never exceeds 300 \mev. 

\begin{figure}[ht!]
    \includegraphics[width=\columnwidth]{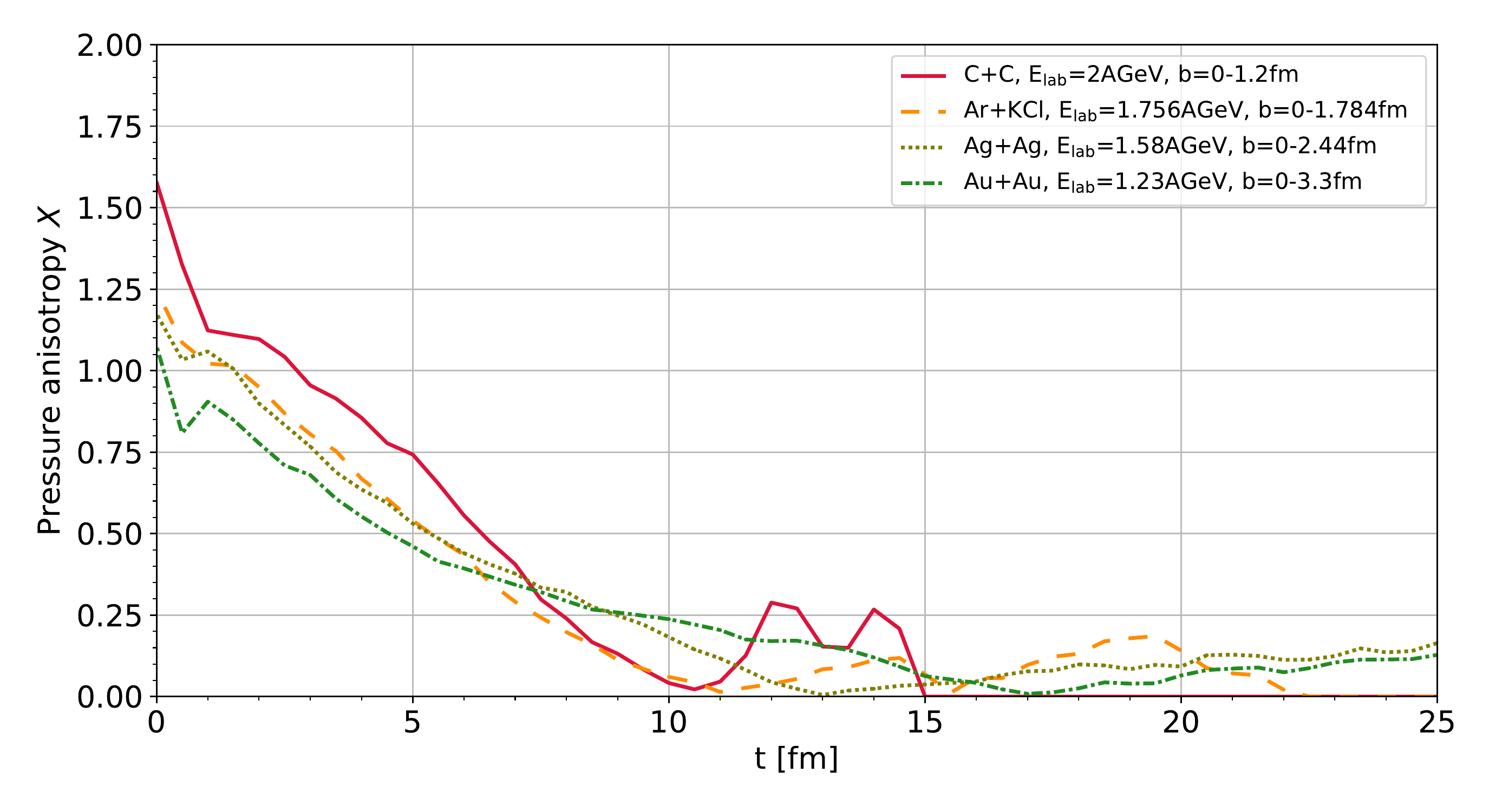}\\
    \includegraphics[width=\columnwidth]{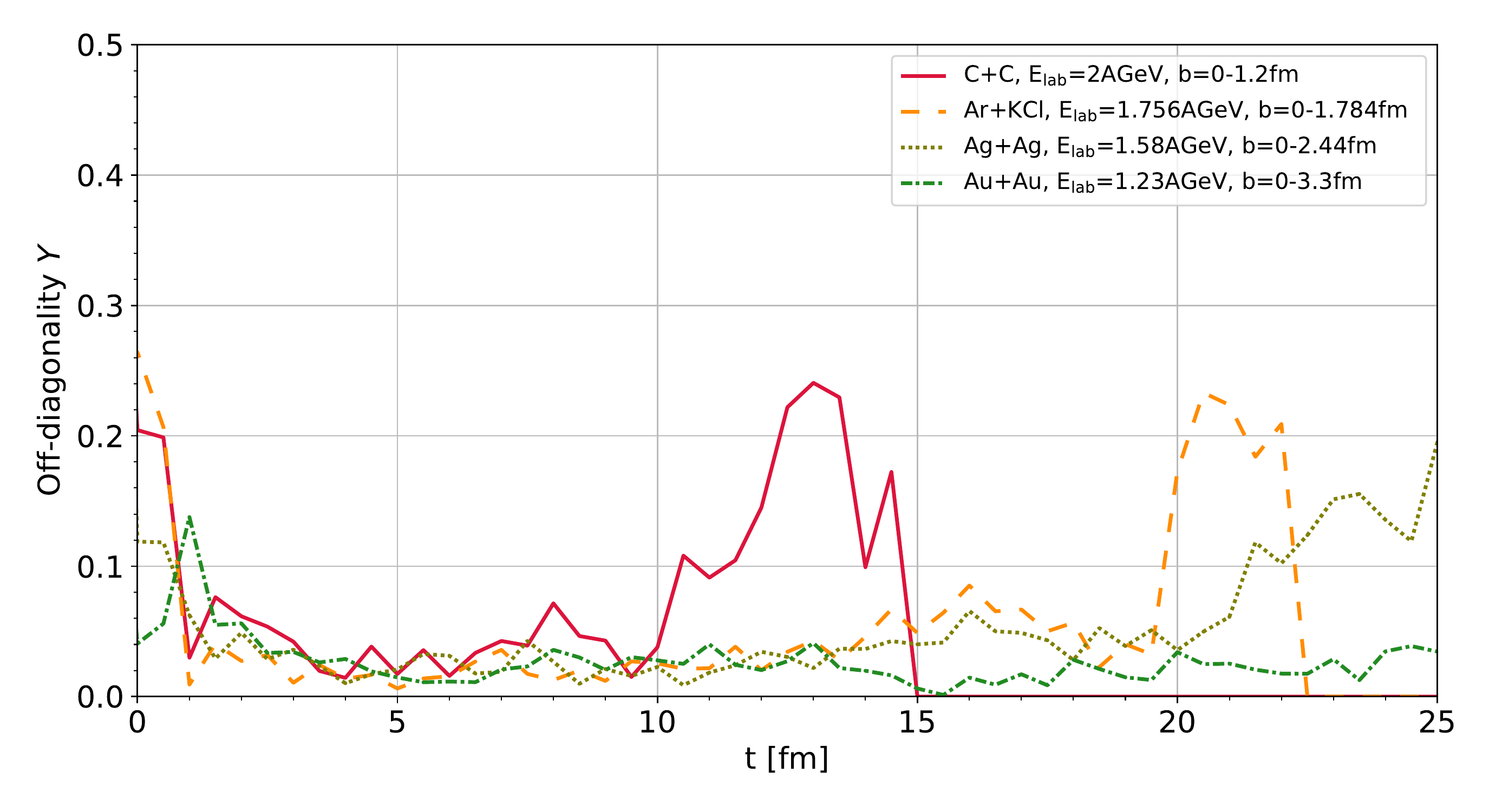}
    \caption{Time evolution of the pressure anisotropy $X$ and of the off-diagonality $Y$ in the center of the system for the reactions listed in Table ~\ref{table:bE}.}
    \label{fig:XY_evo_avgT_lh}
\end{figure}

If we look at the time evolution of the average pressure anisotropy $X$, shown in the upper half of Fig.~\ref{fig:XY_evo_avgT_lh}, after the initial equilibration process the values of this parameter in most of the systems are within the reach of viscous hydrodynamics, with the noticeable exception of C+C collisions, which create a smaller and more quickly dissolving system, that at 15 fm is already very diluted, as one can deduce from Fig.~\ref{fig:edens_evo_center_lh}. The off-diagonality $Y$, shown in the lower half of Fig.~\ref{fig:XY_evo_avgT_lh}, has on average a smaller magnitude, staying below 0.1 for all ion species until the system starts going out of equilibrium, a process that for Carbon ions happens already between 10 fm and 15 fm, while for Ar+KCl between 20 fm and 23 fm. Overall we observe a clear dependence of the degree of the system equilibrium on the atomic number of the colliding ions, that is on the number of collisions within the system, which, for small ions like Carbon, seems hardly sufficient to reach a good level of thermalization before the system dissolves. On the other hand, the good agreement between the predictions of the invariant mass spectrum of dielectrons produced in C+C collisions made with SMASH~\cite{Staudenmaier:2017vtq} in vacuum and the experimental data~\cite{HADES:2009jwc} already suggested that no thermal medium is formed in such a small system at HADES.

\subsubsection{Fluctuations in event by event simulations}
\begin{figure}
    \includegraphics[width=\columnwidth]{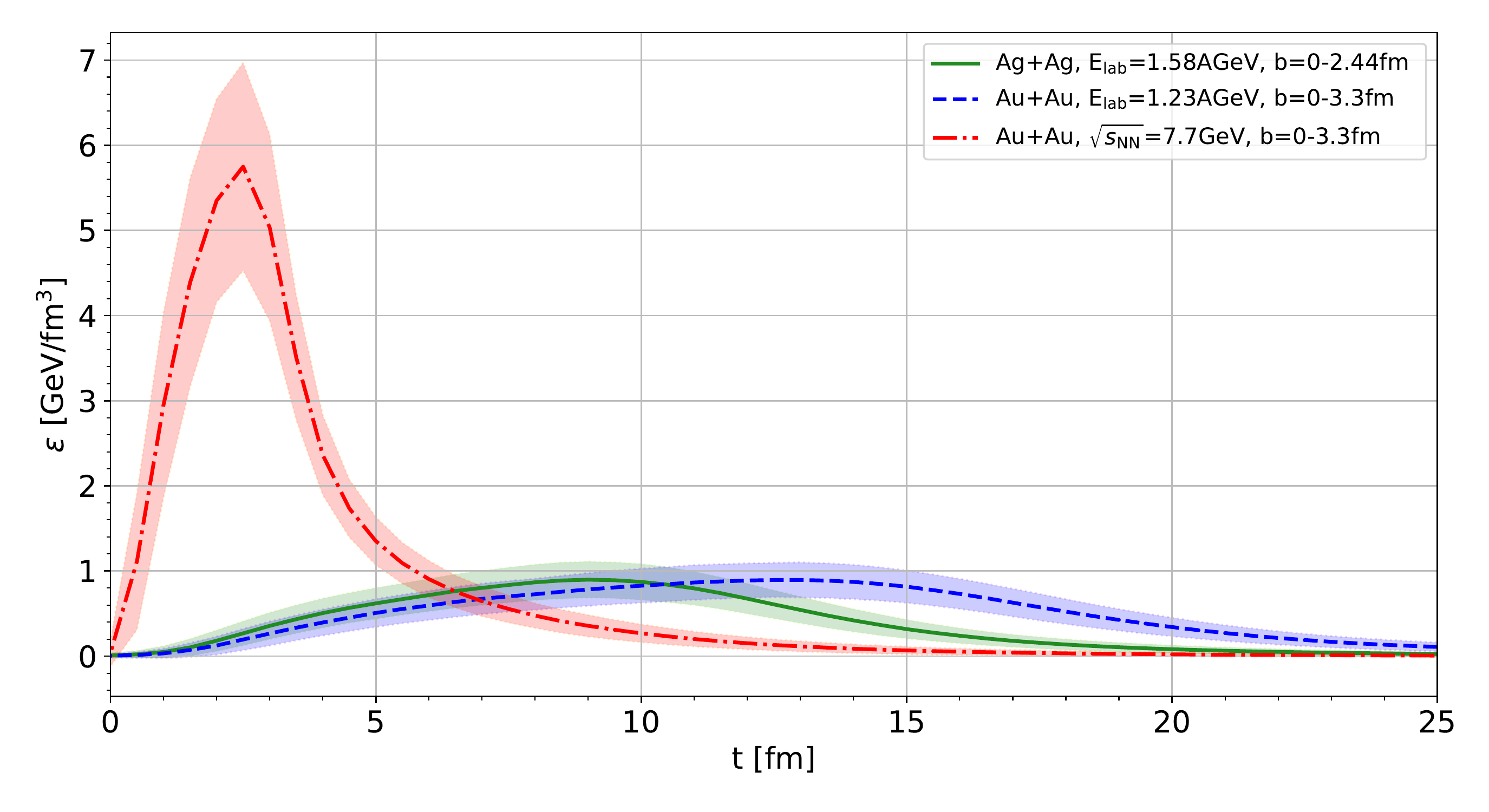}
    \caption{Time evolution of the energy density in the center of the system for central Ag+Ag at \ela = 1.58 AGeV, Au+Au at \ela = 1.23 AGeV and Au+Au at \snn = 7.7 GeV collisions. The colored error bands correspond to the standard deviations.}
    \label{fig:ebe_edens}
\end{figure}

\begin{figure}
 \includegraphics[width=\columnwidth]{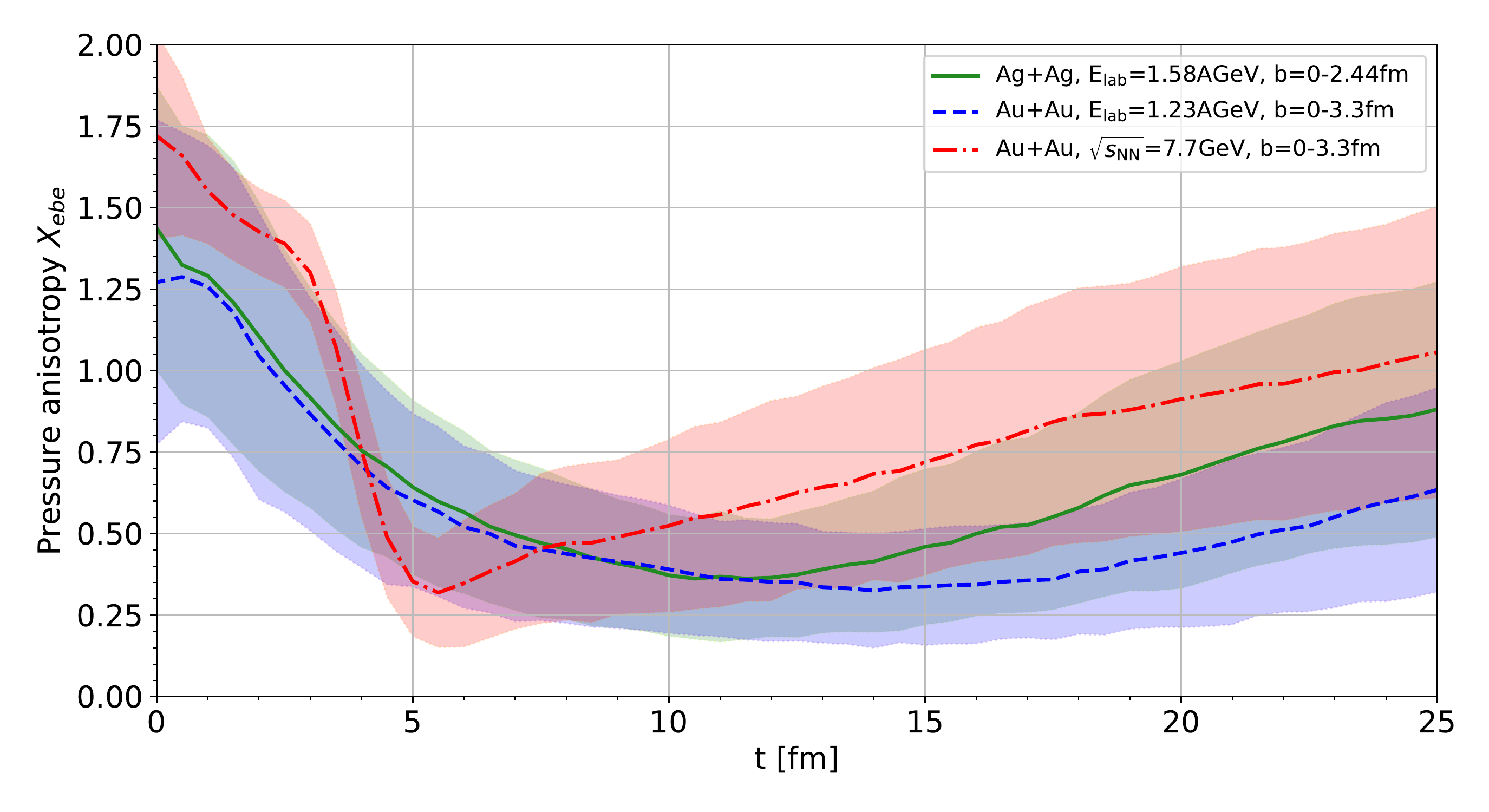}\\
 \includegraphics[width=\columnwidth]{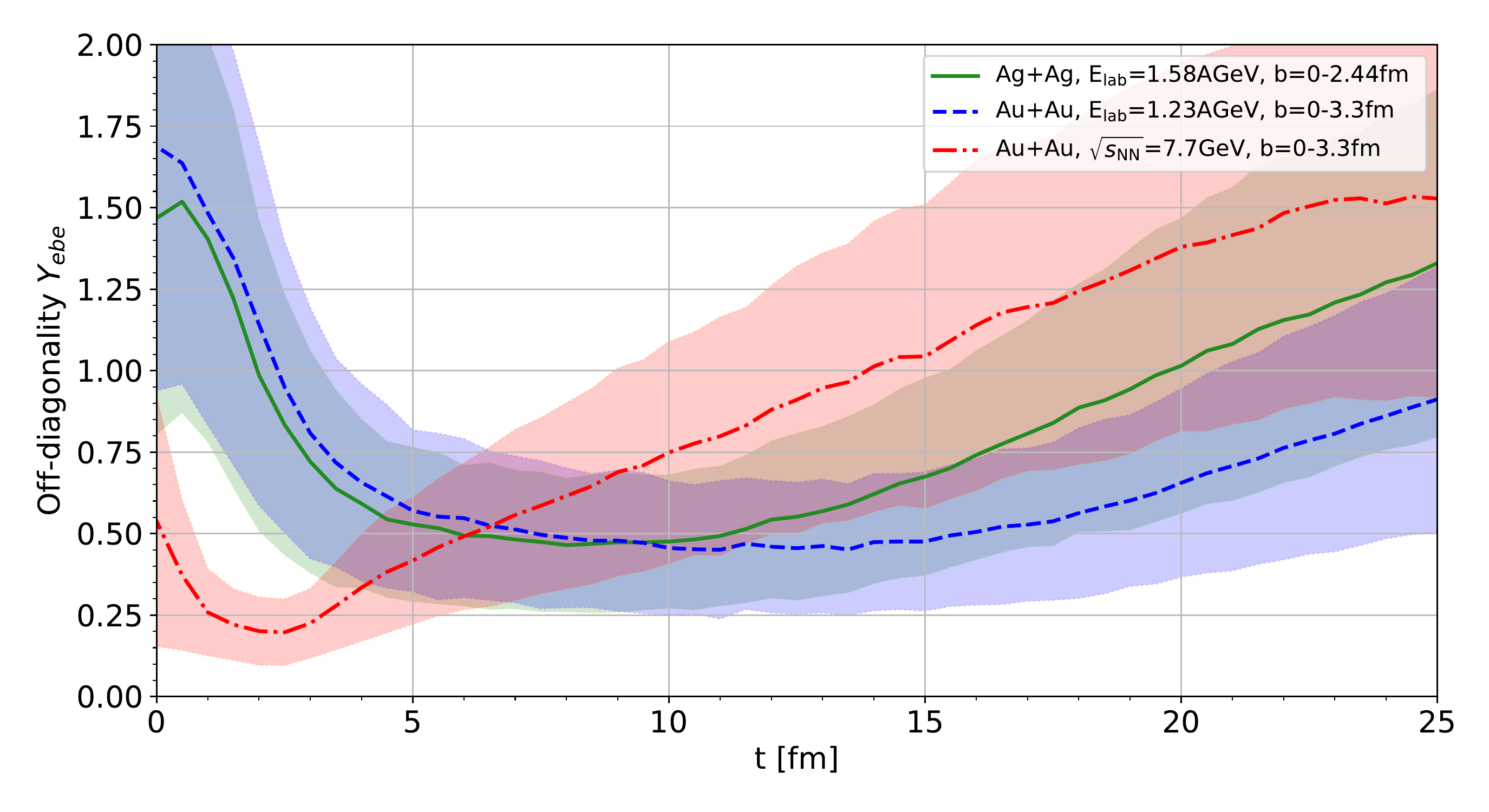}
\caption{Time evolution of $X_{ebe}$ and $Y_{ebe}$ in the center of the system for central Ag+Ag at \ela = 1.58 AGeV, Au+Au at \ela = 1.23 AGeV and Au+Au at \snn = 7.7 GeV collisions. The colored error bands correspond to the standard deviations.}
\label{fig:ebe_XY}
\end{figure}

Our preliminary survey restricted to the center of the system indicates that, on an event by event basis, it is difficult to reach a high degree of thermalization. However,  fluctuations\cite{Stephanov:1999zu,Bluhm:2020mpc,Petersen:2012qc} in the positions of the colliding nuclei might allow the formation of regions with more isotropic conditions. We consider the case of central collisions for Ag+Ag at \ela = 1.58 AGeV, Au+Au at \ela = 1.23 AGeV and Au+Au at \snn = 7.7 GeV and, in addition to the average value, we consider also the standard deviation from the average. In Fig.~\ref{fig:ebe_edens} we show the time evolution of the energy density, noticing that typical fluctuations can deviate from the average value up to roughly 20\%. In the case of $X_{ebe}$ and $Y_{ebe}$, shown in Fig.~\ref{fig:ebe_XY}, fluctuations are much broader, allowing, in some events, to achieve favorable conditions for the applicability of hydrodynamics, nevertheless, as we will see later, typically only within rather small and fragmented volumes.

\subsection{Conditions on the coordinate planes}
We now take a step forward from the case of a single cell and we explore the spatial distribution of the energy density, the pressure anisotropy and the off-diagonality on the orthogonal planes parallel to the coordinate axes of the Cartesian frame of reference passing through the center of the system.\\

\begin{figure*}[ht!]
    \vspace*{-3mm}
    \begin{subfigure}{\textwidth}         
    \centering
    \includegraphics[width=0.78\textwidth, trim={0 1.5 0 1},clip]{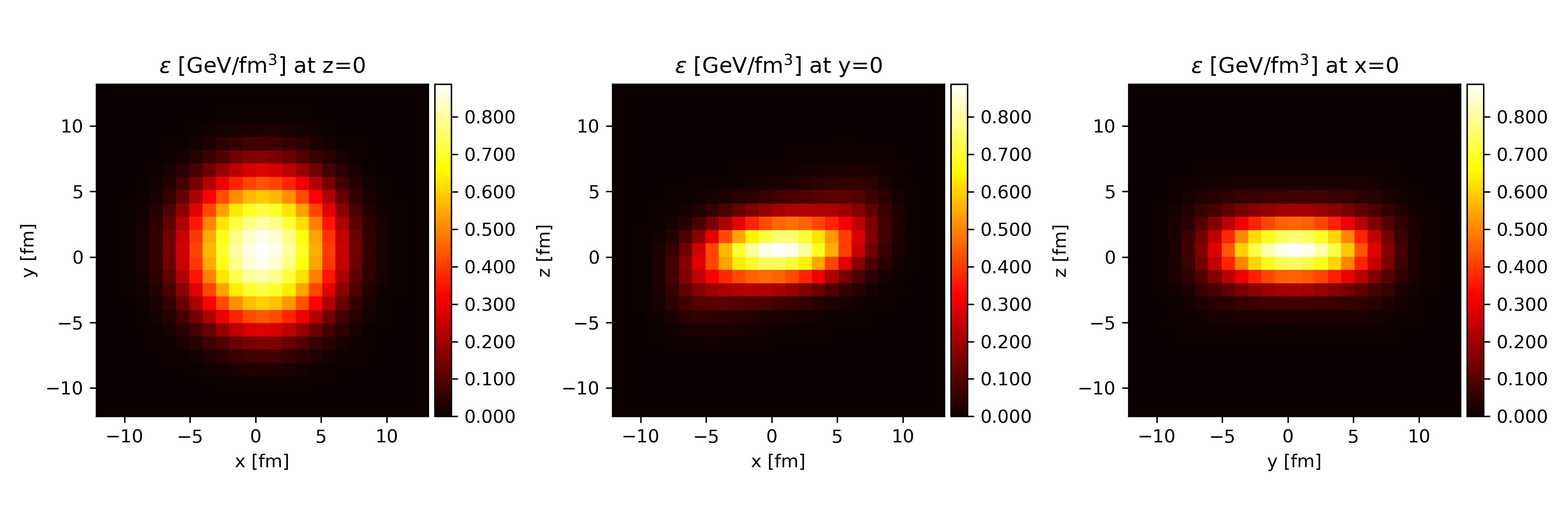}            
   \end{subfigure}
\vspace*{-3mm}
\begin{subfigure}{\textwidth} 
    \centering
    \includegraphics[width=0.78\textwidth, trim={0 1 0 1},clip]{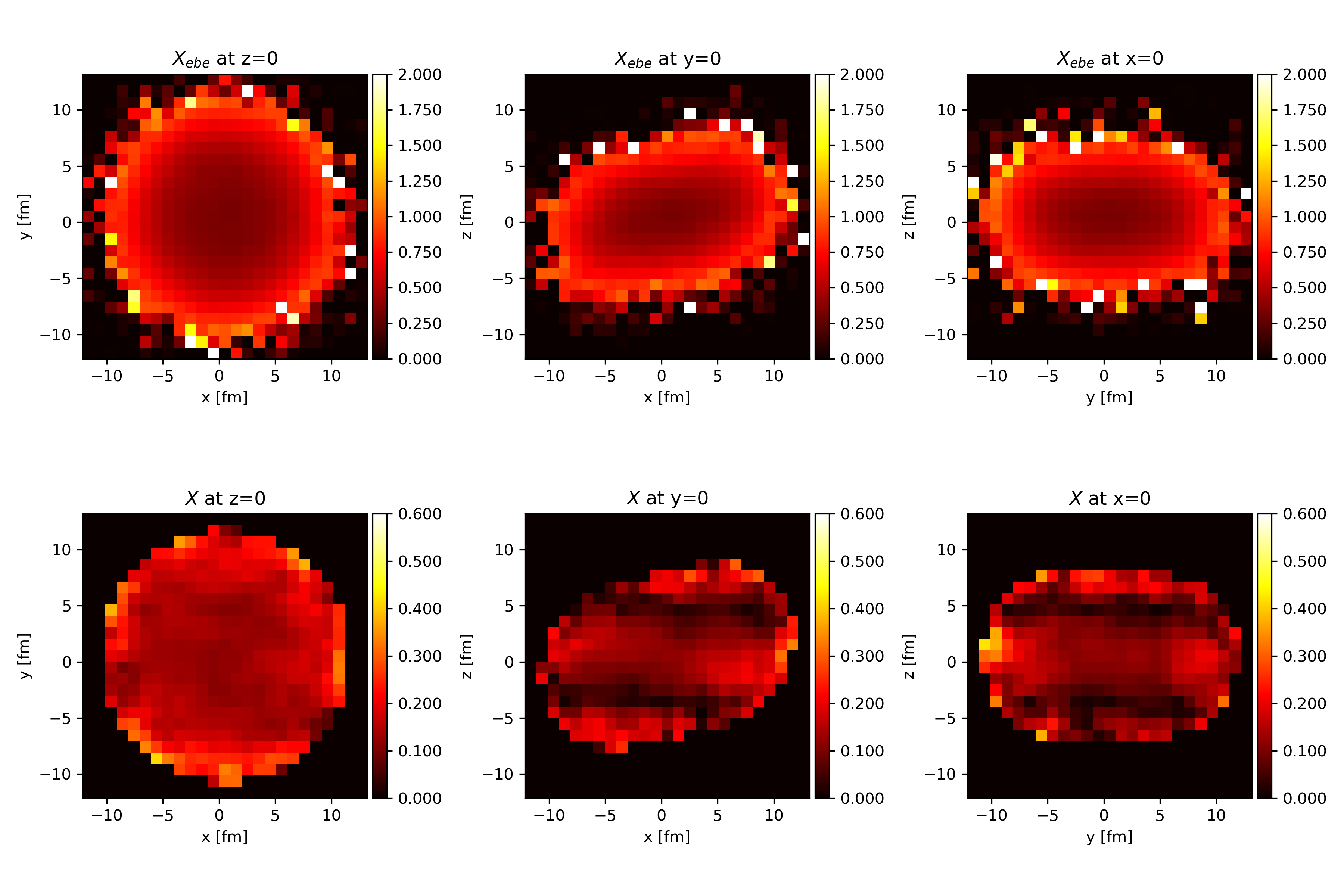}
   \end{subfigure}
\vspace*{-3mm}
\begin{subfigure}{\textwidth} 
    \centering
    \includegraphics[width=0.78\textwidth, trim={0 1 0 1},clip]{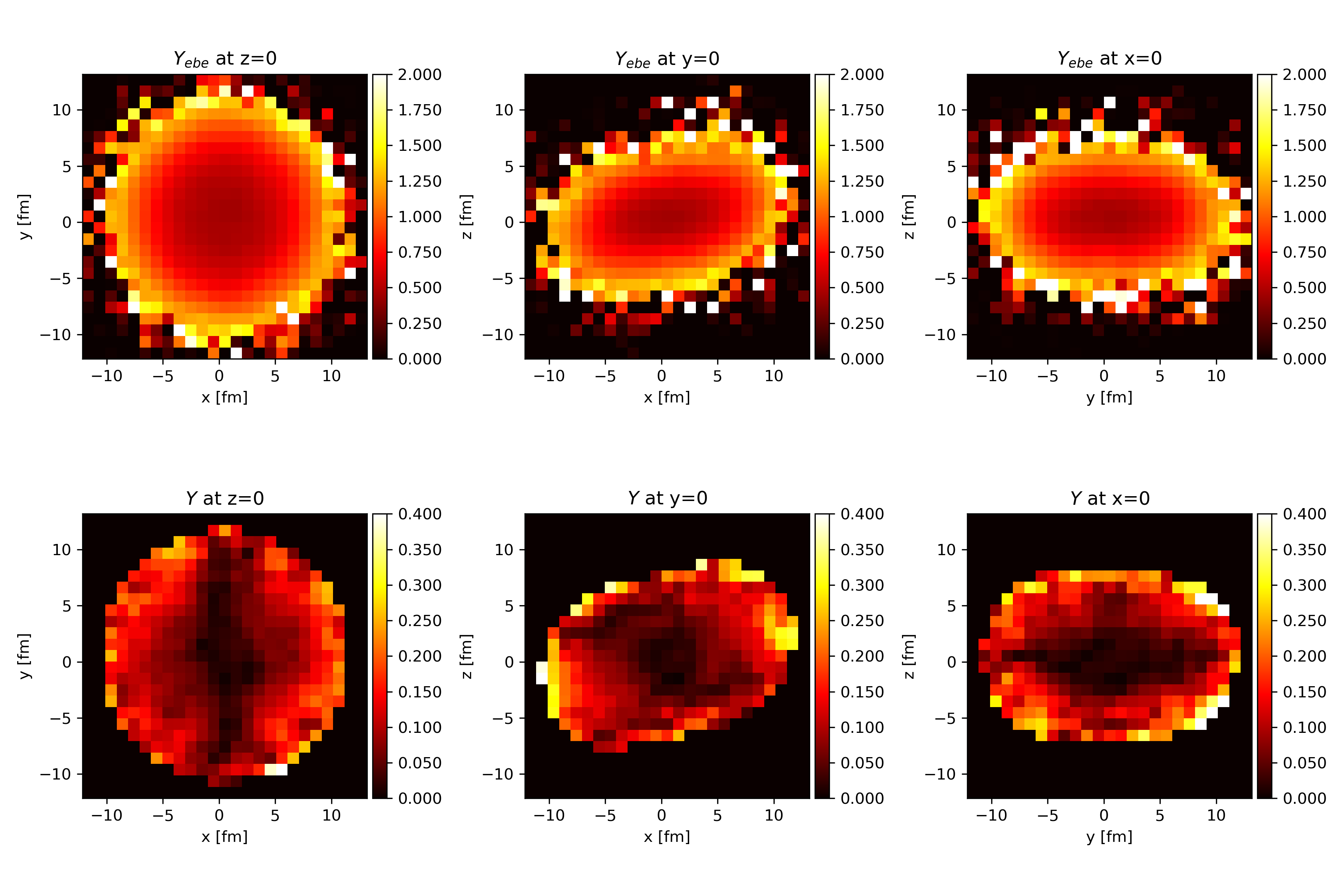}
    \end{subfigure}
\caption{Au+Au collisions at \ela = 1.23 AGeV, t = 14 fm, 0-5\% centrality class, average of 1080 events, distributions along the planes z = 0, y = 0, x = 0 of (from top to bottom): energy density $\varepsilon$, $X_{ebe}$, $X$, $Y_{ebe}$, $Y$.}
\label{fig:2D_imgs}
\end{figure*}

 The upper part of Fig.~\ref{fig:2D_imgs} shows the average energy density distribution in Au+Au collisions at \ela = 1.23 AGeV at t=14 fm in the 0-5\% centrality class. The chosen time approximately corresponds to the maximum in Fig.~\ref{fig:edens_evo_center_Au_snn_dep}, when the initial compression phase ends. The rest of Fig.~\ref{fig:2D_imgs}, from top to bottom, shows the spatial distributions of $X_{ebe}$, $X$, $Y_{ebe}$ and $Y$. We remind that we consider as part of the system the regions with the sum of the diagonal energy momentum tensor components larger than 0.1 \mev, a rather broad acceptance threshold that includes many ``cold'' cells with very low energy density, so in $\varepsilon$ plots a relevant part of the system fades into the background. We notice that both $X_{ebe}$ and $Y_{ebe}$ have considerably larger values than the $X$ an $Y$, thus demonstrating that the smoothing effect due to the averaging process is not limited to the central points, but significantly affects the whole system, including the peripheral regions. However, we remark that in general the peripheral regions are still farther away from isotropic equilibrium than the central ones. As already emerged in the case of the central point and as already noted in Ref.~\cite{Oliinychenko:2015lva}, the averaging process has a somehow larger impact on the off-diagonality than on the pressure anisotropy. The likely explanation of this effect is that the averaging process smooths down the statistical fluctuations due to the different initial positions and momenta of the nucleons of the colliding nuclei. Nevertheless the system has a natural special direction given by the beam axis and while the average of the components of the initial momenta along the $x$ and $y$ directions is $\approx$ 0, this is not the case of the $z$ component. 
 
\subsection{Correlations between pressure anisotropy, off-diagonality and energy density}
\label{sec:correlations}
\begin{figure*}[!ht]
\begin{subfigure}{0.45\textwidth}
    \centering
    \includegraphics[width=\columnwidth]{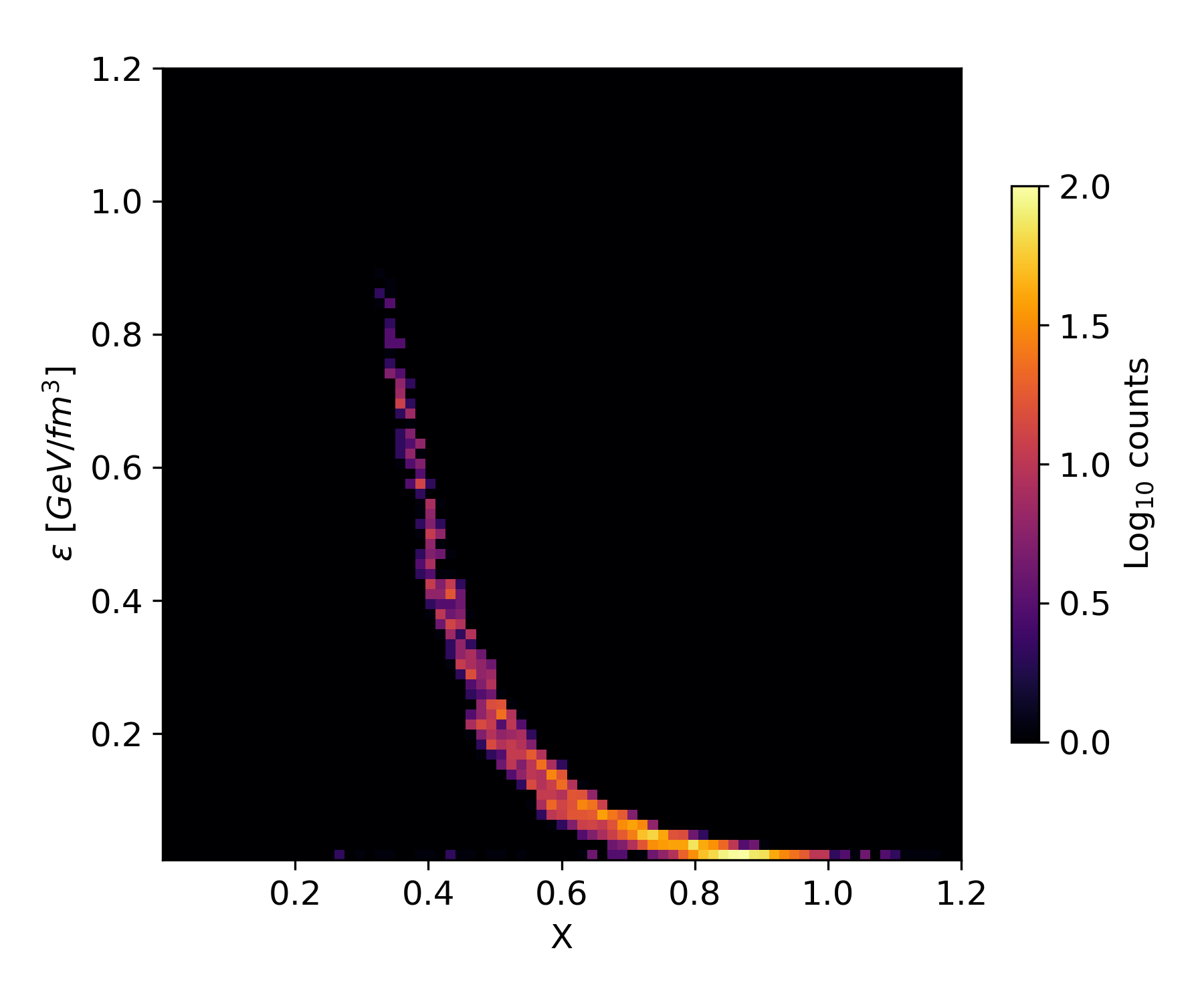}
    \caption{Volume elements w.r.t. $\varepsilon$ and $X_{ebe}$.}
    \label{fig:2D_hist_edens_vs_X_t_14_ebe}
\end{subfigure}\hspace*{\fill}
\begin{subfigure}{0.45\textwidth}
    \centering
    \includegraphics[width=\columnwidth]{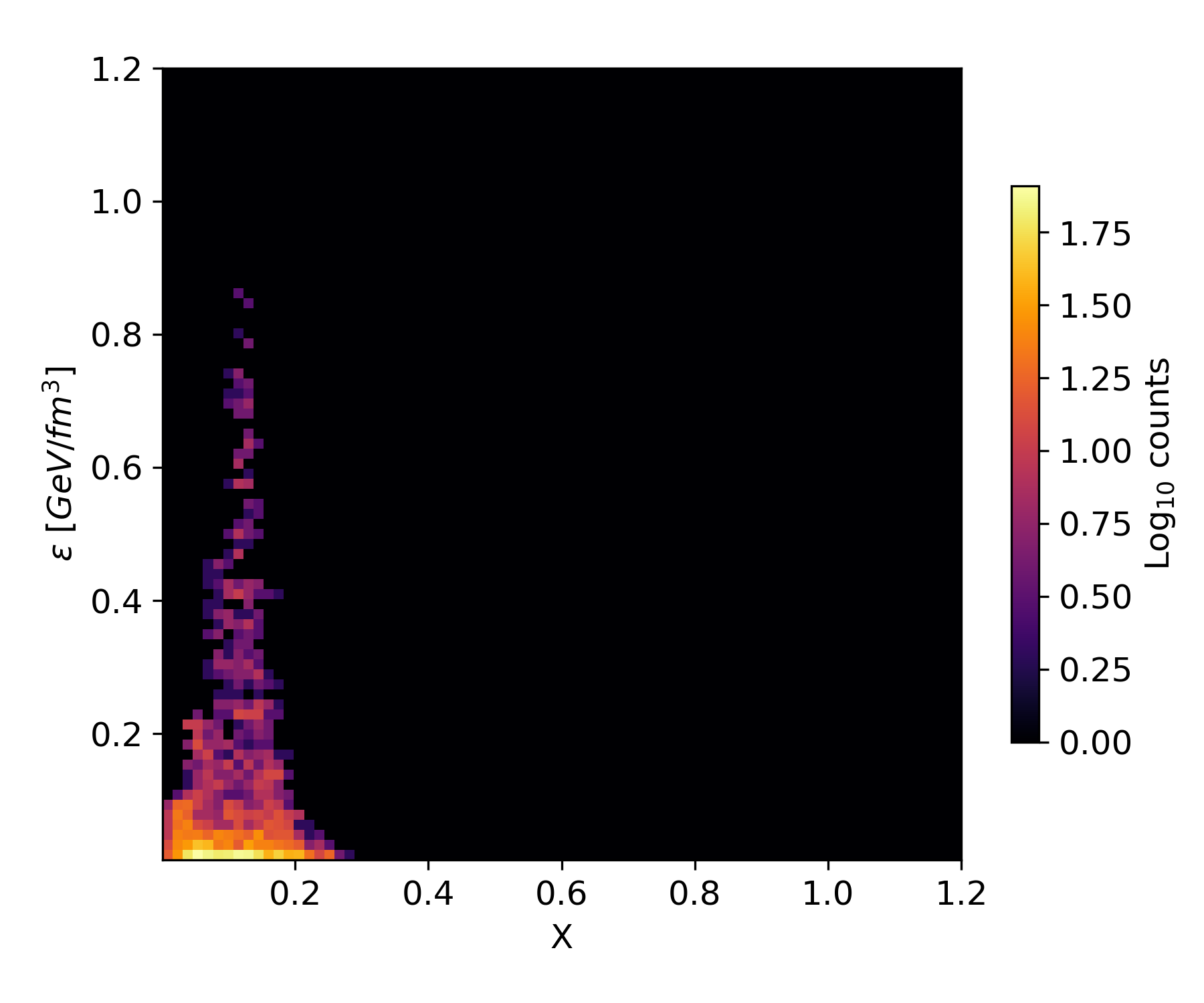}
    \caption{Volume elements w.r.t. $\varepsilon$ and $X$.}
    \label{fig:2D_hist_edens_vs_X_t_14_avg}
\end{subfigure}
\smallskip
\begin{subfigure}{0.45\textwidth}
    \centering
    \includegraphics[width=\columnwidth]{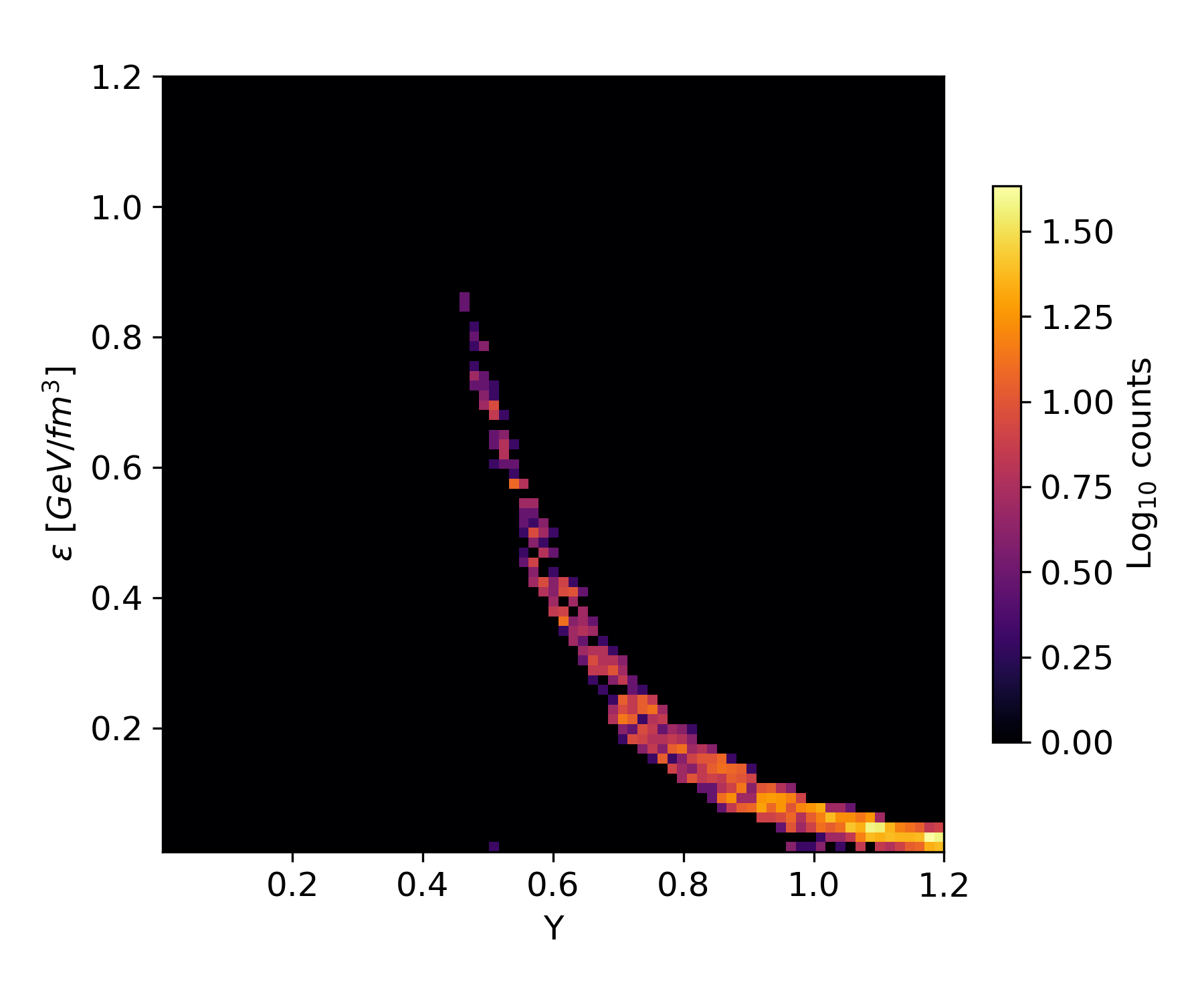}
    \caption{Volume elements w.r.t. $\varepsilon$ and $Y_{ebe}$.}
    \label{fig:2D_hist_edens_vs_Y_t_14_ebe}
\end{subfigure}\hspace*{\fill}
\begin{subfigure}{0.45\textwidth}
    \centering
    \includegraphics[width=\columnwidth]{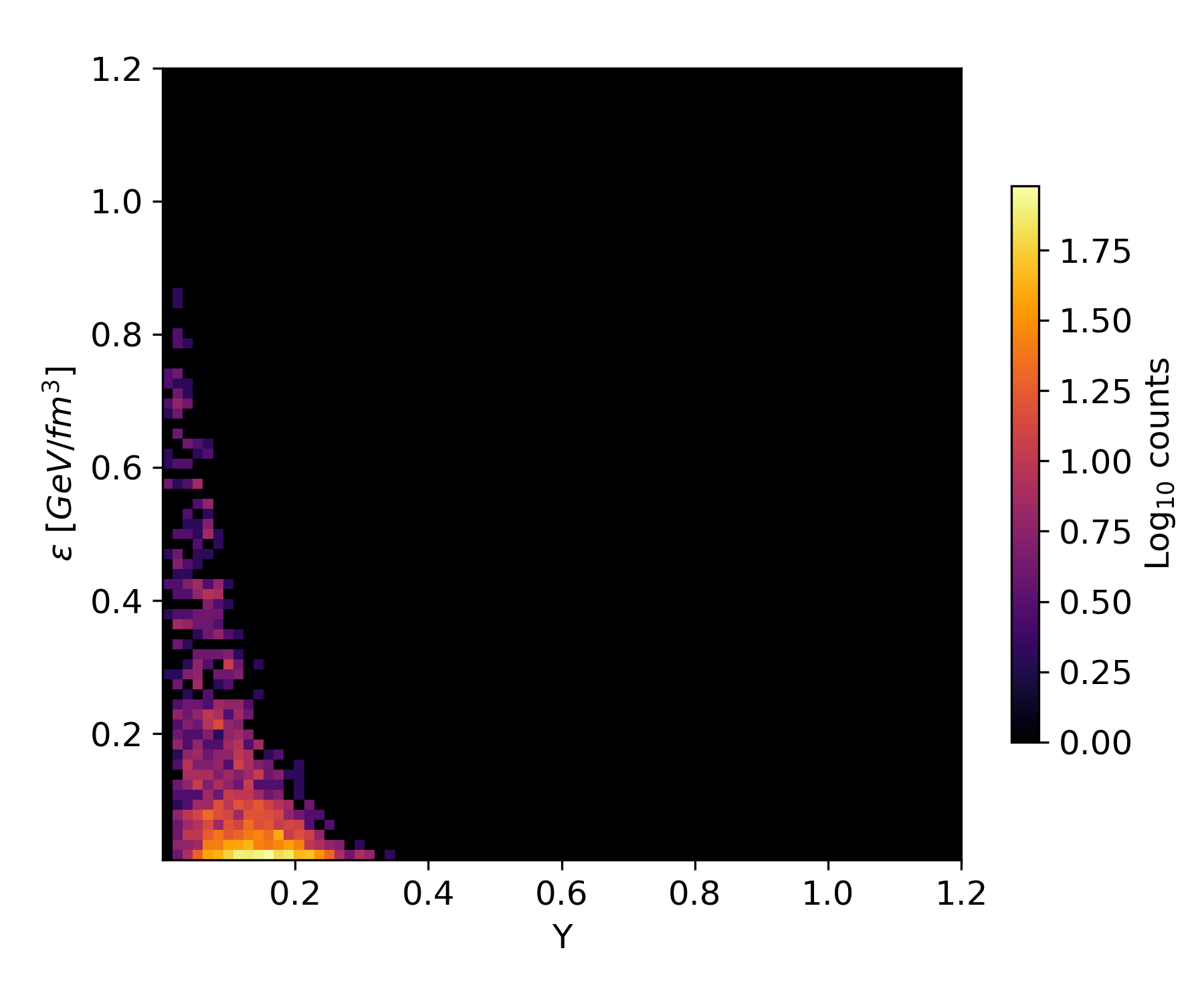}
    \caption{Volume elements w.r.t. $\varepsilon$ and $Y$.}
    \label{fig:2D_hist_edens_vs_Y_t_14_avg}
\end{subfigure}
\smallskip
\begin{subfigure}{0.45\textwidth}
    \centering
    \includegraphics[width=\columnwidth]{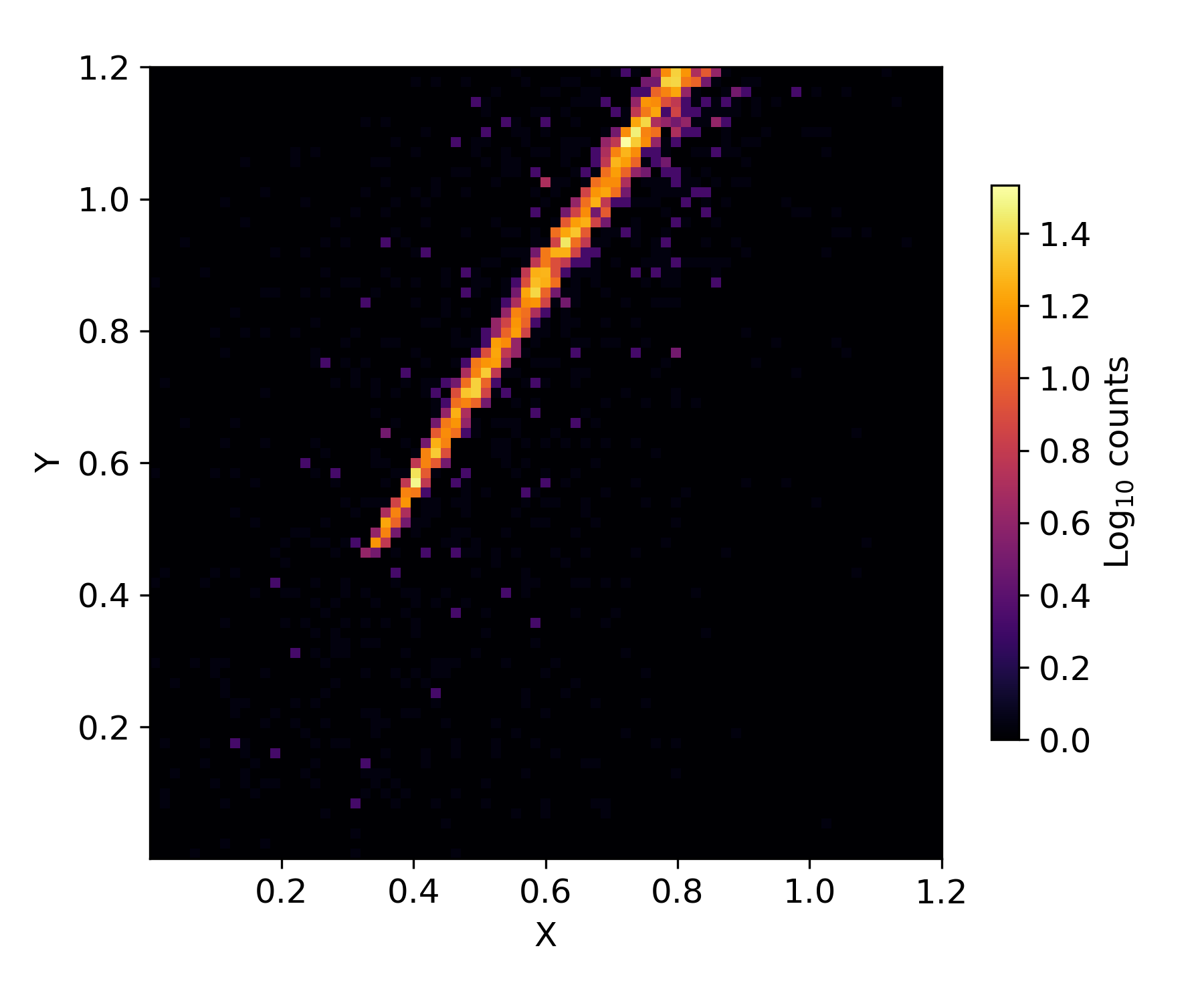}
    \caption{Volume elements w.r.t. $Y_{ebe}$ and $X_{ebe}$.}
    \label{fig:2D_hist_Y_vs_X_t_14_ebe}
\end{subfigure}\hspace*{\fill}
\begin{subfigure}{0.45\textwidth}
    \centering
    \includegraphics[width=\columnwidth]{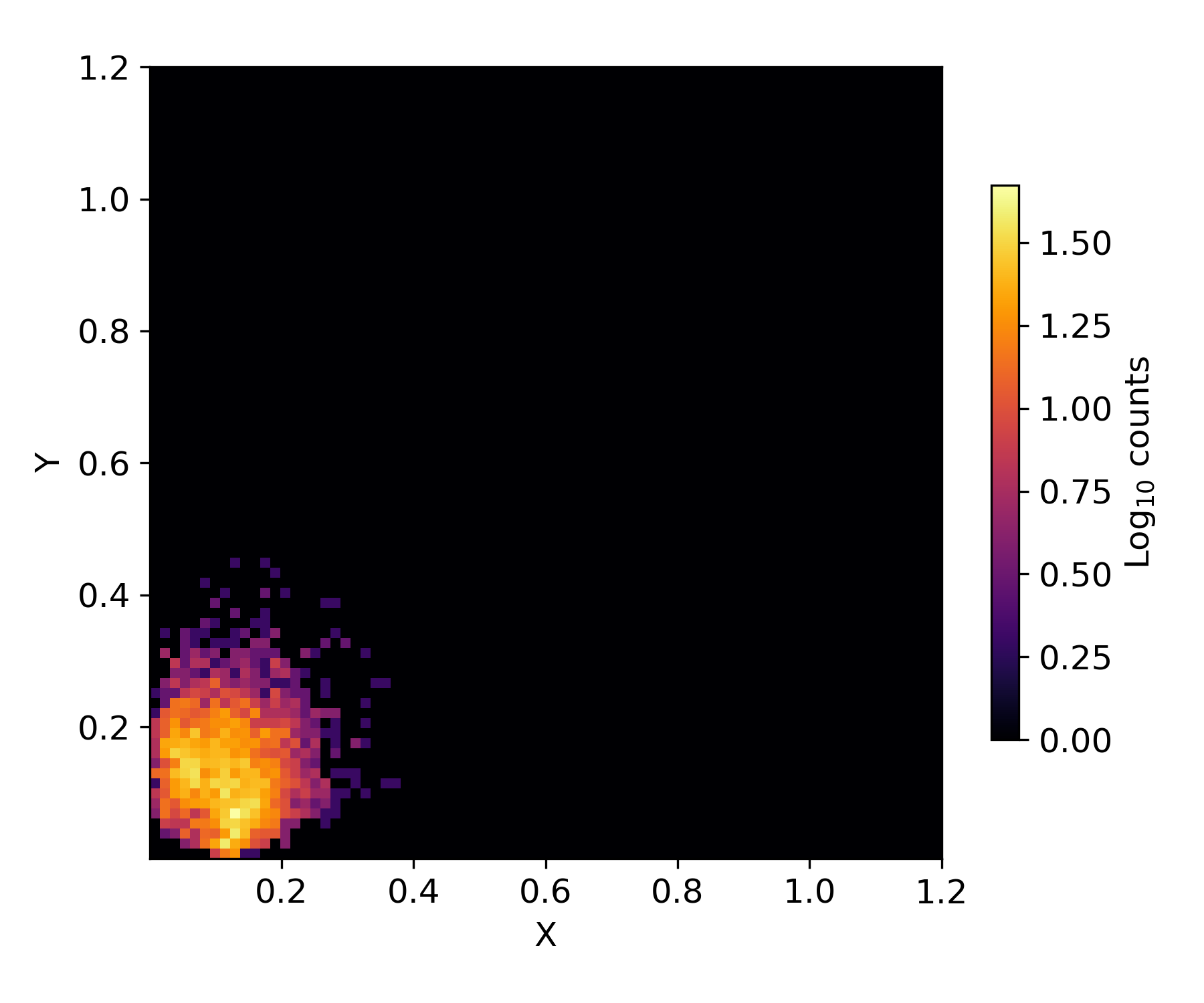}
    \caption{Volume elements w.r.t. $Y$ and $X$.}
    \label{fig:2D_hist_Y_vs_X_t_14_avg}
   \end{subfigure} 
\label{fig:2D_hist}
\caption{2D histograms of volume elements with respect to energy density, pressure anisotropy and off-diagonality. Au+Au collisions at \ela = 1.23 AGeV, 0-5\% centrality class, t = 14 fm.}
\end{figure*}

In addition to evaluate the presence of regions satisfying certain constraints in the systems formed in heavy ion collisions, it is interesting to assess what are the typical values of the constraints that delimit these regions and how they are related to each other. We focus on Au+Au collisions at \ela = 1.23 AGeV, 0-5\% centrality class, and examine the distribution of the values of $\varepsilon$, $X$, $Y$, $X_{ebe}$, $Y_{ebe}$ at t = 14 fm from 1080 events. For the histograms we use the ranges and number of bins reported in Table~\ref{table:bins}. We choose 10 \mev as the lower edge of the energy density bins to avoid some disturbing ``noise'' below this threshold in the event by event case.\\
\begin{table}[h!]
    \centering
    \begin{tabular}{|c|c|c|c|} 
        \hline
        Quantity & Min & Max & Bins\\ 
        \hline\hline
        energy density [GeV/fm$^3$] & 0.01 & 1.2 & 80 \\ 
        $X$, $Y$, $X_{ebe}$, $Y_{ebe}$ & 0.001 & 1.2 & 80\\
        \hline
    \end{tabular}
    \caption{Bins used in the 2D histograms of pressure anisotropy, off-diagonality and energy density }
    \label{table:bins}
\end{table}
Fig.~\ref{fig:2D_hist_edens_vs_X_t_14_ebe} shows the two-dimensional histograms of the volume elements of the system with respect to $\varepsilon$ and $X_{ebe}$. The plot displays a pattern with an increasing number of volume elements with higher pressure anisotropy by decreasing energy density, particularly steep between $X$=0.35 and $X$=0.5. Fig.~\ref{fig:2D_hist_edens_vs_X_t_14_avg} shows how replacing $X_{ebe}$ with $X$ dramatically shifts the distribution towards $X$ values below 0.2, leading to a dense 2D histogram region for $\varepsilon<0.2$ GeV/fm$^3$, with a secondary pattern that extends up to 0.8 GeV/fm$^3$.\\

Fig.~\ref{fig:2D_hist_edens_vs_Y_t_14_ebe} and Fig.~\ref{fig:2D_hist_edens_vs_Y_t_14_avg}, in which $X_{ebe}$ and $X$ are replaced by $Y_{ebe}$ and $Y$, respectively, display characteristics similar to the previous figures, but the damping effect due to the averaging process on the non-diagonal terms of energy momentum tensor is stronger than on the diagonal terms, resulting in a dramatic shift of the dense region of the 2D histogram from $Y_{ebe}>0.5$ to $Y<0.2$ for the most part. 

Fig.~\ref{fig:2D_hist_Y_vs_X_t_14_ebe} shows the two-dimensional histograms of the volume elements of the system with respect to the average $X_{ebe}$ and $Y_{ebe}$. We notice that the area $X_{ebe}<0.3$, $Y_{ebe}<0.4$ is almost completely depleted, while for higher values these quantities are indeed strongly correlated and follow closely an almost linear relation (rough visual estimate without performing a fit: $Y_{ebe}\approx$1.5 $X_{ebe}$), albeit with some sparse points all around. The case of $X$ and $Y$, shown in Fig.~\ref{fig:2D_hist_Y_vs_X_t_14_avg}, is significantly different: the previous pattern of Fig.~\ref{fig:2D_hist_Y_vs_X_t_14_ebe} is replaced by a rather compact cluster at low values, in particular  for $X, Y<0.3$.
In general, the expected trend that high energy densities correlate with cells closer to local equilibrium holds, but one cannot infer a simpler ``thermalization'' criterion from these results. 

\subsection{Volume and fraction of the system satisfying selected constraints}
We now evaluate the volume of the system at a certain time that simultaneously satisfies a set of fixed constraints, summarized in table~\ref{table:constraints}.\\
\begin{table}[h!]
    \centering
    \begin{tabular}{|c|c|} 
        \hline
        Constraint & Values\\ 
        \hline\hline
        minimum energy density & 1, 100, 500 [\mev] \\ 
        maximum pressure anisotropy & 0.1, 0.3, 0.5\\
        maximum off-diagonality & 0.1, 0.3, 0.5\\
        \hline
    \end{tabular}
    \caption{Values of the set of constraints that must be satisfied in the system.}
    \label{table:constraints}
\end{table}
\subsubsection{Constrained fraction of the average system vs time}
\begin{figure}[ht!]
    \centering
    \includegraphics[width=\columnwidth]{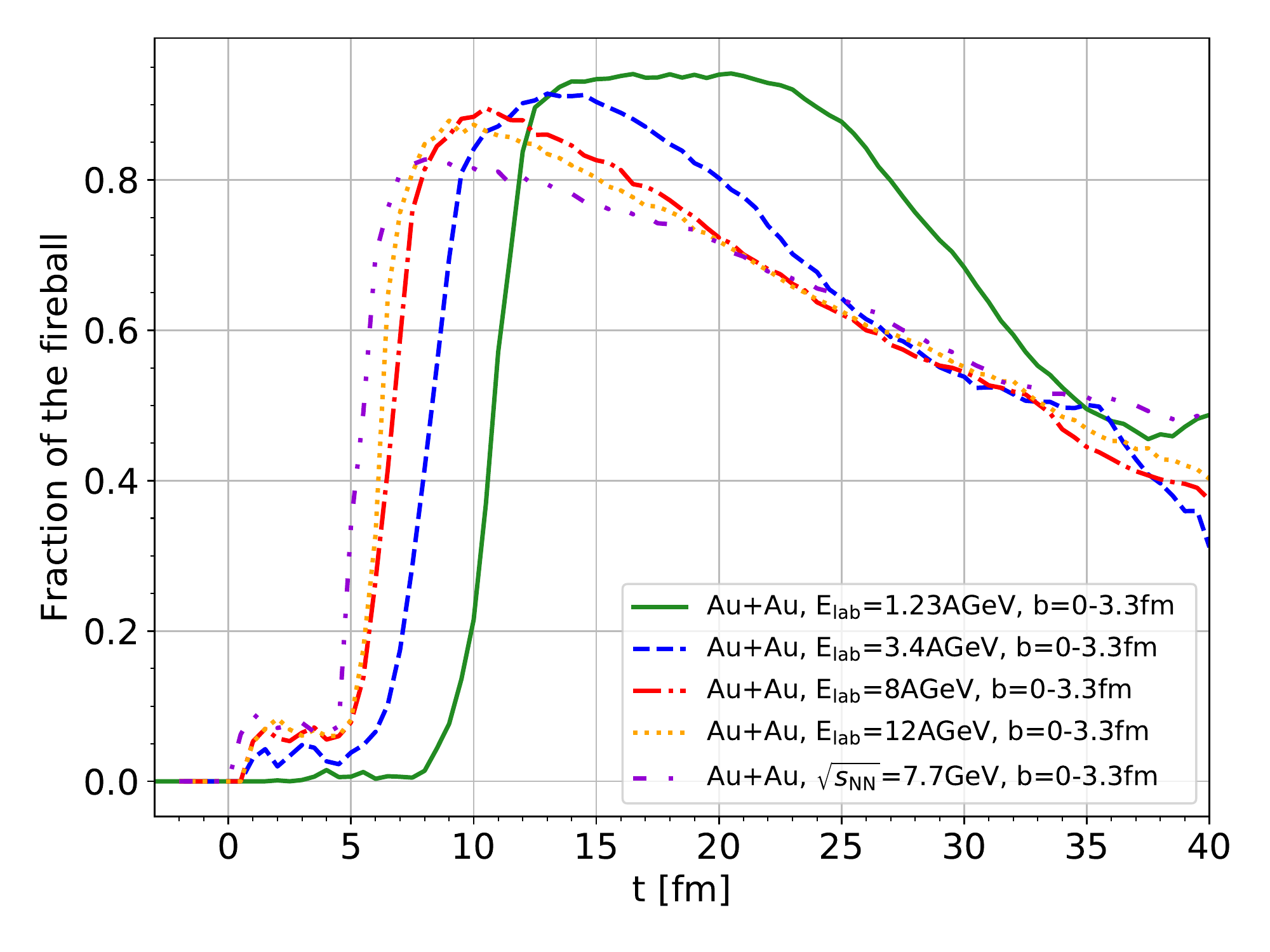}
    \caption{Time evolution of the average fraction of the system satisfying the conditions $X<0.3$,  $Y<0.3$, $\varepsilon>1$ \mev for Au+Au collisions, 0-5\% centrality class, at \ela = 1.23, 3.4, 8, 12 AGeV and \snn = 7.7 GeV.}
    \label{fig:percent_Au_03_03_1_enscan}
\end{figure}

\begin{figure}[ht!]
    \centering
    \includegraphics[width=\columnwidth]{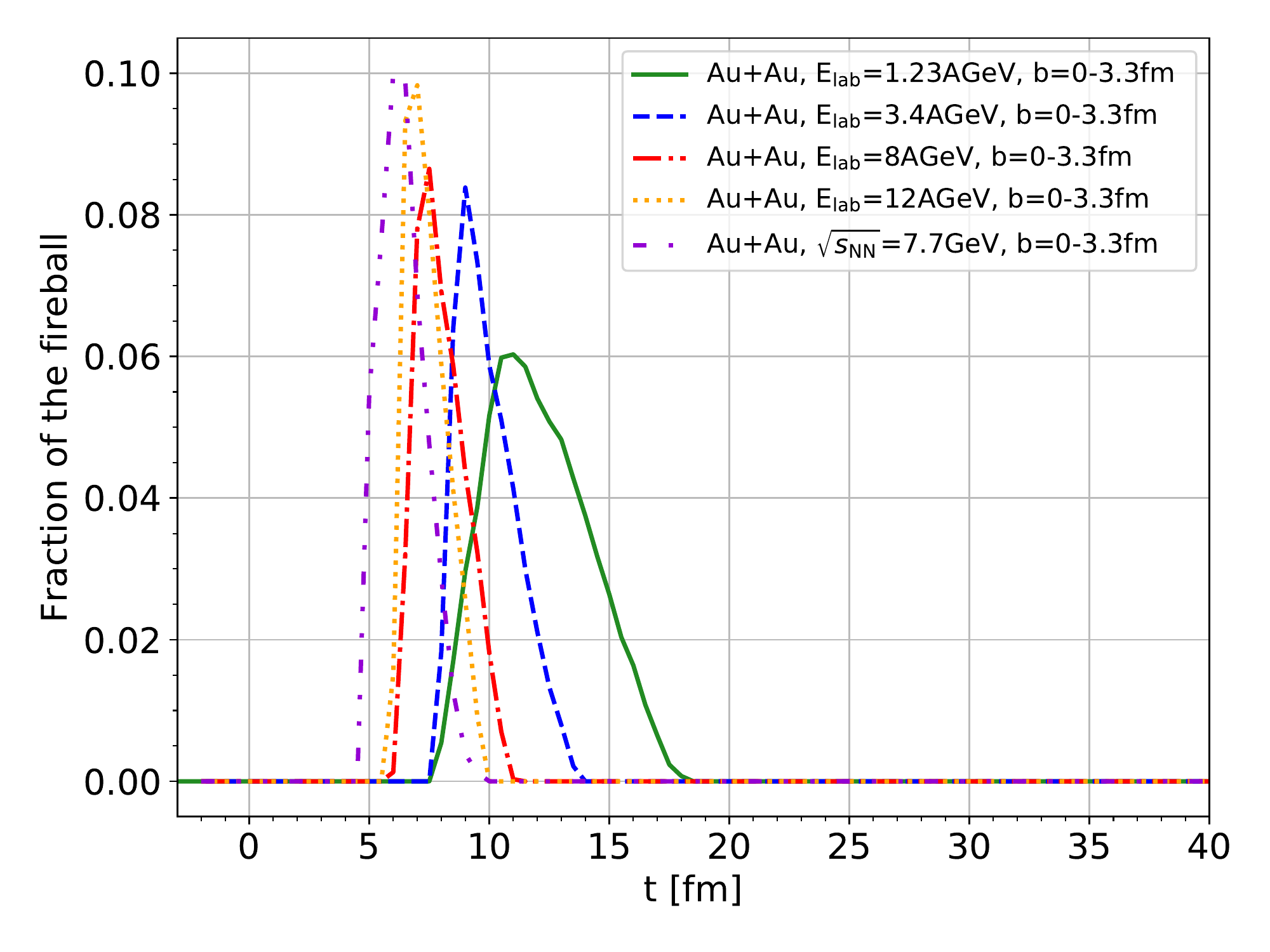}
    \caption{Time evolution of the average fraction of the system satisfying the conditions $X<0.3$,  $Y<0.3$, $\varepsilon>500$ \mev for Au+Au collisions, 0-5\% centrality class, at \ela = 1.23, 3.4, 8, 12 AGeV and \snn = 7.7 GeV.}
    \label{fig:percent_Au_03_03_500_enscan}
\end{figure}

For the sake of brevity, here we report only the results for Au+Au collisions at \ela = 1.23 AGeV in the 0-5\% centrality class. As we anticipated in the first section and as we will discuss again in section \ref{sec:int_volumes}, we expect to see roughly similar qualitative effects for other energies, centralities and system sizes, with some relevant deviations only in the case of the tiny C+C system. We focus on the energy momentum tensor from averaged 1080 events. Fig.~\ref{fig:percent_Au_03_03_1_enscan} shows the time evolution of the fraction of the total system for central Au+Au collisions at \ela = 1.23, 3.4, 8, 12 AGeV and \snn = 7.7 GeV, under the constraints $X, Y < 0.3$ and $\varepsilon > 1$ \mev. We notice that, after an initial equilibration phase which is shorter for more energetic collisions, most of the volume satisfies the imposed constraints. Lower collision energies reach higher maximum percentages than higher collision energies.\\
Fig.~\ref{fig:percent_Au_03_03_500_enscan} shows again, for the same collection of systems, the time evolution of their fraction subjected to the constraints $X, Y < 0.3$ and $\varepsilon > 500$ \mev. We notice that the higher energy density threshold dramatically reduces the eligible fraction of the system, whose peak decreases for decreasing collision energy. Due to the rapid cooling of the system, which is faster for more energetic collisions, already before 20 fm no system has any region that satisfies the given constraints anymore.\\
Fig.~\ref{fig:percent_Au_centr_dep} shows the centrality dependence of the time evolution of the fraction of the system satisfying the constraints $X, Y < 0.3$ and $\varepsilon > 1$ \mev for central Au+Au collisions at \ela = 1.23 AGeV. We notice that there is a rather weak dependence on the centrality class and that in more peripheral collisions, somehow counter intuitively, the fraction of the system that fulfills the given conditions at later times is larger than in the more central ones.\\
Fig.~\ref{fig:percent_lh} shows the time evolution of the fraction of the system satisfying the constraints $X, Y < 0.3$ and $\varepsilon > 1$ \mev for the reactions in Table~\ref{table:bE}. We notice that even in small systems like C+C there is an interval of time in which most of the system satisfies the set of constraints. We also notice that the smaller the system, the shorter the time interval, the earlier the formation time of the thermalized region and the sharper its ending.
\subsubsection{Dependence on the number of events}
Fig.~\ref{fig:percent_n_events} shows, for central Au+Au at \ela = 1.23 AGeV collisions, the fraction of the system satisfying the constraints $X, Y < 0.3$ and $\varepsilon > 1$ \mev depending on the number of averaged events. We notice that there is a strong dependence on the number of events, with a non-uniform, but steady growth of the fraction of the system satisfying the selected constraints directly proportional to their number. Given the huge variety of possible settings by changing the constraints or the number of events, it is not possible to provide unique indications or specific recommendations about how many events one should average to have most part of the system respecting the constraints. At the end of the day, the final choice relies on how much the hydrodynamic model can tolerate deviations from local thermal equilibrium, which are reduced by considering more events, and how important is to take into account fluctuations in the initial conditions, that instead are washed out by the averaging process.\\
From a practical perspective, since anyway many typical experimental observables are obtained as averages of billions of events, using test particles\cite{Weil:2016zrk} to compute the initial conditions for hydrodynamic simulations can be considered, to some extent, as a viable compromise with respect to pure event by event simulations, with multiple instead of single space momentum samplings of nucleons configuration at the beginning of each event. This approach, which is briefly summarized in ~\ref{sec:testparticles} and compared to the average of many events, has already proved to be very useful in Ref.~\cite{Mohs:2020awg} to handle and smooth down the large inhomogeneities of nuclear potentials due to their divergent behavior near the hadrons.\\
Looking at recent alternative approaches, we remark the existence of a few different stochastic hydrodynamics formalisms ~\cite{Akamatsu:2016llw,Martinez:2018wia}, in which off-equilibrium dynamics is incorporated into noise terms, which are collecting growing interest to deal with fluctuations, in particular around the QCD critical point ~\cite{Bluhm:2020mpc,Nahrgang:2018afz}.

\subsubsection{Integrated volumes}
\label{sec:int_volumes}
\begin{figure}[ht!]
    \centering
    \includegraphics[width=\columnwidth]{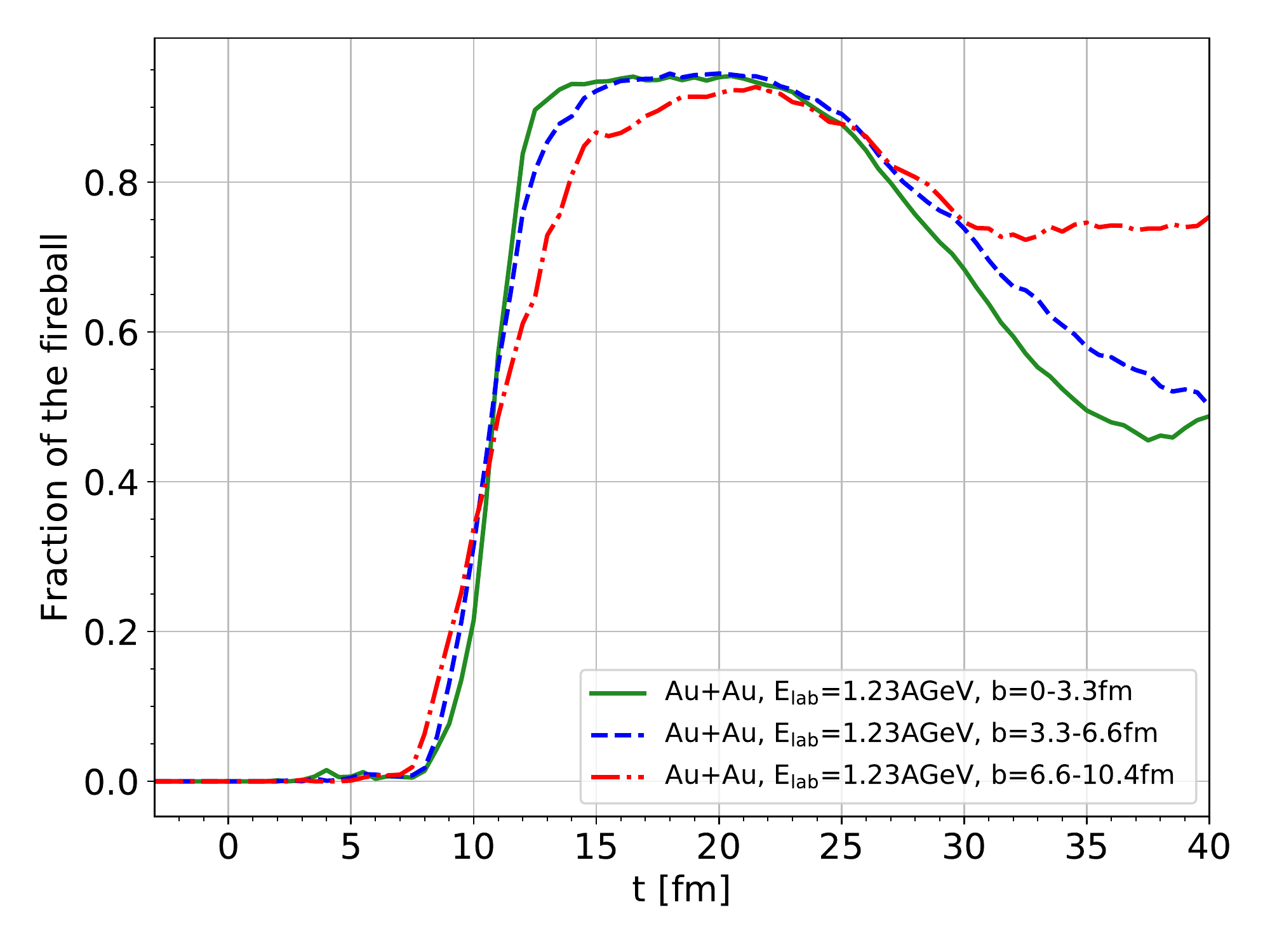}
    \caption{Time evolution of the average fraction of the system satisfying the conditions $X<0.3$,  $Y<0.3$, $\varepsilon>1$ \mev for Au+Au collisions at \ela = 1.23 AGeV, 0-5\%, 5-20\% and 20-50\% centrality classes.}
    \label{fig:percent_Au_centr_dep}
\end{figure}

\begin{figure}[ht!]
    \centering
    \includegraphics[width=\columnwidth]{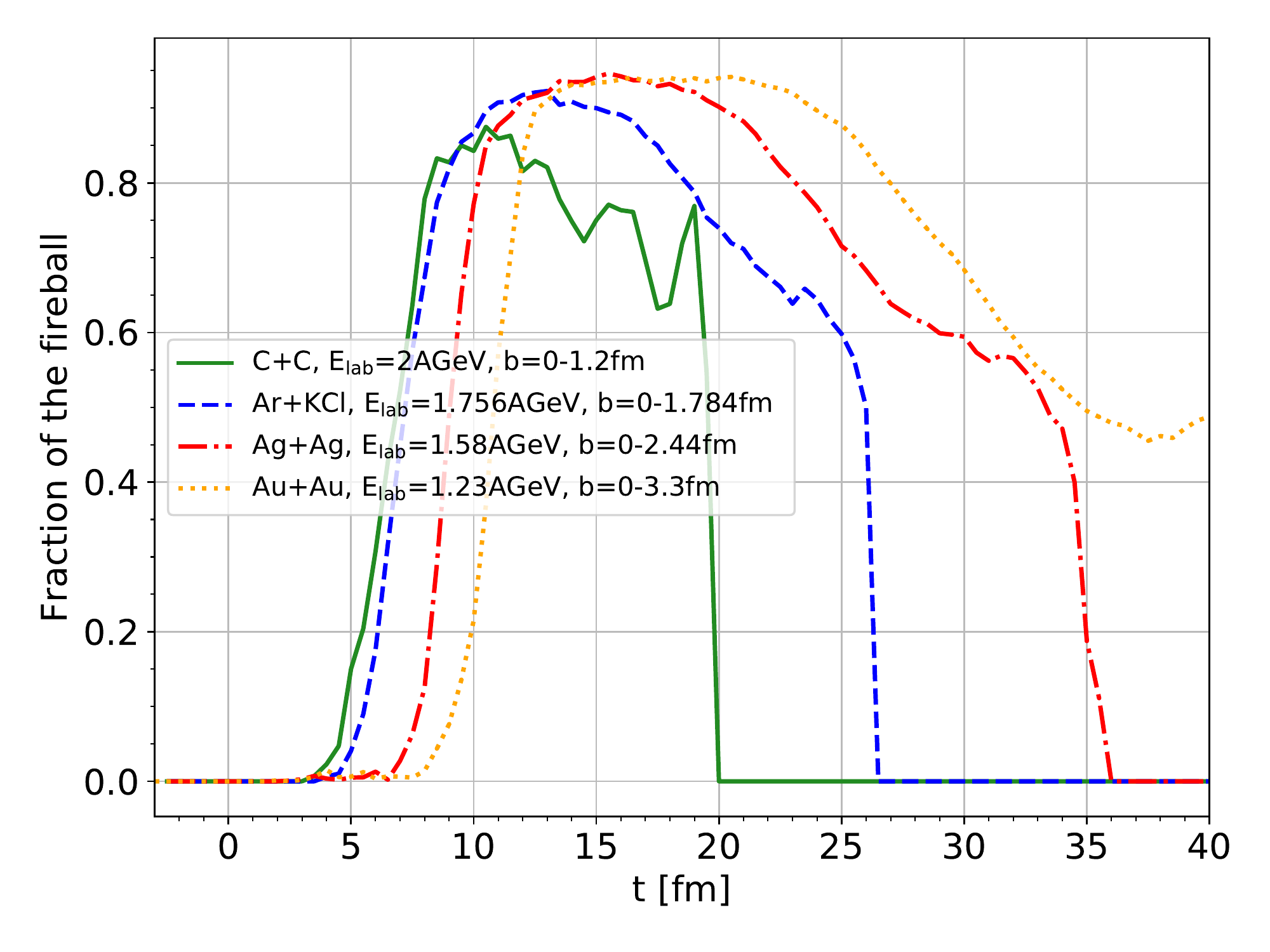}
    \caption{Time evolution of the average fraction of the system satisfying the conditions $X<0.3$,  $Y<0.3$, $\varepsilon>1$ \mev for C+C at \ela = 2 AGeV , Ag+Ag at \ela = 1.58 AGeV, Ar+KCl at \ela = 1.756 AGeV, Au+Au at \ela = 1.23 AGeV collisions, 0-5\% centrality class.}
    \label{fig:percent_lh}
\end{figure}

\begin{figure}[ht!]
    \centering
    \includegraphics[width=\columnwidth]{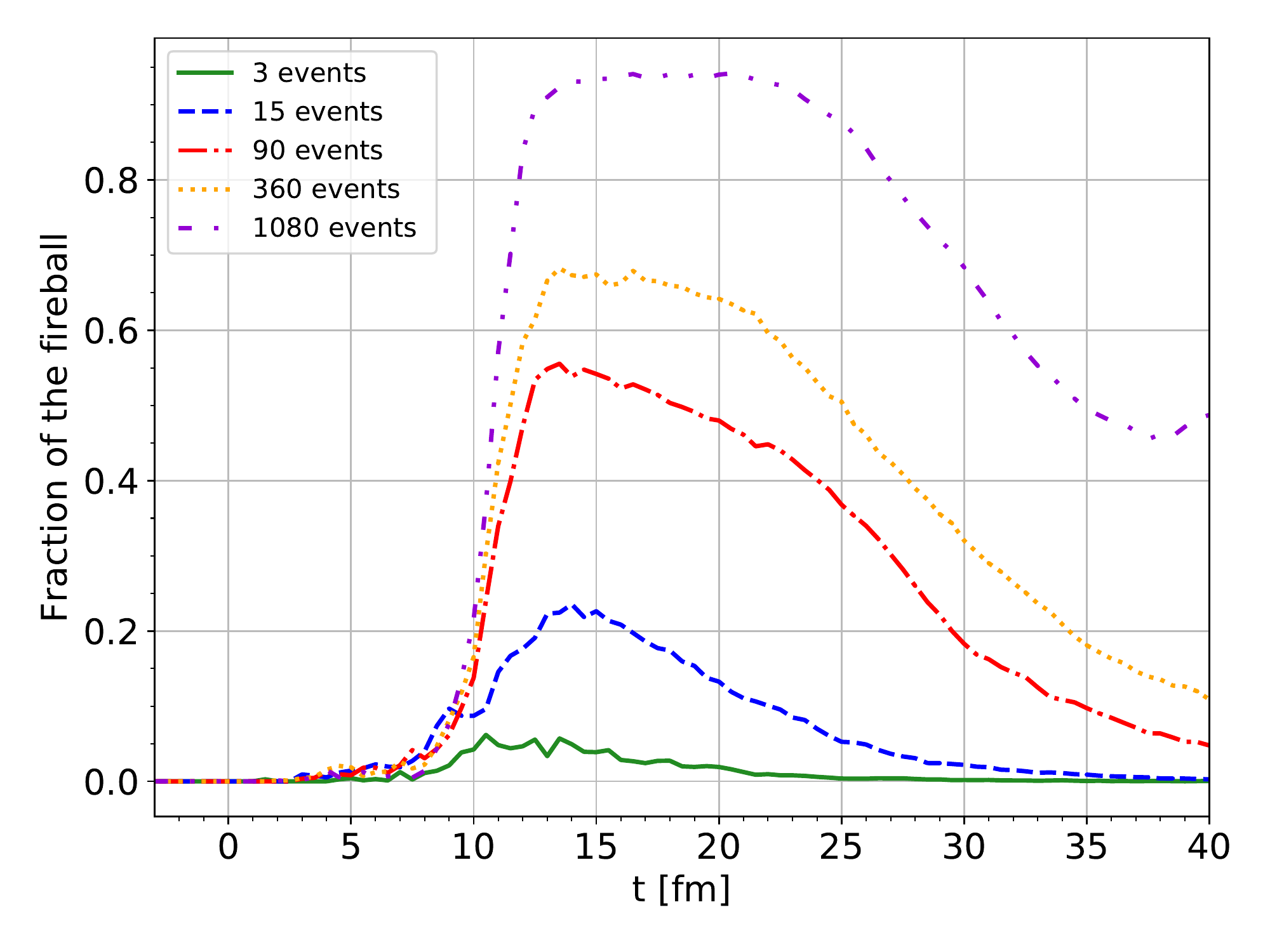}
    \caption{Time evolution of the average fraction of the system satisfying the conditions $X<0.3$,  $Y<0.3$, $\varepsilon>1$ \mev Au+Au at \ela = 1.23 AGeV collisions, 0-5\% centrality class, depending on how many events are taken into account (3, 15, 90, 360 or 1080).}
    \label{fig:percent_n_events}
\end{figure}

\begin{figure}[ht!]
    \centering
    \includegraphics[width=\columnwidth]{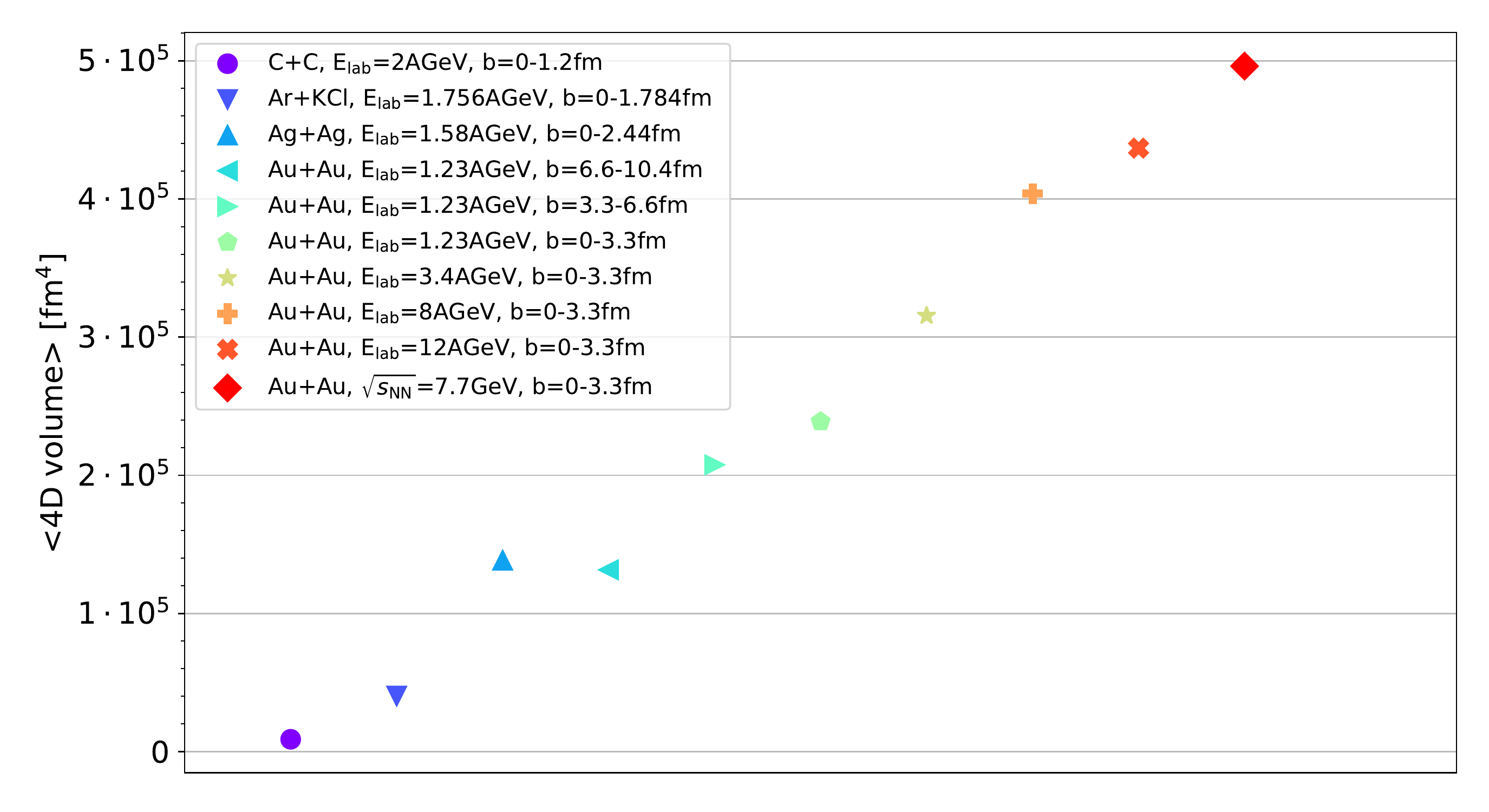}
    \caption{Time integrated volumes of the various systems satisfying the conditions $X<0.3$,  $Y<0.3$, $\varepsilon>1$ \mev.}
    \label{fig:4D_03_03_1}
\end{figure}

\begin{figure}[ht!]
    \centering
    \includegraphics[width=\columnwidth]{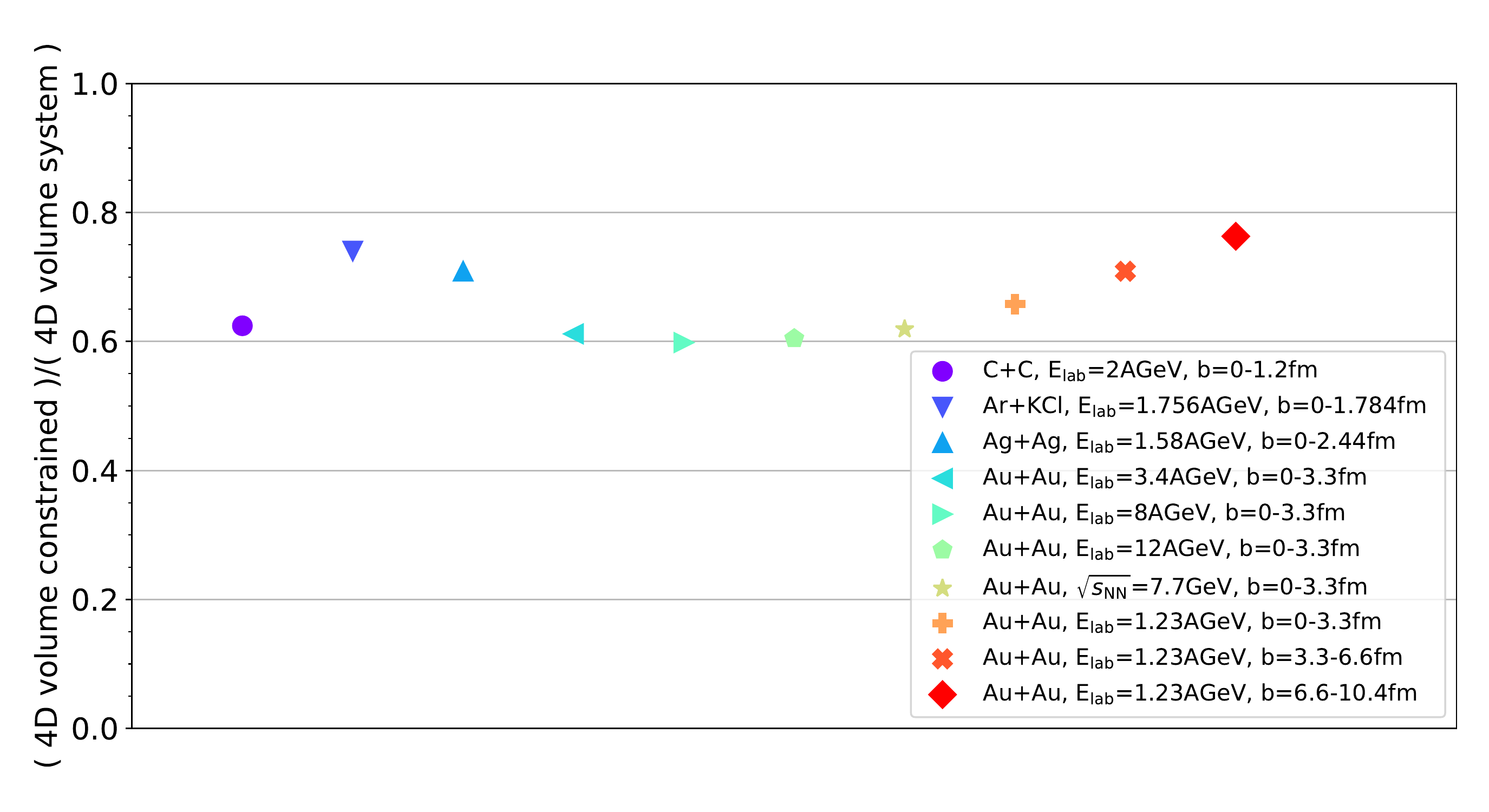}
    \caption{Ratios between time integrated volume satisfying the conditions $X<0.3$,  $Y<0.3$, $\varepsilon>1$ \mev and the time integrated volume of the whole system.}
    \label{fig:4D_03_03_1_fraction}
\end{figure}

To get a more global picture that allows to directly compare different systems to each other, the spatial volumes shown above are now integrated over time. 
Fig.~\ref{fig:4D_03_03_1}, which refers to central Au+Au collisions at \ela = 1.23 AGeV, displays the time integrals (up to t = 40 fm), of the volumes fulfilling the constraints $X, Y < 0.3$ and $\varepsilon > 1$ \mev. Within the same Au+Au ion species the 4D volumes are bigger for more energetic collisions and smaller for more peripheral collisions, in line with the expectations of the system size dependence on beam energy and centrality. Yet we stress that, despite being reasonable, this pattern is not obvious, because we are dealing with volumes subjected to a series of conditions whose effects are hard to foresee. In Fig.~\ref{fig:4D_03_03_1_fraction} we plot the ratio between the time integral of the volume of the system satisfying the chosen constraints and the total unconstrained volume, that presents a few interesting features. First, we notice that all systems lay in the range 0.6-0.8, but with a bit weird pattern. The ratio is already slightly above 0.6 for C+C and it increases when considering the Ar+KCl system, but then decreases for Ag+Ag and Au+Au, possibly suggesting that the effect of the decreasing beam energy per nucleon dominates over the increasing atomic number. Then, consistently with Fig.~\ref{fig:percent_Au_centr_dep}, we notice that the ratio increases for more peripheral collisions. Finally, we observe that at the remaining Au+Au collision energies the ratio remains substantially stable.

\subsection{Morphology of the regions suitable for hydrodynamics}
\begin{figure}[ht!]
    \centering
    \includegraphics[width=\columnwidth]{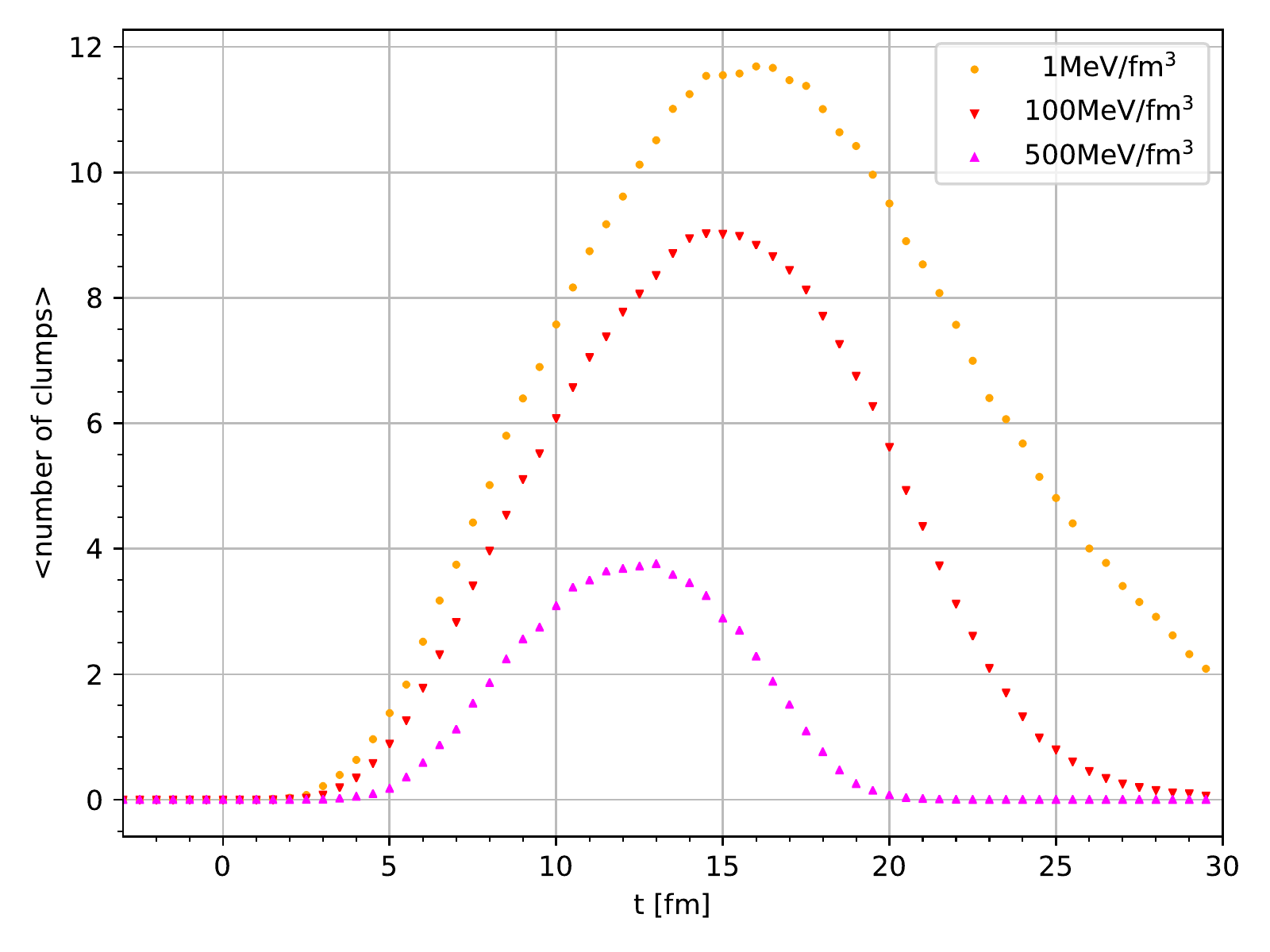}
    \caption{Time evolution of the average number of connected regions (clumps) satisfying the constraints $X_{ebe}<0.3$ and $Y_{ebe}<0.3$ for the energy density thresholds $\varepsilon >$  1, 100 or 500 \mev, in Au+Au collisions at \ela = 1.23 AGeV, 0-5\% centrality class.}
    \label{fig:number_of_clumps}
\end{figure}

\begin{figure}[ht!]
    \centering
    \includegraphics[width=\columnwidth]{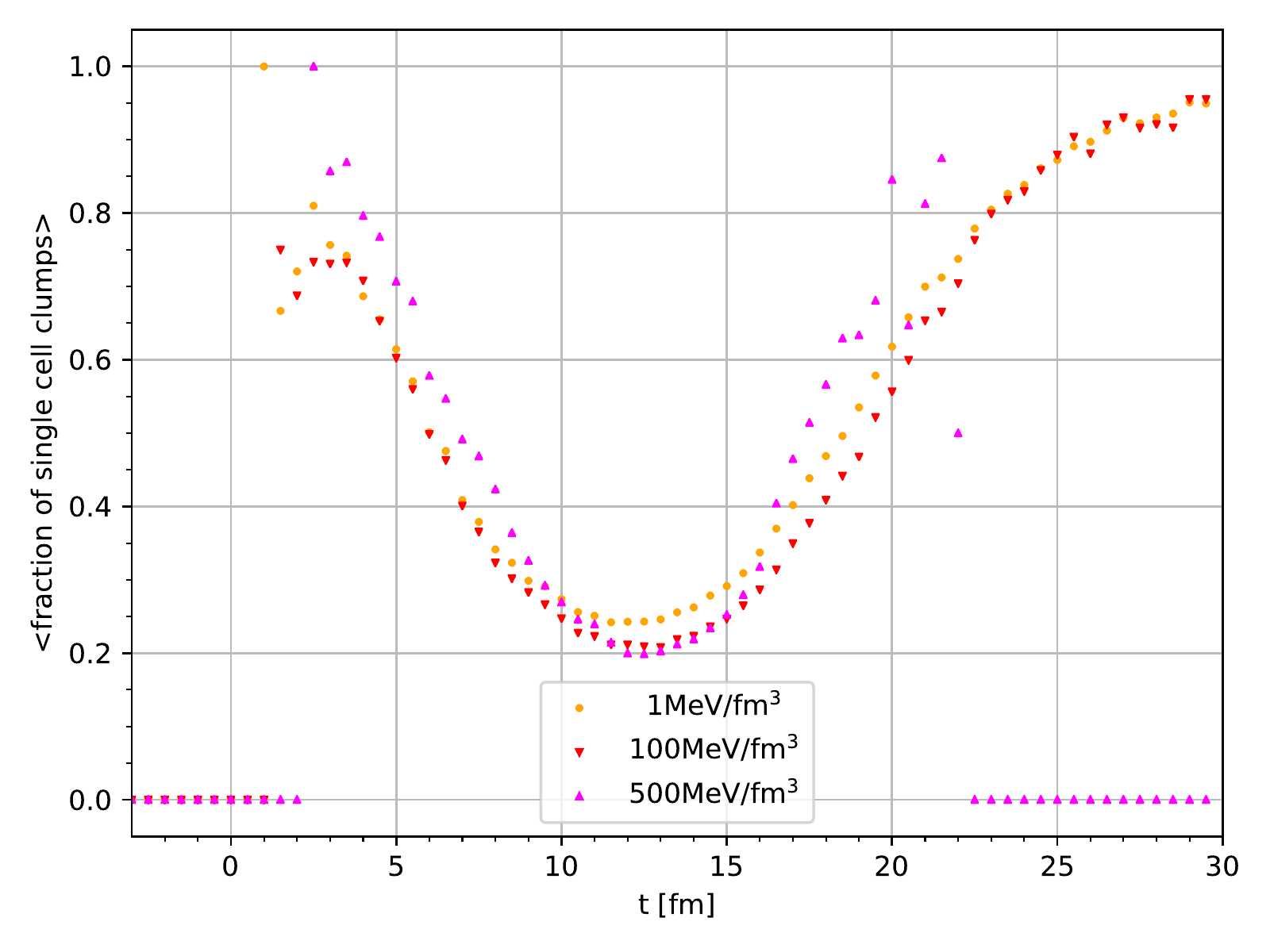}
    \caption{Time evolution of the average compactness of the connected regions (clumps) satisfying the constraints $X_{ebe}<0.3$ and $Y_{ebe}<0.3$ for the energy density thresholds $\varepsilon >$  100 \mev or 500 \mev, in Au+Au collisions at \ela = 1.23 AGeV, 0-5\% centrality class.}
    \label{fig:fraction_of_single_cell_clumps}
\end{figure}

\begin{figure}[ht!]
    \centering
    \includegraphics[width=\columnwidth]{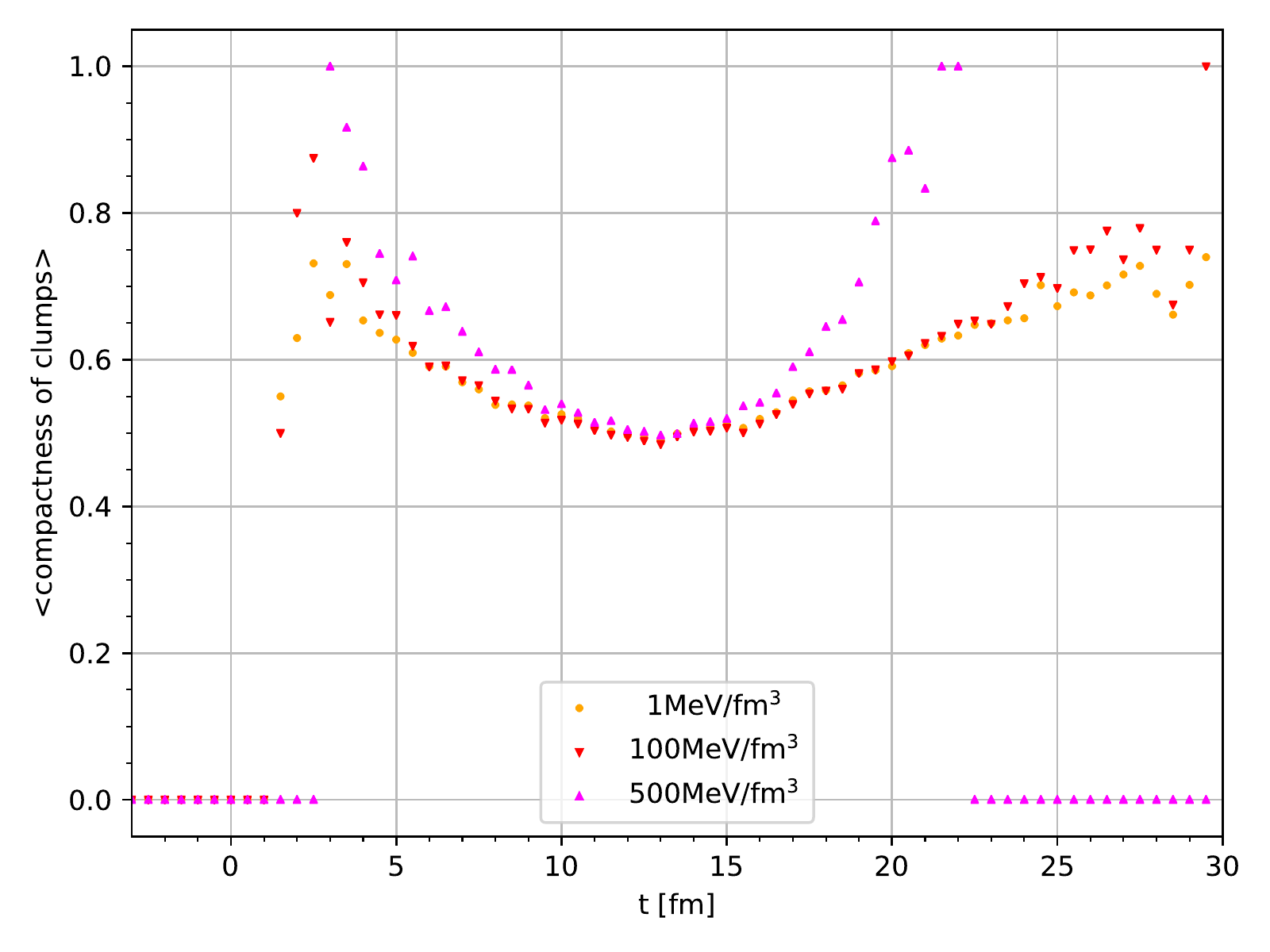}
    \caption{Time evolution of the average compactness of the connected regions (clumps) satisfying the constraints $X_{ebe}<0.3$ and $Y_{ebe}<0.3$ for the energy density thresholds $\varepsilon >$  1, 100 or 500 \mev, in Au+Au collisions at \ela = 1.23 AGeV, 0-5\% centrality class.}
    \label{fig:compactness_of_clumps}
\end{figure}

We now investigate how much fragmented or compact the connected regions of the system that satisfy a certain set of constraints, from now on simply called ``clumps'', are. This investigation is motivated by practical purposes involving the implementation of hydrodynamic codes. For example, the reliability of the results could be limited when modeling a large cubic volume which contains ``bubbles'' or with internal spinodal instabilities\cite{Danielewicz:2019mvp} (not treated in this work) by assuming that fluid description applies everywhere inside it. Moreover, in general numerical hydrodynamics need specific assumptions at the borders of the grid in which the fluid is discretized and evolved. In particular, in the context of heavy ion collisions, depending on the details of the  particlization process\cite{Huovinen:2012is}, the morphology of the clumps might have practical implications when dealing with transitions between hydrodynamic and transport descriptions. In fact, a proper handling of the interfaces between the two regimes is more challenging in the case of a fluid with a complex topology, with highly fragmented clumps and cavities, very irregular shapes and small volume over surface ratios than in the case of regular, smooth hypersurfaces.\\
It is worth to mention that the size and the shape of a homogeneous thermal region can be related to fluctuations and causal correlations, which can be assessed with two particle correlation analysis and Hanbury Brown - Twiss (HBT) interferometry\cite{Heinz:1999rw,Takahashi:2009na,Li:2021day}.\\
We limit this morphological investigation to the event by event case, because we have just seen in Section~\ref{sec:int_volumes} that when considering the average energy momentum tensor most of the system satisfies the constraints $X, Y < 0.3$ and $\varepsilon > 1$ \mev, so in that case fragmentation should be a less relevant phenomenon.  Given the complexity of this topic, we limit our analysis to three basic properties: the average number of clumps that form in the system, the percentage of clumps which is composed by just one cell and a very rough estimate of their average \emph{compactness}. The compactness is defined simply as the ratio between the volume of the smallest parallelepiped with faces parallel to the coordinate planes that encloses the clump and the volume of the cells of the clump. It is clear that this definition has several shortcomings, for example it does not provide any information about the presence of cavities and it can be misleading in the case of clumps having irregular shapes, for example in the case of filaments, with very different outcomes depending on how much they are straight or bent. Nevertheless, a value of the compactness close to zero should suggest heavily fragmented, irregular clumps, while a value above 0.5 should qualitatively indicate the formation of relatively compact regions, ruling out structures like bubbles with big cavities inside or shapes like curly filaments. In determining this quantity, we consider only clumps composed by at least 2 cells, so to avoid that single cell clumps, which have by definition compactness 1 despite their small dimensions, lead to an overestimate the average compactness. In Fig.~\ref{fig:number_of_clumps} we show the number of clumps subjected to the constraints $X_{ebe},Y_{ebe} < 0.3$ versus time in the case of central Au+Au collisions at \ela = 1.23 AGeV. We notice that the fireball contains on average up to 9 distinct clumps at t $\approxeq$ 15 fm when considering a minimum energy density $\varepsilon$ \textgreater 100 \mev. This amount is reduced to roughly 4 clumps at t $\approxeq$ 12.5 fm for $\varepsilon$ \textgreater 500 \mev. The formation of clumps starts earlier and ends later when the energy density constraint is lower, as expected in a system that transits from an initial phase of compression and heating to a phase of expansion and cooling. From a practical point of view, the simultaneous presence of several separated clumps should encourage to pay a special attention in the particlization process when a freeze-out hypersurface for a phase transition in a hydro code is defined at these low energy densities, not only regarding a precise determination of the hypersurface\cite{Huovinen:2012is} and the enforcement of conservation laws\cite{Schwarz:2017bdg}, but possibly also from the influx of hadrons emitted by another clump that adds to the negative Cooper-Frye contributions\cite{Oliinychenko:2014tqa}. Fig.~\ref{fig:fraction_of_single_cell_clumps} shows the average fraction of clumps composed by just one cell of 1 $\mathrm{fm^3}$ volume. We notice that clumps consisting of single cells dominate when clumps start to form and when they are dissolving. However, between 10 and 15 fm, more or less when the number of clumps is at its highest, the fraction of small clumps represents only 20\% of the total. Fig.~\ref{fig:compactness_of_clumps} show the average compactness of clumps, which tends to be larger at the beginning and the end of their presence, when they are smaller, than between 10 and 15 fm, when it goes down to 0.5.



\section{Conclusions}
We used the SMASH hadronic transport approach to compute the energy momentum tensor of the systems of hadrons created in heavy ion collisions, neglecting nuclear potentials and taking into account only the particles that collided at least once. The average pressure anisotropy (Eq.\ref{eq:press_aniso}) and off-diagonality (Eq.\ref{eq:off_dia}) have been evaluated either from the average energy momentum tensor of all runs of a certain type of system ($X, Y$ parameters) or from the energy momentum tensor of individual event by event simulations ($X_{ebe}, Y_{ebe}$ parameters). We focused mostly on Au+Au collisions at $E_{lab}$ = 1.23 AGeV in the 0-5\% centrality class, but we also considered (see Table \ref{table:summary}) C+C, Ar+KCl, Ag+Ag and Au+Au collisions in the energy range $E_{lab}$=1.23 AGeV - \snn=7.7 GeV, exploring the differences between heavy and light ions, the impact of the centrality class and the beam energy dependence.\\
In general, we found that, on an event by event basis, on average only a small fraction of the system volume satisfies the constraints on $X_{ebe}$ and $Y_{ebe}$ that can be considered suitable for the application of hydrodynamics consistent with kinetic theory. The situation changes dramatically when considering the average energy momentum tensor of many events. In this work we focused on 1080 events and we found that, after a relatively short initial phase in which the nuclei have not completely penetrated each other, for some time most of the system with energy density above 1 \mev is close to local equilibrium, albeit this percentage then steadily decreases with time. When considering higher energy density thresholds, like 100 or 500 \mev, the fraction of the system suitable for hydrodynamics is considerably smaller and, due to the rapid cooling of the fireball, it dissolves faster.\\
As one would naively expect, we found that the volume satisfying given $X$, $Y$ constraints decreases from central to peripheral collisions, nevertheless, when considering the ratio with the total unconstrained volume, the fraction of the system suitable for hydrodynamics can be even larger. We also found that light ions central collisions almost never meet the criterion for hydrodynamic applicability in the event by event case, in particular for small systems like C+C, nevertheless, when considering the average energy momentum tensor of many events, the situation is more favorable, albeit these systems tend to dissolve quickly.\\
In conclusions, our investigations suggest that, when considering a single event, in simulations at HADES - CBM - (low end) BESII energies based on hydrodynamic approaches it might be challenging to get results consistent with transport models, like SMASH, because in most cases a very large part of the system significantly deviates from the local thermal equilibrium condition. However, physical observables are extracted from a collection of many events and this circumstance does not seem to hinder the possibility of obtaining bulk dynamics predictions from hydrodynamics in agreement with coarse grained transport models even during the ions compression stage if a proper EoS is chosen\cite{OmanaKuttan:2022the}. On the other hand, after an initial phase of equilibration the average energy momentum tensor across many ($\approx$ 1 K) events seems to posses an overall good degree of thermalization that can extend to most of the system volume (provided that the constraints on $X$, $Y$ are not too tight), albeit it tends to rapidly wane as the system expands and dilutes. \\
We briefly address what we think might be the most relevant concerns about this work. 1) We based our assessments on the pressure anisotropy $X (X_{ebe})$ and the off-diagonality $Y (Y_{ebe})$ because we considered these quantities as the most significant. However, $X (X_{ebe})$ and $Y (Y_{ebe})$ represent just the basic pre-requisites and many aspects which are still relevant for a successful implementation of a hydrodynamic model were left out. 2) All the results are based only on SMASH and they would be more robust if confirmed by other transport codes like UrQMD or PHSD, which have different conceptual and implementation details. We present a very limited comparison with previous results~\cite{Oliinychenko:2015lva} obtained with UrQMD in \ref{sec:comp_w_urqmd}.
 3) The impact of lattice resolution has not been fully investigated, neither we assessed what could be the optimal discretization scale for hydrodynamic simulations, that commonly adopt grid spacing around 0.2 fm.\\ Other minor points, like the impact of spectators and the alternative use of test particles are treated in \ref{sec:spectators} and \ref{sec:testparticles}, respectively, together with a few tables with the time integrated volumes of the system satisfying a certain set of $X (X_{ebe})$, $Y (Y_{ebe})$,  $\varepsilon$ criteria. We also recall that it is possible to visualize many more plots, covering the full combination of the parameters chosen in this study, on the website \url{\ws}.\\

\begin{acknowledgements}
The authors acknowledge funding by the Deutsche Forschungsgemeinschaft (DFG, German Research Foundation) – Project number 315477589 – TRR 211 and the support by the State of Hesse within the Research Cluster ELEMENTS (Project ID 500/10.006). G. Inghirami thanks Dmytro Oliinychenko and all the members of the research group, in particular Justin Mohs, for discussions and fruitful suggestions.\\ The computational resources were provided by the Center for Scientific Computing (CSC) of the Goethe University, Frankfurt am Main, Germany and by the IT Division at the GSI Helmholtzzentrum für Schwerionenforschung, Darmstadt, Germany.
\end{acknowledgements}
\vspace*{1mm}

\appendix

\section{Test particles}
\label{sec:testparticles}
\begin{figure*}[!ht]
    \centering
    \begin{minipage}{.49\textwidth}    
        \includegraphics[width=\textwidth]{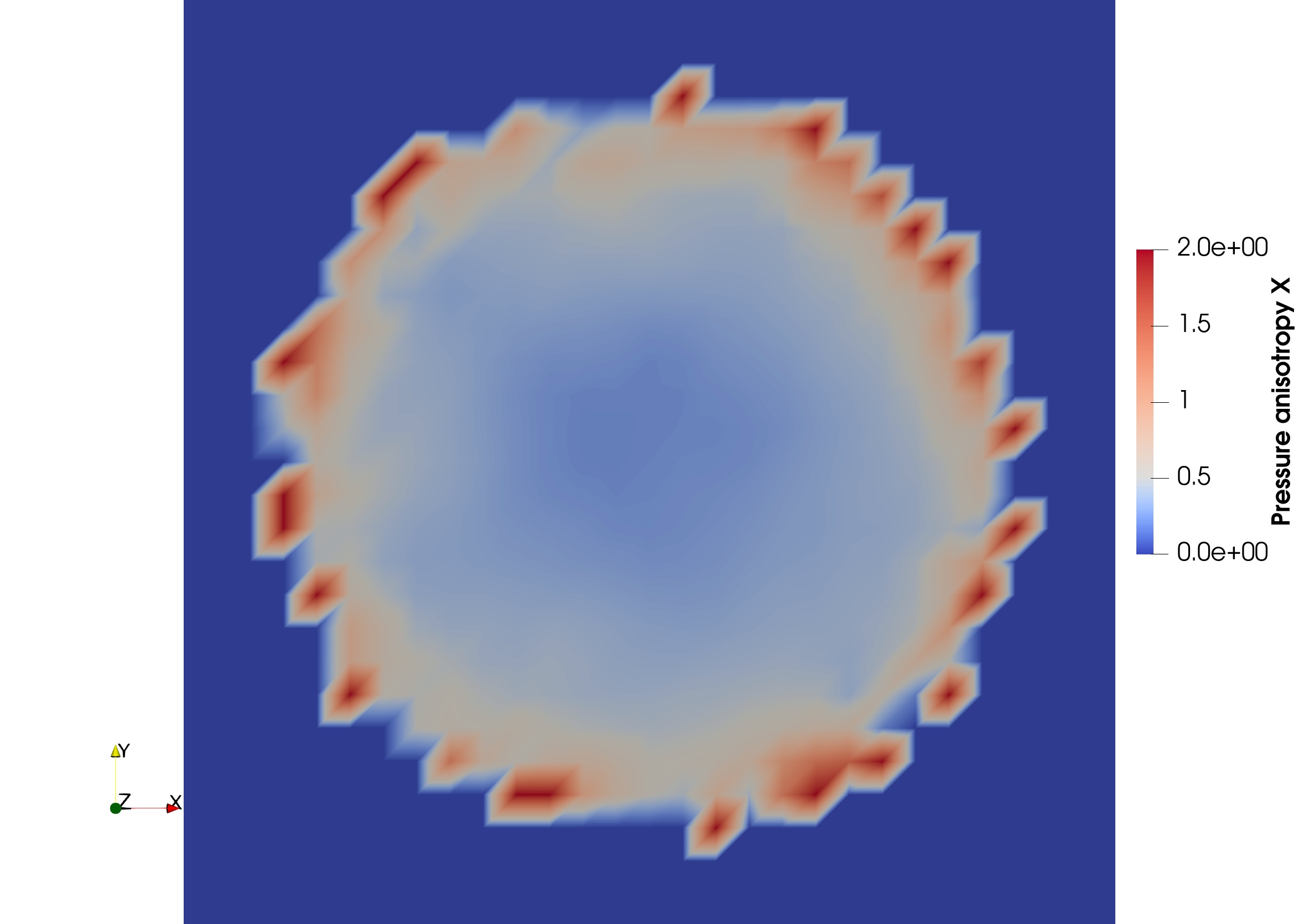}   
    \end{minipage}
    \begin{minipage}{.49\textwidth}
        \includegraphics[width=\textwidth]{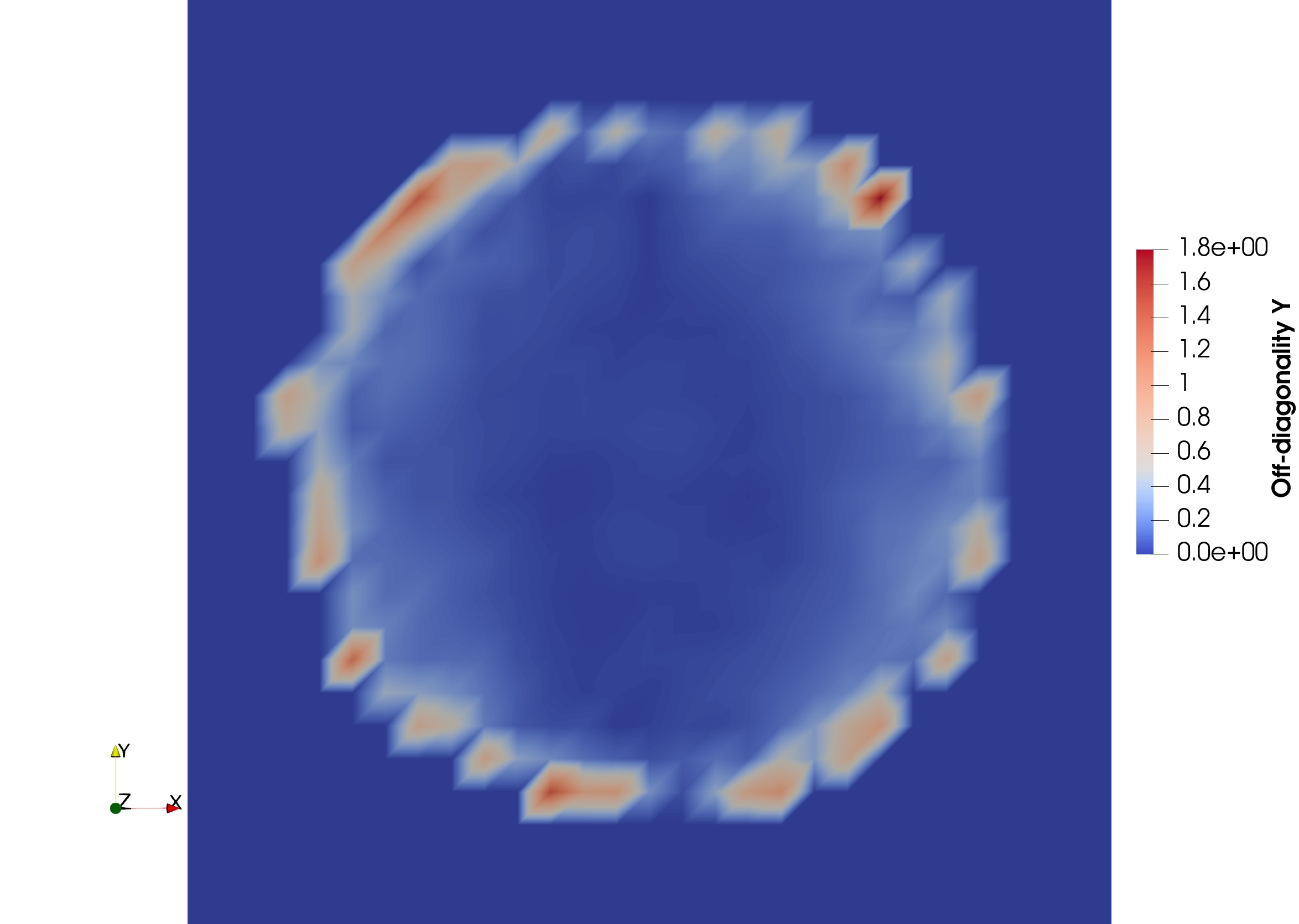}
    \end{minipage}
    \caption{Au+Au collisions at \ela = 1.23 AGeV, 0-5\% centrality class, 1080 events, t=10 fm, z=0. Average energy momentum tensor evaluated taking into account only participant hadrons. Left: pressure anisotropy $X$, right: off-diagonality $Y$.}
    \label{fig:paraview_XY_avg}
\end{figure*}

\begin{figure*}[ht!]
    \centering
     \begin{minipage}{.49\textwidth}    
    \includegraphics[width=\textwidth]{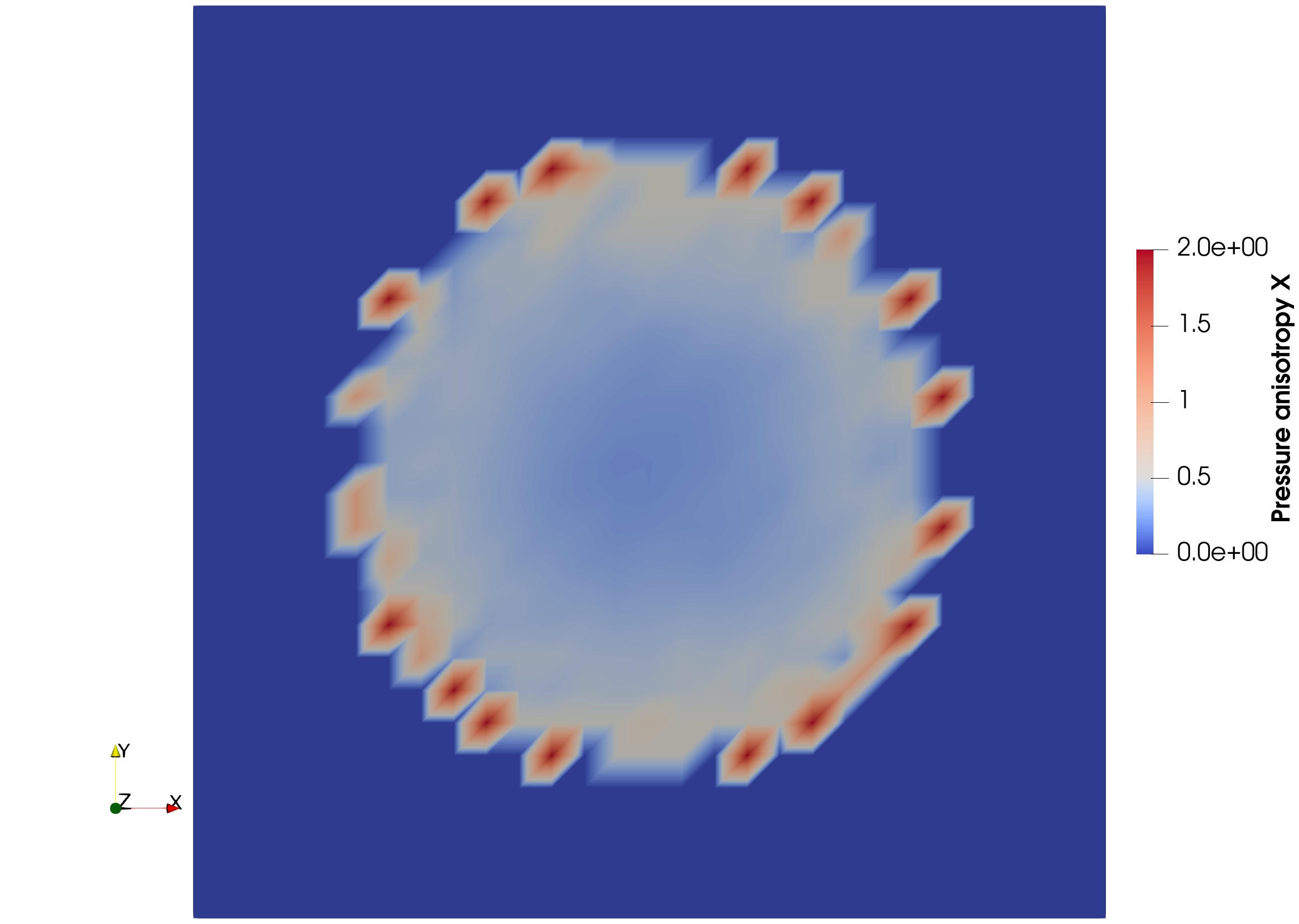}
    \end{minipage}
    \begin{minipage}{.49\textwidth}
    \includegraphics[width=\textwidth]{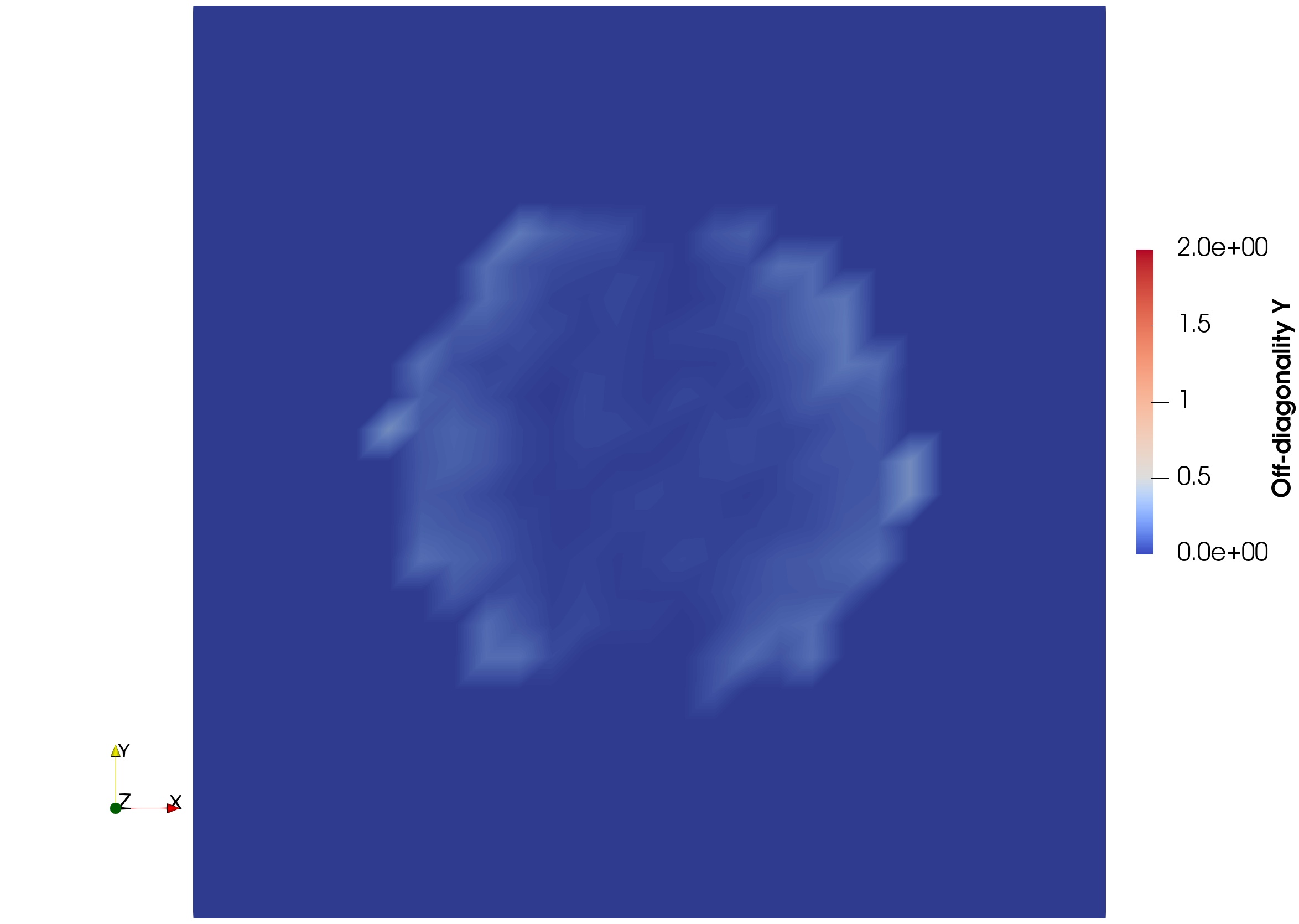}
     \end{minipage}
    \caption{Au+Au collisions at \ela = 1.23 AGeV, 0-5\% centrality class, 1 event with 1080 test particles, t=10 fm, z=0. Average energy momentum tensor evaluated taking into account only participant hadrons. Left: pressure anisotropy $X$, right: off-diagonality $Y$.}
    \label{fig:paraview_XY_tp}
\end{figure*}

SMASH is able to run simulations by using test particles, i.e. by sampling $N_{tp}$ times the number $N$ of initial nucleons, but scaling at the same time the cross section of all interactions by the same factor $N_{tp}$\cite{Weil:2016zrk}:
\begin{equation}
    N\rightarrow N N_{tp},\qquad
    \sigma\rightarrow \sigma/N_{tp}.
\end{equation}
This replacement does not change the scattering rate, i.e. the number of interactions per particle per unit of time, nevertheless it makes the system smoother and similar to the average system obtained by running $N_{tp}$ simulations with the same impact parameter. We validate this statement for Au+Au collisions at \ela = 1.23 AGeV in the 0-5\% centrality class by comparing the distributions of the pressure anisotropy $X$ and the off-diagonality $Y$ on the plane z = 0 at t = 10 fm. The variability of the impact parameter hinders precise quantitative assessments, nevertheless the small range of the sampling still allows for qualitative comparisons. Fig.~\ref{fig:paraview_XY_avg} shows the distributions of $X$ and $Y$ computed from the average energy momentum tensor of 1080 events, while Fig.~\ref{fig:paraview_XY_tp} shows the distributions of $X$ and $Y$ computed from the energy momentum tensor of a single event with 1080 test particles. We notice that the magnitudes of $X$ and $Y$ are roughly the same in both contexts. In general, SMASH needs more time to simulate one event with $N_{tp}$ test particles than $N_{tp}$ events without this approach, however in some circumstances test particles might provide a convenient method to initialize the energy momentum tensor in hydrodynamic simulations.

\section{Impact of spectators}
\label{sec:spectators}
\begin{figure*}[!ht]
    \centering
    \begin{minipage}{.49\textwidth}  
    \includegraphics[width=\columnwidth]{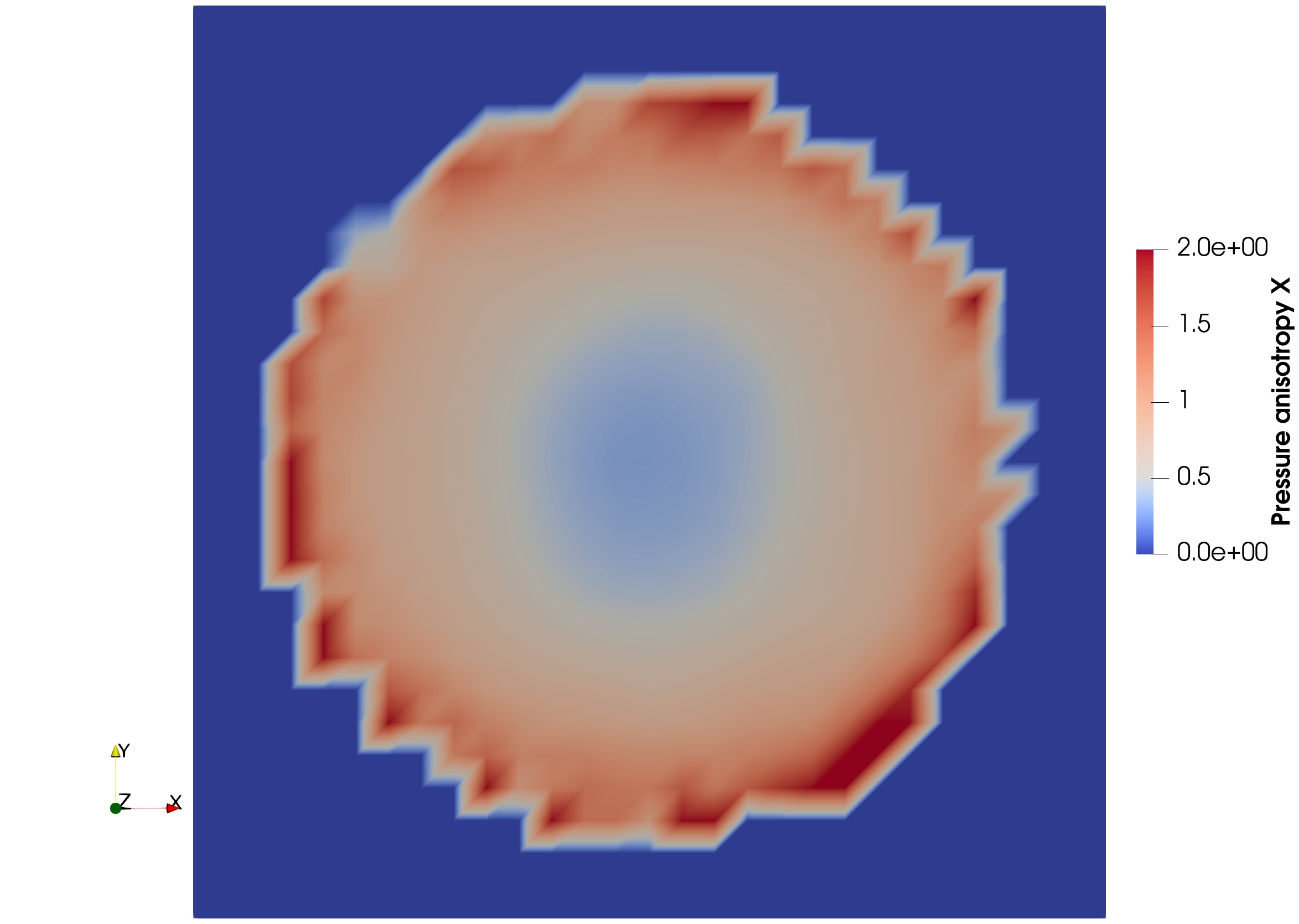}
    \end{minipage}
    \begin{minipage}{.49\textwidth}
    \includegraphics[width=\columnwidth]{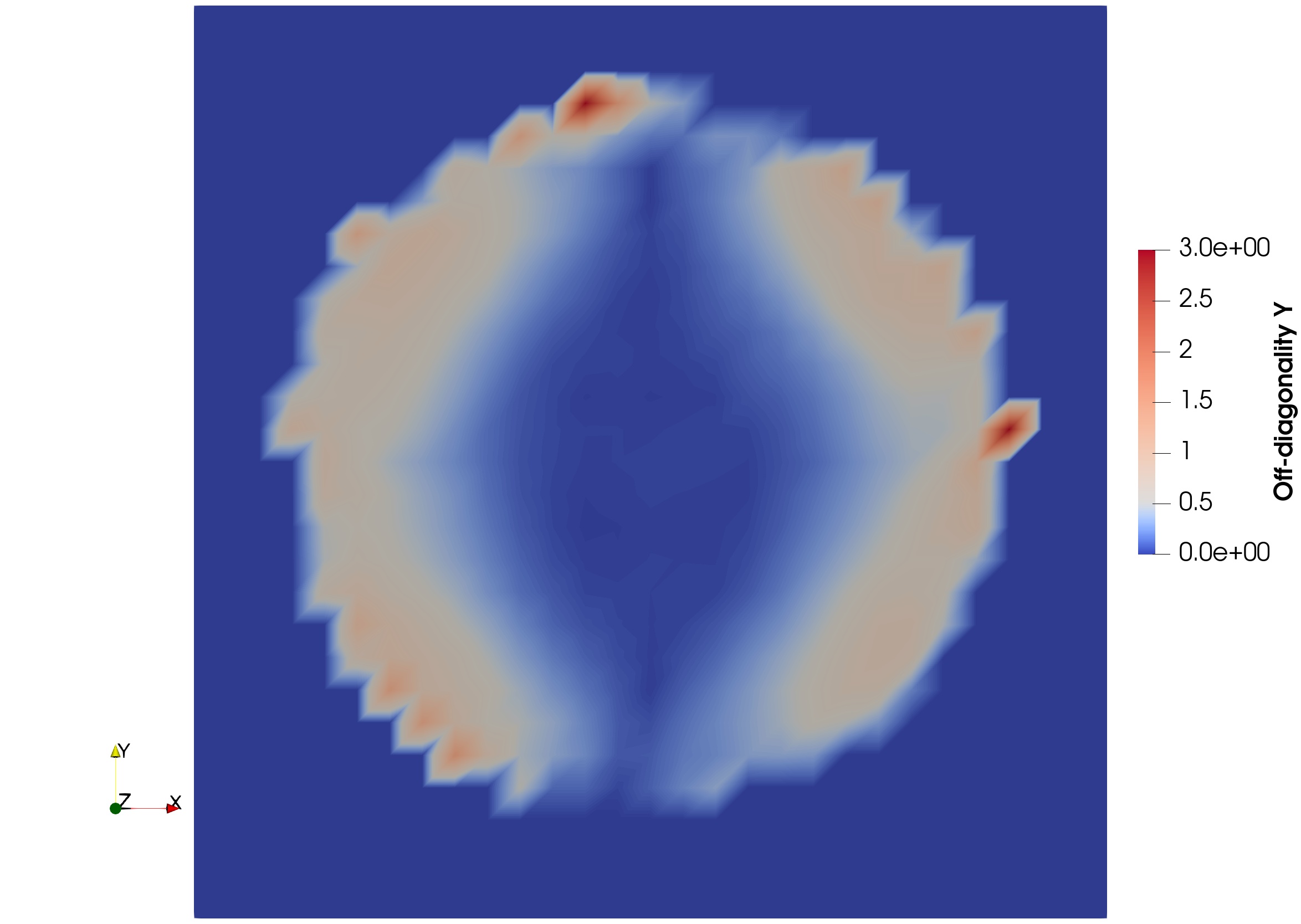}
    \end{minipage}
    \caption{Au+Au collisions at \ela = 1.23 AGeV, 0-5\% centrality class, 1080 events, t=10 fm, z=0. Average energy momentum tensor evaluated taking into account both participant and spectator hadrons. Left: pressure anisotropy X, right: off-diagonality $Y$.}
    \label{fig:paraview_XY_sp}
\end{figure*}

As already mentioned in the description of the procedures followed in this study, we compute the energy momentum tensor of the system by taking into account only the participant hadrons, i.e. the hadrons that had at least one collision after the starting time of the simulation. Since we are neglecting potentials, interactions between hadrons happen only via collisions, therefore the particles that do not collide, called spectators, do not influence the evolution of the system. Nevertheless, we believe that it is interesting to explore what is their impact on the components of the energy momentum tensor. To this aim, we consider again 1080 Au+Au collisions at \ela = 1.23 AGeV, in the 0-5\% centrality class, and we inspect the distributions at time t = 10 fm on the plane z = 0 of the pressure anisotropy $X$ and the off-diagonality $Y$ of the average energy momentum tensor. Fig.~\ref{fig:edens_evo_center_Au_snn_dep} shows that at time t = 10 fm the energy density in the center has not reached the peak, yet, suggesting that the main collision process between the nuclei is not yet completed, so it is expected a detectable effect by the spectators that, because the lack of interactions, are likely to have an higher level of anisotropy than the rest of the system. Actually, as Fig.~\ref{fig:paraview_XY_sp} shows, the inclusion of the spectators in the computation of the energy momentum tensor leads to a significant increase both of the pressure anisotropy $X$ and the off-diagonality $Y$ with respect to the case in which only participants are taken into account (Fig.~\ref{fig:paraview_XY_avg}).

\section{Comparison with UrQMD results}
\begin{figure}[ht!]
    \centering
    \includegraphics[width=\columnwidth]{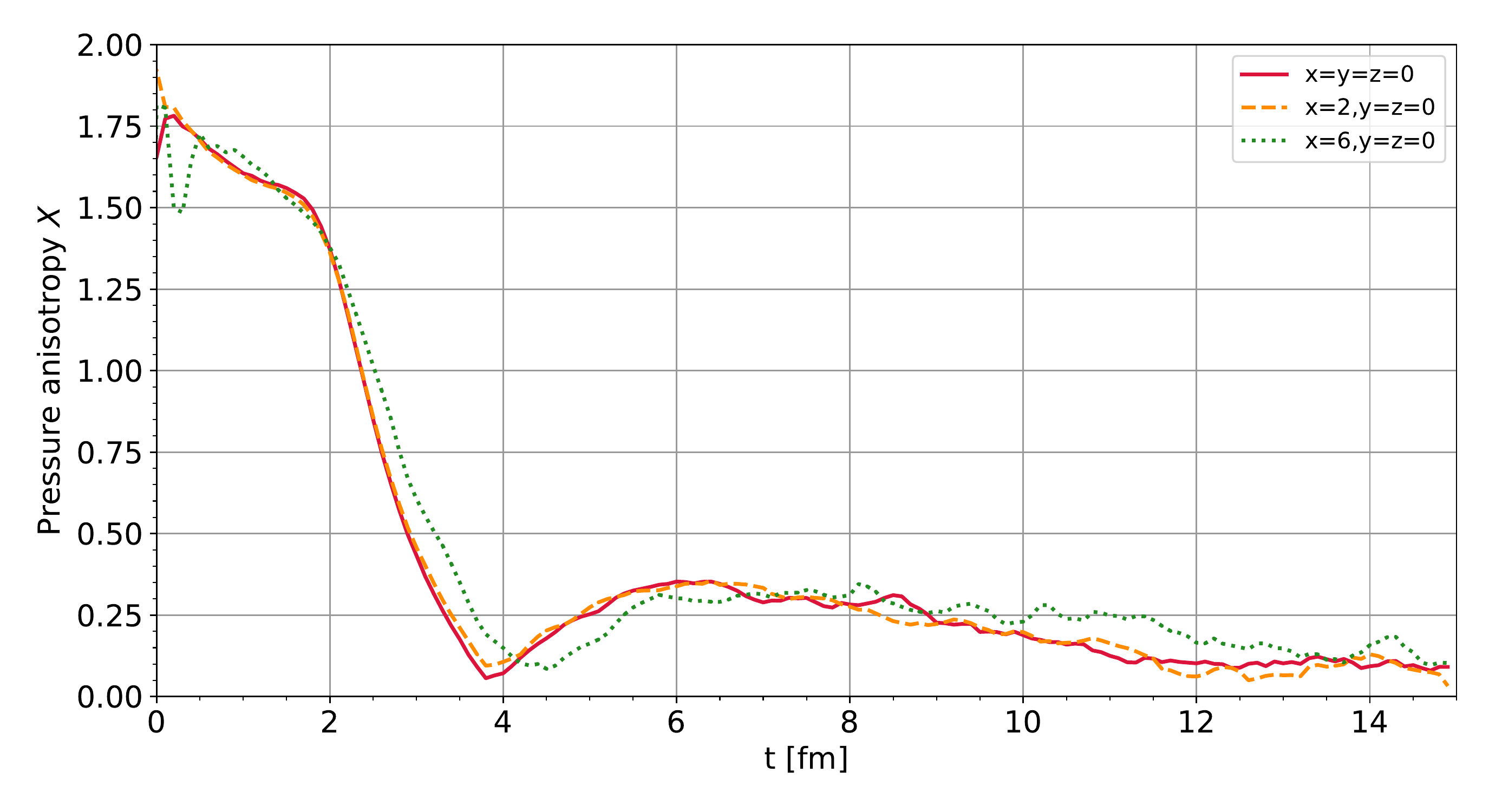}
    \caption{Time evolution of $X$ at x={0,2,6} fm and y=z=0 fm in Au+Au collisions at \ela = 80 AGeV, b=6 fm.}
    \label{fig:comp_urqmd_1D}
\end{figure}

\begin{figure}[ht!]
    \centering
    \includegraphics[width=\columnwidth]{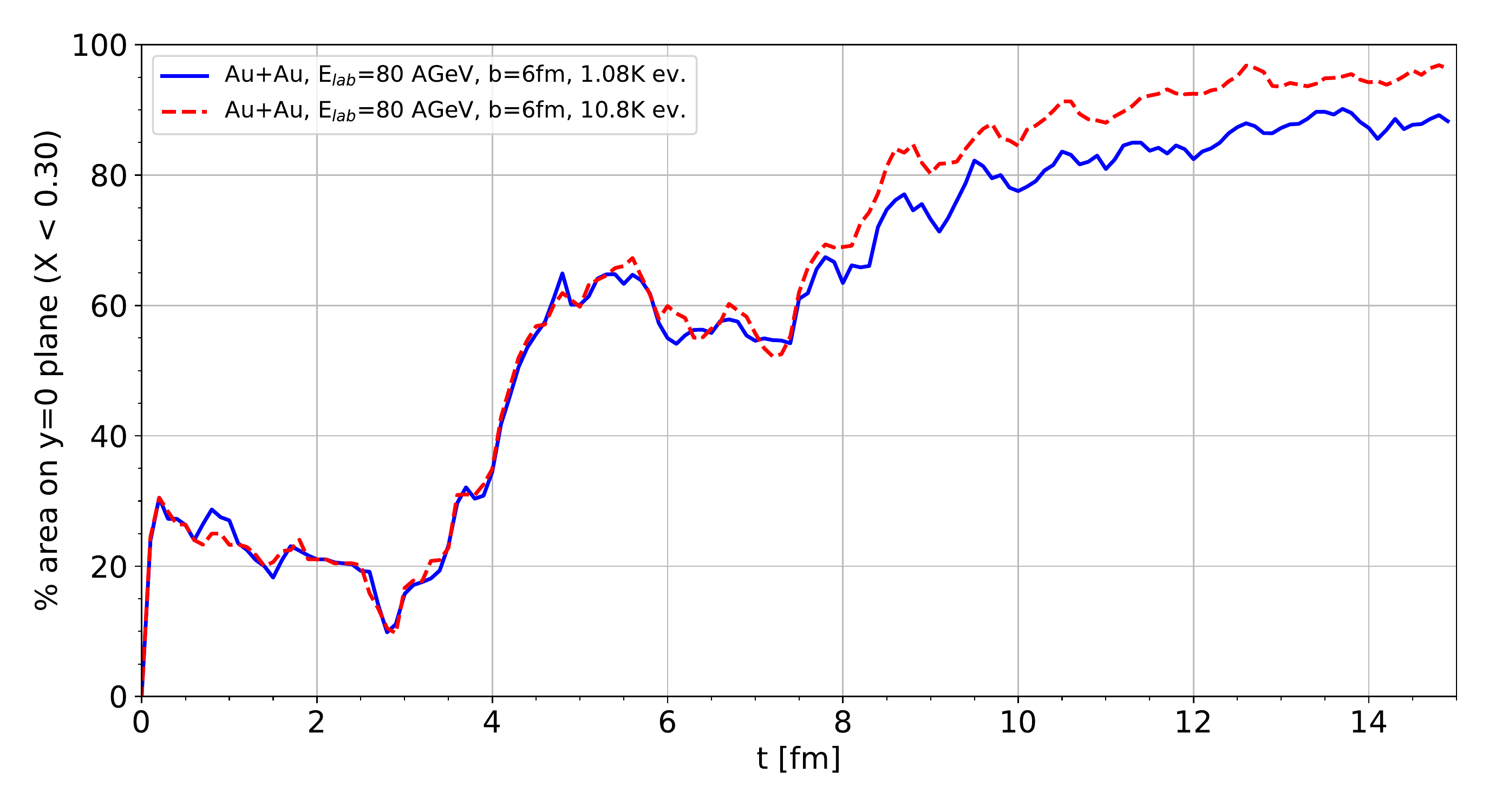}
    \caption{Comparison between the time evolution of the percentage of the system laying on the xz plane that fulfills the condition $X<0.3$ when considering 1080 or 10800 events. The Figure refers to Au+Au collisions at \ela = 80 AGeV, b=6 fm.}
    \label{fig:comp_urqmd_2D}
\end{figure}

Here we perform a brief comparison with the results obtained with UrQMD in Ref.~\cite{Oliinychenko:2015lva} for Au+Au collisions at \ela = 80 AGeV with b = 6 fm, by using approximately the same method as in the present work.\\
Fig.~\ref{fig:comp_urqmd_1D}, to be compared with Fig. 6 in Ref.~\cite{Oliinychenko:2015lva}, shows the time evolution of $X$ at x = 0, 2, 6 fm and y = z = 0 fm. The number of events is slightly different, 1080 in our case and 1000 in the reference, and we also remind that the definitions of the initial time t = 0 fm are not the same in the two codes, nevertheless in the present context these differences do not impede a qualitative comparison. We notice that the most relevant features of the pressure anisotropy, like an initial maximum close to the theoretical limit of 2, the minimum at $3<t<4$ fm, the relative maximum at $t\approxeq6$ fm and the following decrease are common to both figures. However, in our case $X$ seems to be slightly smaller and in the point at x = 6 fm we do not observe any increase after the minimum when compared with the two other points, as it happens in the plot in the Ref.~\cite{Oliinychenko:2015lva}, but instead a small delay shift in the profile. The fact that SMASH results produce smaller $X$ values probably explains why in Fig.~\ref{fig:comp_urqmd_2D} we obtain, for $t\gtrapprox 10$ fm, a percentage of the system larger than 80\% that fulfills the condition $X<0.3$, while in the left part of the Fig. 3 of the Reference it does not exceed 60\% until t = 15 fm, except for an initial peak at around t = 5 fm that in our case is barely visible. We also notice that in Fig.~\ref{fig:comp_urqmd_2D} there is not much difference between the curves based on 1080 or 10800 events only until t = 8 fm, while at later times 1080 events are not sufficient to saturate the results, as it is the case in Ref.~\cite{Oliinychenko:2015lva}.\\
Overall, our new results based on SMASH seem to be in broad qualitative agreement with the old ones based on UrQMD, however it is also clear that there are non negligible differences in the dynamics of the system between the two codes, which, in the case of the second plot, are probably emphasized by fixing a precise threshold, thus leading to differences in accounting up to 20\% of the system volume below that threshold.

\label{sec:comp_w_urqmd}

\section{Integrated volume tables}
\label{sec:integrated_volumes_tables}
Here we provide some tables with the time integrated volume of the system simultaneously satisfying the constraints on the pressure anisotropy: $X (X_{ebe}) < 0.1, 0.3, 0.5$ and on the off-diagonality:   $Y (Y_{ebe}) < 0.1, 0.3, 0.5$, for various minimum energy density thresholds, i.e 1, 100 and 500 \mev. The systems under consideration are listed in table \ref{table:summary}. The results in tables \ref{table:X01Y01}-\ref{table:X05Y05} refer to the average volumes $V_1$, $V_{100}$, $V_{500}$, corresponding $\varepsilon > 1, 100, 500$ \mev respectively, with their standard deviations $\sigma(V_1)$ , $\sigma(V_{100})$ and $\sigma(V_{500})$, subjected on the conditions on $X_{ebe}$ and $Y_{ebe}$ reported in the captions (on an event by event basis). Tables \ref{table:AX01Y01}-\ref{table:AX05Y05} refer to the volumes subjected on the conditions on $X$ and $Y$ reported in the captions (applied to the average energy momentum tensor). All the results presented in this section refer to sets of 1080 events. The unit of measurement for the time integrated volume is $\mathrm{fm^4}$.\\

\begin{table}
    \centering
    \begin{tabular}{|l|l|c|c|c|c|c|c|} 
        \hline
        Abbrev. & Ions & Coll. energy & b [fm] & Centr.\\ 
        \hline
        CC, 2, c & C+C &   $E_{l}$=2 AGeV & 0-1.2 & 0-5\%\\
        ArKCl, 1.76, c & Ar+KCl &  $E_{l}$=1.756 AGeV & 0-1.784 & 0-5\%\\
        AuAu, 1.58, c & Ag+Ag  &  $E_{l}$=1.58 AGeV & 0-2.44 & 0-5\%\\
        AuAu, 1.23, c & Au+Au  &  $E_{l}$=1.23 AGeV & 0-3.3 & 0-5\%\\ 
        AuAu, 1.23, s & Au+Au   &  $E_{l}$=1.23 AGeV & 3.3-6.6  & 5-20\%\\ 
        AuAu, 1.23, p & Au+Au   &  $E_{l}$=1.23 AGeV & 6.6-10.4  & 20-50\%\\ 
        AuAu, 3.4, c & Au+Au   &  $E_{l}$=3.4 AGeV & 0-3.3 & 0-5\%\\   
        AuAu, 8, c & Au+Au   &  $E_{l}$=8 AGeV & 0-3.3 & 0-5\%\\ 
        AuAu, 12, c & Au+Au   &  $E_{l}$=8 AGeV & 0-3.3 & 0-5\%\\ 
        AuAu, 7.7, c   & Au+Au & \snn=7.7 GeV & 0-3.3 & 0-5\% \\
        
        \hline
    \end{tabular}
    \caption{List of the systems under examination: abbreviation, ion species, collision energy, impact parameter b range (in fm) and centrality class.}
    \label{table:summary}
\end{table}

\begin{table}
    \centering
    \begin{tabular}{|l|c|c|c|c|c|c|} 
        \hline
        Ions & $V_1$ & $\sigma_{V_1}$ &  $V_{100}$ & $\sigma_{V_{100}}$ & $V_{500}$ & $\sigma_{V_{500}}$\\ 
        \hline
        CC, 2, c &  0.00 &0.11  &  0.00 &0.02  &  0.00 &0.00\\
        ArKCl, 1.76, c  &  0.07 &1.03  &  0.04 &0.52  &  0.00 &0.08\\
        AgAg, 1.58, c  &  0.59 &3.46  &  0.42 &2.42  &  0.13 &1.06\\
        AuAu, 1.23, c  &  1.70 &6.56  &  1.32 &4.93  &  0.49 &2.50\\
        AuAu, 1.23, s  &  1.02 &5.10  &  0.73 &3.66  &  0.20 &1.50\\
        AuAu, 1.23, p  &  0.33 &2.93  &  0.19 &1.83  &  0.03 &0.48\\
        AuAu, 3.4, c  &  1.41 &5.48  &  1.06 &3.89  &  0.48 &2.15\\
        AuAu, 8, c  &  1.31 &5.29  &  0.95 &3.47  &  0.38 &1.60\\
        AuAu, 12, c  &  1.29 &5.38  &  0.92 &3.48  &  0.34 &1.50\\
        AuAu, 7.7, c  &  1.57 &6.39  &  1.08 &4.13  &  0.39 &1.60\\
        
        \hline
    \end{tabular}
    \caption{Results for $X_{ebe} < 0.1$ and $Y_{ebe} < 0.1$.}
    \label{table:X01Y01}
\end{table}

\begin{table}
    \centering
    \begin{tabular}{|l|c|c|c|c|c|c|} 
        \hline
        Ions & $V_1$ & $\sigma_{V_1}$ &  $V_{100}$ & $\sigma_{V_{100}}$ & $V_{500}$ & $\sigma_{V_{500}}$\\  
        \hline
      CC, 2, c  &  0.10 &1.02  &  0.02 &0.36  &  0.00 &0.00\\
      ArKCl, 1.76, c  &  1.87 &5.62  &  0.96 &3.19  &  0.09 &0.75\\
      AgAg, 1.58, c  &  14.4 &19.0  &  9.86 &13.4  &  2.88 &5.91\\
      AuAu, 1.23, c  &  42.1 &37.2  &  31.6 &28.6  &  11.3 &14.3\\
      AuAu, 1.23, s  &  26.3 &29.8  &  17.9 &21.6  &  4.33 &8.64\\
      AuAu, 1.23, p &  8.68 &17.2  &  4.6 &10.9  &  0.45 &2.42\\
      AuAu, 3.4, c &  34.4 &30.5  &  25.2 &21.9  &  10.9 &12.0\\
      AuAu, 8, c  &  32.7 &28.8  &  22.6 &18.7  &  8.68 &8.86\\
      AuAu, 12, c  &  32.5 &28.9  &  21.9 &18.3  &  7.59 &7.81\\
      AuAu, 7.7, c  &  38.5 &33.3  &  25.3 &21.5  &  8.47 &8.47\\
        \hline
    \end{tabular}
    \caption{Results for $X_{ebe} < 0.1$ and $Y_{ebe} < 0.3$.}
    \label{table:X01Y03}
\end{table}

\begin{table}
    \centering
    \begin{tabular}{|l|c|c|c|c|c|c|} 
        \hline
        Ions & $V_1$ & $\sigma_{V_1}$ &  $V_{100}$ & $\sigma_{V_{100}}$ & $V_{500}$ & $\sigma_{V_{500}}$\\ 
        \hline
      CC, 2, c  &  0.04 &0.63  &  0.01 &0.18  &  0.00 &0.00\\
      ArKCl, 1.76, c  &  0.65 &3.18  &  0.34 &1.87  &  0.04 &0.51\\
      AgAg, 1.58, c  &  4.91 &10.50  &  3.41 &7.27  &  1.03 &3.23\\
      AuAu, 1.23, c  &  14.5 &20.3  &  11.0 &15.5  &  4.02 &7.63\\
      AuAu, 1.23, s  &  9.0 &16.2  &  6.2 &11.6  &  1.57 &4.65\\
      AuAu, 1.23, p  &  3.02 &9.59  &  1.62 &5.92  &  0.17 &1.39\\
      AuAu, 3.4, c  &  12.0 &17.2  &  8.9 &12.2  &  4.06 &6.74\\
      AuAu, 8, c  &  11.4 &16.6  &  8.0 &10.7  &  3.23 &5.11\\
      AuAu, 12, c  &  11.5 &16.9  &  7.8 &10.6  &  2.91 &4.60\\
      AuAu, 7.7, c  &  13.4 &19.1  &  8.95 &12.4  &  3.20 &5.01\\
        \hline
    \end{tabular}
    \caption{Results for $X_{ebe} < 0.3$ and $Y_{ebe} < 0.1$.}
    \label{table:X03Y01}
\end{table}

\begin{table}
    \centering
    \begin{tabular}{|l|c|c|c|c|c|c|} 
        \hline
        Ions & $V_1$ & $\sigma_{V_1}$ &  $V_{100}$ & $\sigma_{V_{100}}$ & $V_{500}$ & $\sigma_{V_{500}}$\\ 
        \hline\vspace*{1mm}
      CC, 2, c  &  0.90 &3.64  &  0.22 &1.43  &  0.00 &0.03\\
      ArKCl, 1.76, c &  16.5 &20.4  &  8.3 &12.1  &  0.89 &2.99\\
      AgAg, 1.58, c &  122 &72  &  81.6 &52.5  &  23.4 &23.5\\
      AuAu, 1.23, c  &  349 &144  &  257 &115  &  88.3 &57.9\\
      AuAu, 1.23, s  &  223 &118  &  148 & 90  &  34.6 &36.4\\
      AuAu, 1.23, p  &  75.7 &72.3  &  39.5 &47.5  &  3.8 &10.3\\
      AuAu, 3.4, c  &  287 &114  &  205 &86  &  86.8 &49.2\\
      AuAu, 8, c  &  275 &102  &  186 &71  &  70.4 &36.2\\
      AuAu, 12, c  &  278 &101  &  183 &67  &  63.0 &31.3\\
      AuAu, 7.7, c  &  329 &114  &  213 &77  &  70.7 &33.3\\
        \hline
    \end{tabular}
    \caption{Results for $X_{ebe} < 0.3$ and $Y_{ebe} < 0.3$.}
    \label{table:X03Y03}
\end{table}

\begin{table}
    \centering
    \begin{tabular}{|l|c|c|c|c|c|c|} 
        \hline
        Ions & $V_1$ & $\sigma_{V_1}$ &  $V_{100}$ & $\sigma_{V_{100}}$ & $V_{500}$ & $\sigma_{V_{500}}$\\ 
        \hline
   CC, 2, c  &  12.0 &19.6  &  2.70 &7.00  &  0.01 &0.24\\
   ArKCl, 1.76, c  &  189 &105  &  85.7 &59.9  &  8.58 &13.1\\
   AgAg, 1.58, c  &  1191 &323  &  709 &228  &  170 &86\\
   AuAu, 1.23, c  &  3134 &625  &  2041 &467  &  568 &200\\
   AuAu, 1.23, s  &  2186 &626  &  1287 &451  &  241 &153\\
   AuAu, 1.23, p  &  832 &506  &  388 &302 &  30.8 &48.5\\
   AuAu, 3.4, c  &  2638 &505  &  1654 &364  &  588 &181\\
   AuAu, 8, c  &  2626 &456  &  1551 &305  &  487 &138\\
   AuAu, 12, c  &  2722 &458  &  1568 &297  &  451 &122\\
   AuAu, 7.7, c  &  3231 &525  &  1853 &345  &  521 &134\\
    \hline
    \end{tabular}
    \caption{Results for $X_{ebe} < 0.5$ and $Y_{ebe} < 0.5$.}
    \label{table:X05Y05}
\end{table}

\begin{table}
    \centering
    \begin{tabular}{|l|c|c|c|} 
        \hline
        Ions & $V_1$ &  $V_{100}$ & $V_{500}$\\ 
      CC, 2, c &       788 &         0 &         0\\
      ArKCl, 1.76, c &      5371 &       253 &         0\\
      AgAg, 1.58, c &     11736 &      2597 &        63\\
      AuAu, 1.23, c &     14674 &      5768 &       269\\
      AuAu, 1.23, s &     20367 &      4258 &        15\\
      AuAu, 1.23, p &     15422 &       874 &         0\\
      AuAu, 3.4, c &     21650 &      4393 &       504\\
      AuAu, 8, c &     37773 &      3436 &       444\\
      AuAu, 12, c &     46736 &      3504 &       400\\
      AuAu, 7.7, c &     56812 &      3205 &       354\\
        \hline
    \end{tabular}
    \caption{Results for $X < 0.1$ and $Y < 0.1$.}
    \label{table:AX01Y01}
\end{table}

\begin{table}
    \centering
    \begin{tabular}{|l|c|c|c|} 
        \hline
        Ions & $V_1$ &  $V_{100}$ & $V_{500}$\\ 
        \hline                                   
       CC, 2, c &      3670 &         0 &         0\\
       ArKCl, 1.76, c &     15632 &       314 &         0\\
       AgAg, 1.58, c &     35262 &      3118 &        63\\
       AuAu, 1.23, c &     45257 &      7067 &       269\\
       AuAu, 1.23, s &     53572 &      5782 &        15\\
       AuAu, 1.23, p &     44426 &      1514 &         0\\
       AuAu, 3.4, c &     86487 &      6120 &       529\\
       AuAu, 8, c &    143924 &      5270 &       468\\
       AuAu, 12, c &    164281 &      5134 &       424\\
       AuAu, 7.7, c &    189200 &      4678 &       366\\
        \hline
    \end{tabular}
        \caption{Results for $X < 0.1$ and $Y < 0.3$.}
    \label{table:AX01Y03}
\end{table}

\begin{table}
    \centering
    \begin{tabular}{|l|c|c|c|} 
        \hline
        Ions & $V_1$ &  $V_{100}$ & $V_{500}$\\
        \hline
        CC, 2, c &      1630 &         0 &         0\\
        ArKCl, 1.76, c &     11410 &       540 &         1\\
        AgAg, 1.58, c &     40814 &      3860 &       294\\
        AuAu, 1.23, c &     72740 &     10778 &      1248\\
        AuAu, 1.23, s &     59498 &      6881 &       452\\
        AuAu, 1.23, p &     36399 &      1698 &         0\\
        AuAu, 3.4, c &     77057 &      8301 &       919\\
        AuAu, 8, c &     96920 &      9641 &       760\\
        AuAu, 12, c &    108759 &     10696 &       742\\
        AuAu, 7.7, c &    129356 &     13201 &       888\\
        \hline
    \end{tabular}
    \caption{Results for $X < 0.3$ and $Y < 0.1$. }
    \label{table:AX03Y01}
\end{table}

\begin{table}
    \centering
    \begin{tabular}{|l|c|c|c|} 
        \hline
        Ions & $V_1$ &  $V_{100}$ & $V_{500}$\\ 
        \hline
       C+C, 2, centr. &      8891 &         0 &         0\\
       ArKCl, 1.76, c &     40262 &       733 &         1\\
       AgAg, 1.58, c &    138436 &      4760 &       294\\
       AuAu, 1.23, c &    238815 &     12736 &      1248\\
       AuAu, 1.23, s &    207558 &      9630 &       452\\
       AuAu, 1.23, p &    131517 &      3225 &         0\\
       AuAu, 3.4, c &    315494 &     11140 &      1006\\
       AuAu, 8, c &    403849 &     12911 &       930\\
       AuAu, 12, c &    436772 &     13868 &       908\\
       AuAu, 7.7, c &    496124 &     16754 &      1046\\
        \hline
    \end{tabular}
    \caption{Results for $X < 0.3$ and $Y < 0.3$. }
    \label{table:AX03Y03}
\end{table}

\begin{table}
    \centering
    \begin{tabular}{|l|c|c|c|} 
        \hline
        Ions & $V_1$ &  $V_{100}$ & $V_{500}$\\ 
      \hline
     C+C, 2, centr. &     11350 &        26 &         0\\
     ArKCl, 1.76, c &     49164 &      1028 &        16\\
     AgAg, 1.58, c &    181903 &      6040 &       575\\
     AuAu, 1.23, c &    343389 &     15751 &      1566\\
     AuAu, 1.23, s &    279168 &     11982 &       639\\
     AuAu, 1.23, p &    163872 &      4635 &         8\\
     AuAu, 3.4, c &    468022 &     12386 &      1496\\
     AuAu, 8, c &    600408 &     13939 &      1276\\
     AuAu, 12, c &    642704 &     14988 &      1248\\
    AuAu, 7.7, c &    716222 &     18220 &      1486\\
        \hline
    \end{tabular}
    \caption{Results for $X < 0.5$ and $Y < 0.5$. }
    \label{table:AX05Y05}
\end{table}

\bibliographystyle{spphys}
\bibliography{bibliography}

\end{document}